\documentclass[elsart-num]{elsart}
\bibstyle{elsart-num}
\usepackage[numbers,square,sort&compress,comma]{natbib}
\usepackage{graphicx}
\usepackage{bm}
\usepackage{amsmath}
\usepackage{amsfonts}
\newlength{\twocolumnwidth}

\newlength{\auxlv}

\begin{document}

\begin{frontmatter}

\title{Causal signal transmission by quantum fields.%
\\ V: Generalised Keldysh rotations and electromagnetic response of the Dirac sea.%
}

\author[a1]{L.\ I.\ Plimak} 
\author[a1,a2,a3]{and S.\ Stenholm} 
\address[a1]{Institut f\"ur Quantenphysik, Universit\"at Ulm, 89069 Ulm, Germany.} 
\address[a2]{Physics Department, Royal Institute of Technology, KTH, Stockholm, Sweden.} 
\address[a3]{Laboratory of Computational Engineering, HUT, Espoo, Finland.} 

\begin{abstract}
The connection between real-time quantum field theory (RTQFT) [see, e.g., A.\ Kamenev and A.\ Levchenko, Advances in Physics {58} (2009) 197] and phase-space techniques [E.\ Wolf and L.\ Mandel, {\em Optical Coherence and Quantum Optics\/} (Cambridge, 1995)] is investigated. 
The Keldysh rotation that forms the basis of RTQFT is shown to be a phase-space mapping of the quantum system based on the symmetric (Weyl) ordering. Following this observation, we define generalised Keldysh rotations based on the class of operator orderings introduced by Cahill and Glauber [Phys.\ Rev.\ {177} (1969) 1882]. Each rotation is a phase-space mapping, generalising the corresponding ordering from free to interacting fields. In particular, response transformation [L.P.\ and S.S., Ann.\ Phys. (N.Y.) {323} (2008) 1989] extends the normal ordering of free-field operators to the time-normal ordering of Heisenberg\ operators. Structural properties of the response transformation, such as its association with the nonlinear quantum response problem and the related causality properties, hold for all generalised Keldysh rotations. 

Furthermore, we argue that response transformation is especially suited for RTQFT formulation of spatial, in particular, relativistic, problems, because it extends cancellation of zero-point fluctuations, characteristic of the normal ordering, to interacting fields. As an example, we consider quantised electromagnetic\ field in the Dirac sea. In the time-normally-ordered representation, dynamics of the field looks essentially classical (fields radiated by currents), without any contribution from zero-point fluctuations. For comparison, we calculate zero-point fluctuations of the interacting electromagnetic\ field under orderings other than time-normal. The resulting expression is physically inconsistent: it does not obey the Lorentz condition, nor Maxwell's equations. 

\end{abstract}


\end{frontmatter}
\newlength{\pwd} 
\settowidth{\pwd}{$/$} 
\section{Introduction}\label{ch:IntV}
In this paper we continue our investigation of dynamical response properties of quantum systems. In papers \cite{API,APII,APIII}, we introduced {\em response transformation\/} of quantum kinematics. In paper \cite{WickCaus}, response transformation was extended to the key technical tool of quantum field theory (QFT), Wick's theorem \cite{Wick,Hori,Schweber,VasF}. 
The next step is to apply it to the standard perturbative techniques of QFT \cite{Schweber,SchwingerC,Perel,Keldysh}. As a practically important example we consider electromagnetic interactions of light and matter. 

Papers \cite{API,APII,APIII} were intended predominantly for the quantum-optical community; our goal was in particular to ``market'' QFT methods to quantum opticians. Here our goal is the opposite: we wish to apply wisdom acquired in quantum optics to QFT. The result of this paper in a nutshell is that, firstly, {\em the nonequilibrium real-time QFT is nothing but the nonlinear quantum response problem formulated in phase-space terms\/}, and, secondly, that {\em the most natural physical picture emerges if using the phase-space mapping based on the so-called time-normal operator ordering\/} \cite{KelleyKleiner,GlauberTN,MandelWolf,APII}. Moreover, in relativistic quantum electrodynamics\ (QED), mappings based on other orderings (e.g., the Keldysh rotation \cite{Keldysh,KamenevLevchenko}) lead to inconsistencies, due to one's well-known inability to impose the Lorentz condition on the operator of the electromagnetic\ potential. Imposing this condition on quantum states of the electromagnetic\ field \cite{Gupta,Bleuler} is not sufficient to cancel unphysical contributions to its {\em fluctuations\/}, except in the time-normally-ordered representation (termed in \cite{API,APII,APIII} response representation). 

Both backgrounds (real-time QFT and phase-space techniques) are well covered in the literature. Relatively recent review articles on the real-time QFT are those of Kamenev and Levchenko \cite{KamenevLevchenko} and Rammer and Smith \cite{RammerSmith}, see also \cite{HaugJauho}. 
An excellent introduction into the topic remains Kadanoff-Baym's classic \cite{KadanoffBaym}, see also \cite{KuboTodaHashitsumeII}. 
The so-called thermofield dynamics, with emphasis on the relation between real-time and imaginary-time (Matsubara's \cite{Matsubara}) techniques, is summarised in monographs of le Bellac \cite{Bellac} and Umezawa, Matsumoto and Tachiki \cite{UmezawaMatsumotoTachiki}. A covariant formulation was introduced by Niemi and Semenoff \cite{NiemiSemenoffAnnPh,NiemiSemenoffNucPh}. Functional techniques was employed in \cite{SemenoffUmezawa,Bellac}. The Dyson equation for the retarded propagator was derived, e.g., by Aurenche and Becherrawy \cite{Therm1}. 
Furthermore, a general introduction to phase-space concepts may be found in the monograph of Mandel and Wolf \cite{MandelWolf}; for a discussion concentrated on the symmetric (Weyl's) operator ordering see the text \cite{Schleich}. The connection between the closed-time-loop formulation and Weyl's ordering has been made by a number of authors, see, e.g., \cite{BuotJensen,PolkReview} and references therein. An attempt to apply methods of real-time QFT to quantum optics was made
by Vinogradov and Stenholm \cite{VinogradovStenholm}. 
Generalisation of the conventional time-normal operator ordering \cite{KelleyKleiner,GlauberTN,MandelWolf} beyond the resonance approximation, making it applicable in relativity, was introduced in \cite{APII} for bosons and \cite{APIII} for fermions.

Conventional real-time QFT \cite{%
MartinSchwinger,KuboIrrevI,KadanoffBaym,KamenevLevchenko,%
RammerSmith,UmezawaMatsumotoTachiki,Bellac,vanLeeuwenEtAl,%
LandsmanVanWeert} relies on the {\em closed-time-loop\ formulation\/} \cite{SchwingerC,Perel,Keldysh} as a general framework, the {\em Keldysh rotation\/} \cite{Keldysh,KamenevLevchenko,RammerSmith} as means of introducing space-time field propagation, and {\em Dyson equations\/} for relevant Green functions solved under suitable approximations as a computational tool. This schedule is adhered to more or less closely in more recent publications \cite{KamenevLevchenko,RammerSmith,Therm1,UmezawaMatsumotoTachiki,Bellac,vanLeeuwenEtAl,LandsmanVanWeert}, but may be only implicit in older ones \cite{MartinSchwinger,KuboIrrevI,KadanoffBaym}. 

In this paper we also adhere to this schedule so far as the closed-time-loop\ techniques and Dyson equations are concerned, while revising the Keldysh rotation. The key observation is that the latter may be alternatively defined as a {\em phase-space mapping\/} based on the symmetric, or Weyl's, ordering of the creation and annihilation operators. An extension of the Keldysh rotation to a class of generalised rotations associated with a class of operator orderings then suggests itself. This class contains, in particular, the conventional Keldysh rotation which is a mapping based on the symmetric ordering, and the {\em response transformation\/} of the closed-time-loop\ techniques \cite{API,APII,APIII}, which is a generalised rotation (mapping) based on the normal ordering \cite{Schweber,Bogol,Itzykson}. 

The normal ordering is the ordering of choice for quantisation of free fields \cite{Itzykson,Schweber,Bogol}, because it warrants zero quantum numbers of vacuum (energy, momentum, etc.). One may say that it warrants an empty quantum vacuum free of ``zero-point fluctuations'', analogous to vacua of classical theories \cite{endZero}. 
Response transformation, which is a generalisation of the normal ordering from free to Heisenberg\ operators, extends this special property of the normal ordering to interacting fields. 

A word of caution is in place here. Since an ordering applied to a single operator leaves this operator intact, everything associated with the average field does not depend on the ordering. In particular, formulae for linear (Kubo's) and nonlinear response functions are shared by all generalised Keldysh rotations. The difference is only in the definition of quantum noise of interacting systems. {\em Under the response transformation, zero-point fluctuations do not contribute to the quantum noise.\/} Verification of this statement as a general theorem will be a subject of forthcoming papers. Here, we confine our attention to a (relatively) simple example: interaction of quantised electromagnetic\ field with the Dirac vacuum in the first nonvanishing order of perturbation theory. We show that response transformation maps this quantum problem one-to-one to a semiclassical problem of propagation of a classical (c-number) field in a quantum medium. The Dirac vacuum enters the theory {\em only\/} through its linear response (linear susceptibility). Zero-point fluctuations in the Dirac vacuum are eliminated. This ``simple example'' involves divergences and renormalizations, allowing us to demonstrate that these concepts are consistent with both the semiclassical viewpoint and the response transformation. A clear advantage is the possibility to isolate, through ``tuned'' regularizations, a self-contained physical ``subtheory'' without divergences. 

Papers of the present authors \cite{Corresp,Feedback,BWO} aside, the only work we are aware of, that can be seen as a predecessor of this investigation, is the article of 
Aurenche and Becherrawy \cite{Therm1}. These authors replace the Keldysh rotation by diagonalisation of the \mbox{$2\times 2$} matrix propagator. Our approach differs from that of Aurenche and Becherrawy by its phase-space origins, by the emphasis on response, operator orderings and classical connotations, and by the use of one retarded propagator rather than a retarded-advanced pair. Nonetheless similarities between this paper and \cite{Therm1} are striking. Both look at restructuring of the Keldysh series following redefinition of the propagator, with emergence of new types of vertices. One also finds separation of the frequency-positive and frequency-negative\ parts as a technical tool, cf.\ Eqs.\ (A3), (A4) in appendix A in \cite{Therm1}. However, different models (spinor QED here {\em versus\/} scalar field theories with self-action in \cite{Therm1}) make direct comparison of our results to \cite{Therm1} somewhat difficult. Such comparison will be a subject of a separate paper. 

The present paper comprises three logical parts, which are in turn split between the main body and the appendices. The first part (\mbox{Secs \ref{ch:BrS}}--\ref{ch:TX} and appendices \ref{ch:OS}, \ref{ch:T}) deals with the formal techniques. Here we establish the connection of our approach with the real-time QFT, and generalise results of papers \cite{API,APII,APIII} beyond the time-normal operator ordering. We start from a brief summary in \mbox{Sec.\ \ref{ch:BrS}}. 
In \mbox{Sec.\ \ref{ch:A}}, we establish the connection between orderings of free-field operators and generalised Keldysh rotations. 
In \mbox{Sec.\ \ref{ch:TX}}, generalised Keldysh rotations and the corresponding operator orderings are introduced for interacting fields. 
In appendices \ref{ch:OS} and \ref{ch:T}, analyses of \mbox{Secs \ref{ch:A}} and \ref{ch:TX} are generalised beyond Gaussian systems. 

General implications of our analyses are the subject of the second part of the paper (\mbox{Secs \ref{ch:QN}}--\ref{ch:M} and appendix \ref{ch:ZZZ}).
In \mbox{Sec.\ \ref{ch:QN}}, we define ``quantum noise'' of quantised fields, and discuss its classical connotations. 
The latter are put to use in \mbox{Sec.\ \ref{ch:M}}, where the key dynamical results of the paper are formulated in a conjectural way, following the analogy between classical stochasticity and ``quantum noise'' in response representation. 
Verification of conjectures of \mbox{Sec.\ \ref{ch:M}}, which involves tedious analyses of diagrammatic structures, is the subject of appendix \ref{ch:ZZZ}. 

In the third part of the paper (\mbox{Sec. \ref{ch:RD}} and appendices \ref{ch:FAB}, \ref{ch:O}), we concern ourselves with the ``simple example''---quantised electromagnetic field in the Dirac sea. 
We calculate the linear susceptibility of the Dirac vacuum, including the unavoidable renormalisation, and show consistency of our results with more traditional approaches. We also calculate the ``quantum noise'' of the electromagnetic field, and show that in the time-normal representation it vanishes, while an attempt to use other orderings leads to inconsistencies.
In appendix \ref{ch:FAB}, we touch upon the choice of signs and dimensional factors in formulae (we use SI units), and rederive response transformations of Green functions of free fields by conventional QFT means (integrals in the complex plane of energy). 
Appendix \ref{ch:O} outlines details of the calculations omitted in \mbox{Sec. \ref{ch:RD}}. 
\section{Quick summary}%
\label{ch:BrS}
We assume familiarity of the reader with basics of phase-space and closed-time-loop\ techniques, including the concepts of symmetric (Weyl's), time and reverse-time orderings \cite{MandelWolf,Itzykson,Bogol,Schweber,SchwingerC,Perel,Keldysh,KamenevLevchenko,RammerSmith}. These orderings are denoted as \mbox{$W$}, \mbox{$T_+$} and \mbox{$T_-$}, respectively (the \mbox{$T_{\pm}$} orderings are often denoted \mbox{$T$} and \mbox{$\bar T$}). For formal definitions see \mbox{Secs \ref{ch:G}} and \ref{ch:AA} below. 
 
Let \mbox{$\hat q(t)$} be an arbitrary Hermitian bosonic operator, free or Heisenberg\ one; we omit all its arguments except time. In the closed-time-loop\ formalism, one defines the kernels, 
\begin{align} 
\begin{aligned} 
\protect\big \langle 
\hat q(t)\hat q(t')
 \big \rangle, \quad 
\protect\big \langle 
T_{\pm}\hat q(t)\hat q(t')
 \big \rangle, 
\end{aligned} 
\label{eq:8VU} 
\end{align}%
where the averaging is over the Heisenberg\ state of the field. The Keldysh rotation \cite{Keldysh,KamenevLevchenko,RammerSmith} replaces them by two kernels: the average retarded commutator, 
\begin{align} 
\begin{aligned} 
\protect\big \langle 
T_+\hat q(t)\hat q(t')
 \big \rangle - \protect\big \langle 
\hat q(t')\hat q(t)
 \big \rangle &= \protect\big \langle 
\hat q(t)\hat q(t')
 \big \rangle - \protect\big \langle 
T_-\hat q(t)\hat q(t')
 \big \rangle \\ &= \theta(t-t')\protect\big \langle 
\protect\big [ 
\hat q(t),\hat q(t')
 \big ] 
 \big \rangle, 
\end{aligned} 
\label{eq:59PK} 
\end{align}%
and the average anticommutator, 
\begin{align} 
\begin{aligned} 
\protect\big \langle 
T_+\hat q(t)\hat q(t')
 \big \rangle + \protect\big \langle 
T_-\hat q(t)\hat q(t')
 \big \rangle &= \protect\big \langle 
\hat q(t)\hat q(t')
 \big \rangle + \protect\big \langle 
\hat q(t)\hat q(t')
 \big \rangle \\ &= \protect\big \langle 
\hat q(t)\hat q(t')
+\hat q(t')\hat q(t)
 \big \rangle . 
\end{aligned} 
\label{eq:89LK} 
\end{align}%
The retarded commutator is associated with linear response of the field \cite{KuboIrrevI} and thus affords a macroscopic interpretation, but what about the anticommutator?

The key formal observation is that, in the interaction picture, the anticommutator coincides with the symmetrically (Weyl) ordered product of free-field operators, 
\begin{align} 
\begin{aligned} 
\hat q(t)\hat q(t')
+\hat q(t')\hat q(t) = 2W\hat q(t)\hat q(t') \quad \textrm{(free field)}. 
\end{aligned} 
\label{eq:13AH} 
\end{align}%
This relation is nontrivial, because Weyl's ordering is defined not for field operators, but for the underlying creation and annihilation operators. 

Equation (\protect\ref{eq:13AH}) raises two questions: 
\begin{itemize}
\item
Can one define generalised Keldysh rotations with other types of ordering of the creation and annihilation operators\ in place of Weyl's one? For instance, with the normal ordering?
Can generalised Keldysh rotations be extended to Heisenberg\ operators?
\item
What are the properties of emerging representations of QFT?
\end{itemize}
The answer to the first question is unqualified ``yes''. To the second question we provide two partial answers: 
\begin{itemize}
\item
Response transformation analysed in \cite{API,APII,APIII} is a generalised Keldysh rotation related to the normal ordering. Its structural properties, in particular, the association with the response problem and causality, extend to all generalised Keldysh rotations. 
\item
Within the simple model mentioned in the introduction, the special property of the response transformation is elimination of zero-point fluctuations from quantum dynamics.
\end{itemize}
The rest remains subject to further work. 

\section{Operator orderings and generalised Keldysh rotations for free-field operators}%
\label{ch:A}
\subsection{Time and closed-time-loop operator ordering}%
\label{ch:G}
Here we summarize definitions of the time and closed-time-loop operator orderings, using this opportunity to introduce notation. Let \mbox{$\protect{\hat{\mathcal X}}_1(t),\protect{\hat{\mathcal X}}_2(t),\cdots,\protect{\hat{\mathcal X}}_m(t)$} be arbitrary bosonic operators, and \mbox{$t_1>t_2>\cdots >t_m$}. Then, 
\begin{align} 
\begin{aligned} 
T_+\protect{\hat{\mathcal X}}_1(t_1)\protect{\hat{\mathcal X}}_2(t_2)\cdots\protect{\hat{\mathcal X}}_m(t_m) 
= \protect{\hat{\mathcal X}}_1(t_1)\protect{\hat{\mathcal X}}_2(t_2)\cdots\protect{\hat{\mathcal X}}_m(t_m), 
\\ 
T_-\protect{\hat{\mathcal X}}_1(t_1)\protect{\hat{\mathcal X}}_2(t_2)\cdots\protect{\hat{\mathcal X}}_m(t_m) 
= \protect{\hat{\mathcal X}}_m(t_m)\cdots\protect{\hat{\mathcal X}}_2(t_2)\protect{\hat{\mathcal X}}_1(t_1). 
\end{aligned} 
\label{eq:40XD} 
\end{align}%
This definition is extended to arbitrary time sequences assuming that bosonic operators commute under the orderings. Furthermore, the Schwinger-Perel-Keldysh {\em closed-time-loop ordering\/} \cite{SchwingerC,Perel,Keldysh,KamenevLevchenko}, denoted $T_C$, is defined as a {\em double time-ordered\/} structure, 
\begin{align} 
 &T_C{\hat{\mathcal X}}_{1-}(t_1)\cdots{\hat{\mathcal X}}_{m-}(t_m) 
{\hat{\mathcal Y}}_{1+}(t'_1)\cdots{\hat{\mathcal Y}}_{n+}(t'_n) 
\nonumber\\ &\quad
= 
T_-{\hat{\mathcal X}}_1(t_1)\cdots{\hat{\mathcal X}}_m(t_m) \, 
T_+{\hat{\mathcal Y}}_1(t'_1)\cdots{\hat{\mathcal Y}}_n(t'_n) . 
\label{eq:10BU} 
\end{align}%
Rather than visually keeping the operators under the $T_{\pm}$-orderings, one marks the operators with the ${}_{\pm}$ indices and allows them to commute freely. These indices serve only for ordering purposes and otherwise should be disregarded. Note that we put the C-contour indices into subscripts (unlike, e.g., Kamenev and Levchenko \cite{KamenevLevchenko}), because superscripts are reserved for other purposes. 
\subsection{The Cahill-Glauber ordering of creation and annihilation operators}%
\label{ch:AA}
For simplicity, consider an harmonic oscillator with frequency \mbox{$
\omega_0 
$}, equipped with the standard bosonic creation/annihilation pair, 
\begin{align} 
\begin{aligned} 
\protect\big [ 
\hat a,\hat a^{\dag}
 \big ] = 1 . 
\end{aligned} 
\label{eq:87LH} 
\end{align}%
According to Cahill and Glauber, {\em ordered products\/} of \mbox{$\hat a,\hat a^{\dag}$} are defined postulating the operator-valued generating function \cite{CahillGlauber,AgarwalWolfI}, 
\begin{align} 
\begin{aligned} 
O_s \exp\big(
\beta \hat a^{\dag}- \beta ^*\hat a
 \big) 
 &= \exp\bigg(
\beta \hat a^{\dag}- \beta ^*\hat a + \frac{s|\beta |^2}{2}
 \bigg) 
, 
\end{aligned} 
\label{eq:23MV} 
\end{align}%
where \mbox{$
\beta 
$} is an auxiliary complex variable and \mbox{$
-1\leq s \leq 1
$} is a real parameter. We shall also have numerous opportunities to use the parameters, 
\begin{align} 
\begin{aligned} 
s_{\pm} = \frac{1\pm s}{2} . 
\end{aligned} 
\label{eq:82QK} 
\end{align}%
Then, 
\begin{align} 
\begin{aligned} 
O_s\hat a &= \hat a, \quad 
O_s\hat a^{\dag}= \hat a^{\dag}, \quad 
O_s\hat a^2 = \hat a^2, \quad 
O_s\hat a^{\dag 2} = \hat a^{\dag 2}, \quad 
O_s\hat a\hat a^{\dag}= 
s_+\hat a^{\dag}\hat a
+
s_-\hat a\hat a^{\dag}
, 
\end{aligned} 
\label{eq:24MW} 
\end{align}%
etc. For \mbox{$s=0,1,-1$} we recover, respectively, the symmetric, or Weyl's, the normal and the antinormal orderings, 
\begin{align} 
\begin{aligned} 
O_0\hat a\hat a^{\dag} &= (1/2)\big(
\hat a^{\dag}\hat a
+
\hat a\hat a^{\dag}
 \big) = W \hat a\hat a^{\dag}
 & &(\mathrm{symmetric}),
\\ 
O_1\hat a\hat a^{\dag} &= \hat a^{\dag}\hat a = {\bf :}
\hat a\hat a^{\dag}
{\bf :} 
 & &(\mathrm{normal}),
\\ 
O_{-1}\hat a\hat a^{\dag} &= \hat a\hat a^{\dag}
= A \hat a\hat a^{\dag}
 & &(\mathrm{antinormal}).
\end{aligned} 
\label{eq:2RE} 
\end{align}%
Of special interest to us will be the symmetric and the normal orderings. 
\subsection{The Cahill-Glauber ordering of free-field operators}%
\label{ch:AF}
By definition, orderings (\ref{eq:23MV}) are extended to products of free-field operators by linearity. As a generic example, consider the displacement operator, 
\begin{align} 
\begin{aligned} 
\hat q(t) = \sqrt{\frac{\hbar }{2}}\big(
\hat a\text{e}^{-i\omega_0t} + 
\hat a^{\dag}\text{e}^{i\omega_0t}
 \big) . 
\end{aligned} 
\label{eq:1RD} 
\end{align}%
This operator is a Heisenberg\ one with respect to the free Hamiltonian, 
\begin{align} 
\begin{aligned} 
\hat H_0 = \hbar \omega _0\hat a ^{\dag}\hat a . 
\end{aligned} 
\label{eq:9RN} 
\end{align}%
Then, e.g.,
\begin{align} 
\begin{aligned} 
 &O_s\hat q(t)\hat q(t') 
= \frac{\hbar }{2}\protect\big [ 
\hat a^2\text{e}^{-i\omega (t+t')} + \hat a^{\dag 2}\text{e}^{i\omega (t+t')} 
+ 2O_s
\hat a\hat a^{\dag}
\cos \omega (t-t') 
 \big ] , 
\end{aligned} 
\label{eq:22SB} 
\end{align}%
etc. Of use to us will be the formula, 
\begin{align} 
\begin{aligned} 
O_{s}\hat q(t)\hat q(t')
- 
O_{s'}\hat q(t)\hat q(t') = \frac{\hbar (s'-s)}{2}\cos\omega _0(t-t') . 
\end{aligned} 
\label{eq:70PW} 
\end{align}%
It readily follows from Eqs.\ (\protect\ref{eq:24MW}) and (\ref{eq:22SB}). 

\subsection{The Keldysh rotation and the symmetric ordering}%
\label{ch:GKB}
Assume that the oscillator is in a Gaussian (thermal, squeezed, etc.) state with zero average displacement,
\begin{align} 
\begin{aligned} 
\protect\big \langle 
\hat q(t)
 \big \rangle = 0 . 
\end{aligned} 
\label{eq:61PM} 
\end{align}%
For an approach free of these limitations see appendix \ref{ch:OS}. 
The closed-time-loop\ formulation of such Gaussian system reduces to kernels (\ref{eq:8VU}), where \mbox{$\hat q(t)$} is now defined by (\ref{eq:1RD}), and the averaging is over the Gaussian state of the oscillator. The same specification applies to Eqs.\ (\protect\ref{eq:59PK}), (\ref{eq:89LK}) for the ``rotated'' kernels. 

The retarded commutator (\ref{eq:59PK}) now reads, 
\begin{align} 
\begin{aligned} 
\theta(t-t')\protect\big \langle 
\protect\big [ 
\hat q(t),\hat q(t')
 \big ] 
 \big \rangle = i\hbar D_{\text{R}}(t-t') , 
\end{aligned} 
\label{eq:23AT} 
\end{align}%
where \mbox{$D_{\text{R}}$} is Kubo's linear response function of the oscillator \cite{KuboIrrevI} (see also \cite{API}). 
The commutator in (\ref{eq:23AT}) is a c-number, making the averaging redundant. This warrants that \mbox{$
D_{\text{R}}
$} does not depend on the state of the oscillator, and that it is a function of time difference; both these properties are artefacts of the linear problem. Explicitly, 
\begin{align} 
\begin{aligned} 
D_{\text{R}}(t-t') &= -\theta(t-t')\sin\omega _0(t-t') 
= \int \frac{d\omega }{2\pi }\text{e}^{-i\omega (t-t')}D_{\mathrm{R}\omega }, 
\end{aligned} 
\label{eq:24WM} 
\end{align}%
where 
\begin{align} 
\begin{aligned} 
D_{\mathrm{R}\omega } = \frac{1}{2}\protect\Big [ 
\frac{1}{\omega -\omega _0 +i0^+} 
- 
\frac{1}{\omega +\omega _0 +i0^+}
 \Big ] . 
\end{aligned} 
\label{eq:34WX} 
\end{align}%
As in papers \cite{API,APII,APIII,WickCaus}, omitted integration limits indicate the maximal possible area of integration: the whole time axis, the whole space, etc. 

Furthermore, for the anticommutator of displacement operators we have, 
\begin{align} 
\begin{aligned} 
 &\hat q(t)\hat q(t')
+\hat q(t')\hat q(t) 
= \hbar \protect\big [ 
\hat a^2\text{e}^{-i\omega (t+t')} + \hat a^{\dag 2}\text{e}^{i\omega (t+t')} 
+ \big(
\hat a\hat a^{\dag}+\hat a^{\dag}\hat a
 \big) \cos \omega (t-t')
 \big ] 
. \end{aligned} 
\label{eq:13RS} 
\end{align}%
Comparing this to Eqs.\ (\protect\ref{eq:24MW}) and (\ref{eq:22SB}) we recover Eq.\ (\protect\ref{eq:89LK}). For free fields in general, Eq.\ (\protect\ref{eq:89LK}) may be verified applying Eq.\ (\protect\ref{eq:13RS}) modewise. 

Equations (\protect\ref{eq:23AT}) and (\ref{eq:13RS}) make it evident that the Keldysh rotation disentangles information about the system response from the information about its quantum state. The latter is expressed in terms of the symmetrically ordered average, which in turn may be written as a c-number quasiaverage over the 
Wigner function \cite{MandelWolf,Schleich}, 
\begin{align} 
\begin{aligned} 
 &\protect\big \langle 
W\hat q(t)\hat q(t')
 \big \rangle = \int d^2\alpha W(\alpha )q_{\mathrm{in}}(t)q_{\mathrm{in}}(t'), 
\end{aligned} 
\label{eq:16RV} 
\end{align}%
 where the classical (c-number) in-field depends on the complex amplitude \mbox{$\alpha $}, 
\begin{align} 
\begin{aligned} 
 &q_{\mathrm{in}}(t) = \sqrt{\frac{\hbar }{2}}\big(
\alpha \text{e}^{-i\omega_0t} + 
\alpha ^*\text{e}^{i\omega_0t}
 \big) . 
\end{aligned} 
\label{eq:20RZ} 
\end{align}%
The integration in (\ref{eq:16RV}) is over the whole complex plane of \mbox{$\alpha $}. Grounds for calling quantity (\ref{eq:20RZ}) {\em classical in-field\/} will become clear in \mbox{Sec.\ \ref{ch:GKD}}, cf.\ Eq.\ (\protect\ref{eq:18MQ}). 

\subsection{The Keldysh rotation as a functional substitution}%
\label{ch:GKS}
Formally, the Keldysh rotation may be introduced as a change of variables in the functional bilinear form, 
\begin{align} 
\begin{aligned} 
 &\Lambda [\eta_+,\eta_-] 
= 
-\frac{1}{2}\eta_+\protect\big \langle T_+ \hat q \hat q \big \rangle\eta_+
-\frac{1}{2}\eta_-\protect\big \langle T_- \hat q \hat q \big \rangle\eta_-
+\eta_-\protect\big \langle \hat q \hat q \big \rangle\eta_+
, 
\end{aligned} 
\label{eq:15MM} 
\end{align}%
where \mbox{$\eta_{\pm}(t)$} are a pair of auxiliary c-number functions. Square brackets signify functional arguments. We use condensed notation, 
\begin{align} 
\begin{aligned} 
fg &= \int dt f(t) g(t), 
\quad
fKg &= \int dt dt'f(t)K(t-t')g(t), 
\end{aligned} 
\label{eq:47XM} 
\end{align}%
where \mbox{$f(t),g(t)$} are c-number or q-number functions and \mbox{$K(t-t')$} is a c-number kernel. The form (\ref{eq:15MM}) contains full information about the oscillator in a Gaussian state. 

In terms of \mbox{$\Lambda $}, the Keldysh rotation (\ref{eq:59PK}), (\ref{eq:89LK}) amounts to the substitution, 
\begin{gather} 
\eta_{\pm}(t) = \frac{j_W(t)}{\hbar }\pm \frac{\eta(t)}{2}. 
\label{eq:61YB} 
\end{gather}%
For \mbox{$\eta ,j_W$} we have, 
\begin{align} 
\eta (t) &= \eta_+(t)-\eta_-(t), 
\label{eq:51BY} 
\\ 
j_W(t) &= \frac{\hbar }{2}\protect\big [ 
\eta_+(t)+\eta_-(t)
 \big ] , 
\label{eq:95LR} 
\end{align}%
In these variables, 
\begin{align} 
\begin{aligned} 
\Lambda [\eta_+,\eta_-] 
\settoheight{\auxlv}{$|$}%
\raisebox{-0.3\auxlv}{$|_{\eta_{\pm}\to\eta ,j_W}$} 
 &= 
-i\eta D_{\text{R}}j_W 
- \frac{1}{2}\eta\protect\big \langle 
W\hat q\hat q
 \big \rangle \eta
, 
\end{aligned} 
\label{eq:17MP} 
\end{align}%
where notice was taken of Eq.\ (\protect\ref{eq:13RS}) for the anticommutator. 

\subsection{Generalised Keldysh rotations}%
\label{ch:GKI}
It suggests itself to generalise the Keldysh rotation to other types of ordering, with the information about the state expressed by the corresponding quasidistributions, 
\begin{align} 
\begin{aligned} 
 &\protect\big \langle 
O_s\hat q(t)\hat q(t')
 \big \rangle = \int d^2\alpha p_s(\alpha )q_{\mathrm{in}}(t) q_{\mathrm{in}}(t'). 
\end{aligned} 
\label{eq:17RW} 
\end{align}%
With \mbox{$s=0,1,-1$} one encounters the Wigner, P and Q-functions \cite{MandelWolf,Schleich,AgarwalWolfI}, 
\begin{align} 
\begin{aligned} 
p_0(\alpha ) = W(\alpha ), \quad 
p_1(\alpha ) = P(\alpha ), \quad 
p_{-1}(\alpha ) = Q(\alpha ). 
\end{aligned} 
\label{eq:18RX} 
\end{align}%
The linear response function, which is insensitive to the ordering, is not changed. 

Following the pattern of Eq.\ (\protect\ref{eq:17MP}), a {\em generalised Keldysh rotation\/} is introduced as a change of variables in the form \mbox{$\Lambda $}, such that, 
\begin{align} 
\begin{aligned} 
\Lambda [\eta_+,\eta_-] 
\settoheight{\auxlv}{$|$}%
\raisebox{-0.3\auxlv}{$|_{\eta_{\pm}\to\eta ,j_s}$} 
 &= 
-i\eta D_{\text{R}}j_s 
- \frac{1}{2}\eta\protect\big \langle 
O_s\hat q\hat q
 \big \rangle \eta
. 
\end{aligned} 
\label{eq:65PR} 
\end{align}%
Rotations differ in variable \mbox{$j_s(t)$}, while variable \mbox{$
\eta(t)
$} is shared by all rotations. We note without going into particulars that this is ultimately due to the property, 
\begin{align} 
\begin{aligned} 
{ 
O_s\hat q(t)
} = { 
T_+\hat q(t)
} = { 
T_-\hat q(t)
} = { 
\hat q(t)
} 
, 
\end{aligned} 
\label{eq:22MU} 
\end{align}%
which holds for all orderings. 
\subsection{Response and reordering of operators}%
\label{ch:RQN}
Before attempting to solve for the variable \mbox{$j_s(t)$}, consider the following question. Can the difference between orderings (\ref{eq:70PW}) be related to some physical property of the field \mbox{$\hat q(t)$}? The natural candidate is the linear response. Indeed, we now show the formula, 
\begin{align} 
\begin{aligned} 
 &\protect\big \langle 
O_s\hat q(t)\hat q(t')
 \big \rangle - 
\protect\big \langle 
O_{s'}\hat q(t)\hat q(t')
 \big \rangle 
= (s'-s) Z(t-t') , 
\end{aligned} 
\label{eq:30BA} 
\end{align}%
where 
\begin{align} 
\begin{aligned} 
Z(t-t') &= (i\hbar/2)\protect\big [ 
D_{\text{R}}^{(+)}(t-t') 
+ 
D_{\text{R}}^{(+)}(t'-t) 
- D_{\text{R}}^{(-)}(t-t') 
- 
D_{\text{R}}^{(-)}(t'-t) 
 \big ] , 
\end{aligned} 
\label{eq:86DM} 
\end{align}%
and \mbox{$D_{\text{R}}^{(\pm)}$} are the frequency-positive and frequency-negative\ parts of \mbox{$D_{\text{R}}$}. 

The operation of separation of the frequency-positive and frequency-negative\ parts of functions plays a central role in our analyses, and we take this opportunity to reiterate its definition and most useful properties. Namely, with \mbox{$f(t)$} being an arbitrary function,
\begin{align} 
f (t) &= f ^{(+)}(t) + f ^{(-)}(t) , 
\label{eq:4JH} 
\\ 
f ^{(\pm)}(t) &= \int\frac{d\omega }{2\pi }\mathrm{e}^{-i\omega t}\theta(\pm\omega )
f _{\omega }, 
\quad 
f _{\omega } = \int dt \mathrm{e}^{i\omega t}f (t) . 
\label{eq:66LW} 
\end{align}%
The ${}^{(\pm)}$ operations are conveniently expressed as integral transformations, 
\begin{align} 
\begin{aligned} 
f^{(\pm)}(t) = \int dt' \delta ^{(\pm)}(t-t') f(t') 
\equiv \mathcal{F}^{(\pm)}_t f(t), 
\end{aligned} 
\label{eq:39KV} 
\end{align}%
where 
\begin{align} 
\begin{aligned} 
\delta ^{(\pm)}(t) = \delta ^{(\mp)}(-t) = \pm\frac{1}{2\pi i(t\mp i0^+)} 
\end{aligned} 
\label{eq:40KW} 
\end{align}%
are the frequency-positive and frequency-negative\ parts of the delta-function. 
Of use will be the formula, 
\begin{align} 
\begin{aligned} 
\mathcal{F}^{(\pm)}_t f(t-t') = \mathcal{F}^{(\mp)}_{t'} f(t-t') , 
\end{aligned} 
\label{eq:39BL} 
\end{align}%
verified by considering its Fourier-transformation. 
A useful observation is also that \mbox{$\mathcal{F}^{(\pm)}_t$} are non-Hermitian orthogonal projectors, 
\begin{align} 
\begin{aligned} 
\mathcal{F}^{(+)}_t+\mathcal{F}^{(-)}_t &= 1, 
\quad {[\mathcal{F}^{(\pm)}_t]}^2 = \mathcal{F}^{(\pm)}_t, \quad
\mathcal{F}^{(+)}_t\mathcal{F}^{(-)}_t = 
\mathcal{F}^{(-)}_t\mathcal{F}^{(+)}_t = 0 .
\end{aligned} 
\label{eq:59XZ} 
\end{align}%
For more details see appendix A in \cite{APII}. Formulae (\ref{eq:4JH})--(\ref{eq:59XZ}) equally apply with \mbox{$
t\to ct=x_0
$}, cf.\ \mbox{Sec.\ \ref{ch:TXC}} below. 
 
As to Eq.\ (\protect\ref{eq:30BA}), it follows from Eq.\ (\protect\ref{eq:70PW}) and the identity, 
\begin{align} 
\begin{aligned} 
 &\cos\omega _0\tau 
= i\protect\big [ 
D_{\text{R}}^{(+)}(\tau ) 
+ 
D_{\text{R}}^{(+)}(-\tau ) - 
D_{\text{R}}^{(-)}(\tau ) 
- 
D_{\text{R}}^{(-)}(-\tau ) 
 \big ] . 
\end{aligned} 
\label{eq:31BB} 
\end{align}%
To prove it, consider its Fourier-transformation, 
\begin{align} 
\begin{aligned} \pi \protect\big [ 
\delta (\omega -\omega _0) + 
\delta (\omega +\omega _0)
 \big ] = i\big(
D_{\mathrm{R}\omega } - 
D_{\mathrm{R}-\omega }
 \big) \operatorname{sign}\omega , 
\end{aligned} 
\label{eq:33BD} 
\end{align}%
which is readily verified by making use of Eq.\ (\protect\ref{eq:34WX}). 

Unlike (\ref{eq:70PW}), Eq.\ (\protect\ref{eq:30BA}) is directly generalised to quantum fields in the true meaning of the word, including interacting ones: its ``Heisenberg'' counterpart is Eq.\ (\protect\ref{eq:26AW}) in \mbox{Sec.\ \ref{ch:QN}} below. As a technical tool, Eq.\ (\protect\ref{eq:30BA}) will be instrumental in deriving generalised Keldysh rotations in \mbox{Sec.\ \ref{ch:GKJ}}. Its physical implications will be discussed in \mbox{Sec.\ \ref{ch:GKX}}. 

\subsection{Explicit formula for $j_{s}(t)$}%
\label{ch:GKJ}
To find \mbox{$j_s(t)$} we subtract Eqs.\ (\protect\ref{eq:17MP}) and (\ref{eq:65PR}). Their left-hand sides (lhs) are by definition identical, although written in different variables. Subtracting them yields a functional equation for \mbox{$
j_s(t)
$}, 
\begin{align} 
\begin{aligned} 
 &
\eta D_{\text{R}}\big(
j_s-j_W
 \big) 
= \frac{s\hbar }{2}\eta D_{\text{R}}\protect\big [ 
\eta^{(+)}-\eta^{(-)}
 \big ]
. 
\end{aligned} 
\label{eq:63PP} 
\end{align}%
The \mbox{$^{(\pm)}$} operations are defined in \mbox{Sec.\ \ref{ch:RQN}}. 
When obtaining Eq.\ (\protect\ref{eq:63PP}), notice was taken of Eq.\ (\protect\ref{eq:30BA}) 
and of the formula, 
\begin{align} 
\begin{aligned} 
\int dt' D_{\text{R}}^{(\pm)}(t-t')\eta (t') = \int dt' D_{\text{R}}(t-t')\eta^{(\pm)}(t'). 
\end{aligned} 
\label{eq:34BE} 
\end{align}%
To verify it, consider its Fourier-transformation. 

Clearly Eq.\ (\protect\ref{eq:63PP}) is satisfied if, 
\begin{align} 
\begin{aligned} 
j_s(t) - j_W(t) = \frac{s\hbar }{2}\protect\big [ 
\eta^{(+)}(t)-\eta^{(-)}(t)
 \big ] . 
\end{aligned} 
\label{eq:25WN} 
\end{align}%
Recalling Eqs.\ (\protect\ref{eq:51BY}), (\ref{eq:95LR}) for \mbox{$\eta $} 
and \mbox{$j_W$} we obtain, 
\begin{align} 
\begin{aligned} 
j_s(t) &= \hbar \protect\big [ 
\eta^{(s+)}_+(t) + \eta^{(s-)}_-(t) 
 \big ] , 
\end{aligned} 
\label{eq:25MX} 
\end{align}%
where the \mbox{${}^{(s\pm)}$} operations are linear combinations of \mbox{${}^{(\pm)}$}, 
\begin{align} 
\begin{aligned} 
f^{(s\pm)}(t) &= s_+ f^{(\pm)}(t) + s_- f^{(\mp)}(t) 
= s f^{(\pm)}(t) + s_- f(t) . 
\end{aligned} 
\label{eq:12VY} 
\end{align}%
The parameters \mbox{$s_{\pm}$} are introduced by Eq.\ (\protect\ref{eq:82QK}). The last formula in (\ref{eq:12VY}) is convenient for transition from the conventional Keldysh rotation ($s=0$, $s_+=s_-=1/2$) to response transformation \cite{API,APII,APIII} ($s=s_+=1$, $s_-=0$). 
 
For the Keldysh rotation, 
Eq.\ (\protect\ref{eq:12VY}) trivialises, 
\begin{align} 
\begin{aligned} 
f^{(0+)}(t)=f^{(0-)}(t)=\frac{1}{2}f(t), 
\end{aligned} 
\label{eq:50BX} 
\end{align}%
and we recover Eq.\ (\protect\ref{eq:95LR}). With this exception, Eqs.\ (\protect\ref{eq:25MX}), (\ref{eq:12VY}) are integral transformations. 

\subsection{Generalised response substitution}%
\label{ch:GKF}
To find the substitution introducing variables \mbox{$\eta,j_s$}---and thus to demonstrate that we deal with a genuine change of functional variables---we break the expressions for them in frequency-positive and frequency-negative\ parts, (omitting arguments for brevity)
\begin{align} 
\eta^{(+)} &= \eta_+^{(+)}- \eta_-^{(+)}, \quad 
j_s^{(+)}= \hbar \protect\big [ 
s_+\eta_+^{(+)}+s_-\eta_-^{(+)}
 \big ] , 
\label{eq:57XX} 
\\ 
\eta^{(-)} &= \eta_+^{(-)}- \eta_-^{(-)}, \quad 
j_s^{(-)}= \hbar \protect\big [ 
s_-\eta_+^{(-)}+s_+\eta_-^{(-)}
 \big ] . 
\label{eq:58XY} 
\end{align}%
These formulae are two independent systems of algebraic equations in the subspaces of frequency-positive and frequency-negative\ functions, cf.\ Eqs.\ (\protect\ref{eq:59XZ}) and comments thereon. Solving Eqs.\ (\protect\ref{eq:57XX}) and (\ref{eq:58XY}) for, respectively, \mbox{$\eta_{\pm}^{(+)}$} and \mbox{$\eta_{\pm}^{(-)}$}, and recombining the frequency-positive and frequency-negative\ parts we find the substitution sought, 
\begin{align} 
\begin{aligned} 
\eta_+(t) = \frac{j_s(t)}{\hbar } + \eta^{(s-)}(t), \quad
\eta_-(t) = \frac{j_s(t)}{\hbar } - \eta^{(s+)}(t) . 
\end{aligned} 
\label{eq:60YA} 
\end{align}%
With the exception of the conventional Keldysh rotation \mbox{($s= 0$)}, these relations are integral transformations. 
They generalise to arbitrary \mbox{$s$} the Keldysh rotation \mbox{($s=0$)} given by Eq.\ (\protect\ref{eq:61YB}), and the response substitution \mbox{($s=1$)}, 
\begin{align} 
\begin{aligned} 
\eta_+(t) = \frac{j_{\mathrm{e}}(t)}{\hbar } + \eta^{(-)}(t), \quad
\eta_-(t) = \frac{j_{\mathrm{e}}(t)}{\hbar } - \eta^{(+)}(t) , 
\end{aligned} 
\label{eq:2LX} 
\end{align}%
introduced in \cite{API,APII,APIII}. 

\section{Generalised Keldysh rotations and time-{\em s\/}-ordered products of Heisenberg\ operators}%
\label{ch:TX}
\subsection{Preliminaries}%
\label{ch:TXA}
In this section, we extend generalised Keldysh rotations to interacting fields. We stay within the Gaussian case (specified shortly). For a general approach see appendix \ref{ch:T}. 

To be specific, we consider the Heisenberg\ operator of the electromagnetic\ 4-potential \mbox{$\protect{\hat{\mathcal A}}_{\nu}(x)$}. As a dynamical quantity it is introduced in \mbox{Sec.\ \ref{ch:CR}} below. All that matters here is that all quantities in the ensuing discussion be defined, directly or by means of a limiting procedure (renormalisation). What follows can readily be adapted to other fields, including fermionic ones. 

We employ conventional 4-vector notation: 
\begin{align} 
\begin{aligned} 
x^{\nu } &= \{x^0,{\bf x}\} = \{ct,{\bf x}\}, &
x_{\nu } &= \{x_0,-{\bf x}\} = \{ct,-{\bf x}\},
\end{aligned} 
\label{eq:48YZ} 
\end{align}%
etc. The metric tensor $g_{\mu \nu }$ is diagonal, with
\begin{align} 
\begin{aligned} 
g_{00}=-g_{11}=-g_{22}=-g_{33}=1. 
\end{aligned} 
\label{eq:49ZA} 
\end{align}%
Summation is implied over pairs of identical co- and contravariant indices. 
 
We use SI units and restore dimensional coefficients omitted in QFT texts---after all, it is pretty awkward to talk about 
quantum-classical correspondences while using units where \mbox{$
\hbar = 1
$}. Dimensions of the 4-vectors of potential and current are chosen to coincide with those of their spatial components: 
\begin{align} 
\protect{\hat{\mathcal A}}^{\nu }(x) = \protect\Big \{ 
\frac{\hat\Phi(x)}{c},\hat{{\bf A}}(x)
 \Big \} , \quad
\protect{\hat{\mathcal J}}^{\nu }(x) = \protect\big \{ 
c\hat\rho(x),\hat{{\bf j}}(x)
 \big \} , 
\label{eq:79WA} 
\end{align}%
where \mbox{$
\hat\Phi(x)
$}, \mbox{$
\hat{{\bf A}}(x)
$}, \mbox{$
\hat\rho(x)
$}, and \mbox{$
\hat{{\bf j}}(x)
$} are, respectively, the scalar and vector potentials and the charge and current densities. More details on the choice of dimensional factors for the electromagnetic\ potential and related quantities may be found in appendix \ref{ch:FA}. 
\subsection{Cumulants of the 4-potential}%
\label{ch:TXB}
We do not assume the average field to be zero, 
\begin{align} 
\begin{aligned} 
\protect\big \langle 
\protect{\hat{\mathcal A}}_{\nu}(x)
 \big \rangle \neq 0 ,
\end{aligned} 
\label{eq:61TU} 
\end{align}%
and therefore have to distinguish the closed-time-loop Green functions and the corresponding cumulants, (identified by commas)
\begin{align} 
\begin{aligned} 
\protect\big \langle 
T_{\pm}\protect{\hat{\mathcal A}}_{\nu}(x)\protect{\hat{\mathcal A}}_{\nu '}(x ')
 \big \rangle 
 &
= \protect\big \langle 
T_{\pm}\protect{\hat{\mathcal A}}_{\nu}(x),\protect{\hat{\mathcal A}}_{\nu '}(x ')
 \big \rangle + \protect\big \langle 
\protect{\hat{\mathcal A}}_{\nu}(x)
 \big \rangle\protect\big \langle 
\protect{\hat{\mathcal A}}_{\nu '}(x ')
 \big \rangle, 
\\ 
\protect\big \langle 
\protect{\hat{\mathcal A}}_{\nu}(x)\protect{\hat{\mathcal A}}_{\nu '}(x ')
 \big \rangle 
 &
= \protect\big \langle 
\protect{\hat{\mathcal A}}_{\nu}(x),\protect{\hat{\mathcal A}}_{\nu '}(x ')
 \big \rangle + \protect\big \langle 
\protect{\hat{\mathcal A}}_{\nu}(x)
 \big \rangle\protect\big \langle 
\protect{\hat{\mathcal A}}_{\nu '}(x ')
 \big \rangle. 
\end{aligned} 
\label{eq:59TS} 
\end{align}%
Generalisation of Eq.\ (\protect\ref{eq:15MM}) to Heisenberg\ fields employs closed-time-loop\ cumulants rather than Green functions, 
\begin{align} 
\begin{aligned} 
\Lambda_2[\eta_+,\eta_-] 
 &= -i(\eta_+-\eta_-)\protect\big \langle 
\protect{\hat{\mathcal A}}
 \big \rangle
\\ &\quad
+\eta_-\protect\big \langle 
\protect{\hat{\mathcal A}},\protect{\hat{\mathcal A}}
 \big \rangle\eta_+ 
-\frac{1}{2}\eta_+\protect\big \langle 
T_{+}\protect{\hat{\mathcal A}},\protect{\hat{\mathcal A}}
 \big \rangle\eta_+
-\frac{1}{2}\eta_-\protect\big \langle 
T_{-}\protect{\hat{\mathcal A}},\protect{\hat{\mathcal A}}
 \big \rangle\eta_- 
, 
\end{aligned} 
\label{eq:60TT} 
\end{align}%
where \mbox{$\eta_{\pm}^{\nu}(x)$} are a pair of auxiliary c-number 4-vector functions \cite{endMayerL}\nocite{tHooft,Mayer}. To emphasise the structure of formulae we use condensed notation, 
\begin{align} 
\begin{aligned} 
fg &= \int d^4x f^{\nu}(x)g_{\nu}(x), 
\quad 
fKg &= \int d^4xd^4x' f^{\nu}(x)K_{\nu \nu '}(x,x')g^{\nu '}(x '), 
\end{aligned} 
\label{eq:62TV} 
\end{align}%
where \mbox{$f^{\nu}(x),g^{\nu}(x)$} are 4-vector functions and \mbox{$K_{\nu \nu '}(x,x')$} is a c-number kernel. The ``Gaussian case'' mentioned above is in the fact that we limit our attention to the one- and two-pole cumulants. {\em It does not constitute a dynamical approximation\/}. 
\subsection{Generalised Keldysh rotations for Heisenberg\ fields}%
\label{ch:TXC}
A generalised Keldysh rotation is defined as a change of variables in the form (\ref{eq:60TT}), 
\begin{align} 
\eta _{+}^{\nu}(x) &= (\hbar c)^{-1} j_{s}^{\nu}(x)
+ \eta^{\nu (s-)}(x) , 
 &
\eta _{-}^{\nu}(x) &= (\hbar c)^{-1} j_{s}^{\nu}(x)
- \eta^{\nu (s+)}(x) . 
\label{eq:27SH} 
\\ 
\eta ^{\nu}(x) &= \eta _+^{\nu}(x)-\eta _-^{\nu}(x), &
j_{s}^{\nu}(x) &= \hbar c\protect\big [ 
\eta ^{\nu (s+)}_+(x)+\eta ^{\nu (s-)}_-(x)
 \big ] . 
\label{eq:28SJ} 
\end{align}%
By definition, the \mbox{${}^{(s\pm)}$} and \mbox{$^{(\pm)}$} operations apply to the time variable, 
\begin{align} 
f^{(s\pm)}(x) &= \mathcal{F}^{(s\pm)}_{x_0}f(x) 
= s_+f^{(\pm)}(x) + s_-f^{(\mp)}(x) 
= s f^{(\pm)}(x) + s_- f(x) . 
\label{eq:10VW} 
\\ 
f^{(\pm)}(x) &= \mathcal{F}^{(\pm)}_{x_0}f(x) 
= \int d^4x' 
\delta^{(\pm)}(x_0-x_0')\delta^{(3)}({\bf x}-{\bf x}')f(x') , 
\label{eq:86UW} 
\end{align}%
where
\begin{align} 
\begin{aligned} 
\delta ^{(\pm)}(x_0-x_0') = \delta ^{(\mp)}(x_0'-x_0) = \pm\frac{1}{2\pi i(x_0-x_0'\mp i0^+)} 
\end{aligned} 
\label{eq:36BH} 
\end{align}%
are the frequency-positive and frequency-negative\ parts of the delta-function, 
cf.\ Eqs.\ (\protect\ref{eq:4JH})--(\ref{eq:40KW}), (\ref{eq:12VY}), and the remark after Eq.\ (\protect\ref{eq:59XZ}). Substitution (\ref{eq:27SH}) is equivalent to introducing variables \mbox{$\eta (t)$} and \mbox{$j_{s}(t)$} of \mbox{Sec.\ \ref{ch:GKF}} modewise. 

The \mbox{${}^{(s\pm)}$} operations inherit the critical properties of \mbox{${}^{(\pm)}$}: 
\begin{align} 
f^{(s+)}(x)+f^{(s-)}(x) &= f(x), 
\label{eq:21AR} 
\\ 
\protect\big [ 
\mathcal{F}^{(s\pm)}_{x_0}f(x)
 \big ]^* &= \mathcal{F}^{(s\mp)}_{x_0}f^*(x) , 
\label{eq:22AS} 
\\ 
\int d^4x f^{(s\pm)}(x) g(x) &= \int d^4x f(x) g^{(s\mp)}(x) . 
\label{eq:11VX} 
\end{align}%
The analyses of \mbox{Refs.\ \protect\cite{APII,APIII}} may therefore be generalised to an arbitrary Keldysh rotation by substituting the \mbox{${}^{(s\pm)}$} operation for \mbox{${}^{(\pm)}$}. It is not difficult to ensure that all results of papers \cite{APII,APIII} pertaining to the ``formally classical structure of quantum response'' persist if replacing response transformation by an arbitrary generalised Keldysh rotation. In particular, causality properties of response transformation in fact hold for arbitrary Keldysh rotations. For details see appendix \ref{ch:T}. In the main body of the paper we confine ourselves to linear media, where the Gaussian case suffices. 
\subsection{Time-{\it s\/}-ordering of operators}%
\label{ch:TXD}
Applying substitution (\ref{eq:27SH}) to \mbox{$\Lambda _2$} we find the ``rotated'' form to be, 
\begin{align} 
\begin{aligned} 
 &\Lambda_2[\eta_+,\eta_-] 
\settoheight{\auxlv}{$|$}%
\raisebox{-0.3\auxlv}{$|_{\eta_{\pm}\to\eta,j_{s}}$} 
= - i\eta \protect\big \langle 
\protect{\hat{\mathcal A}}
 \big \rangle 
- i\eta \mathcal{D}_{\text{R}}j_{s} 
- \frac{1}{2}\eta \protect\big \langle 
\mathcal{T}_s\protect{\hat{\mathcal A}},\protect{\hat{\mathcal A}}
 \big \rangle \eta . 
\end{aligned} 
\label{eq:63TW} 
\end{align}%
It depends on three ``rotated'' cumulants: the {\em average field\/} (\ref{eq:61TU}), 
{\em Kubo's linear response function\/} of the Heisenberg\ field, 
\begin{align} 
\begin{aligned} 
\mathcal{D}_{\mathrm{R}\nu}^{\nu '}(x,x') &= (i\hbar c)^{-1}\theta(x_0-x_0')\protect\big \langle 
\protect\big [ 
\protect{\hat{\mathcal A}}_{\nu }(x),\protect{\hat{\mathcal A}}^{\nu '}(x)
 \big ] 
 \big \rangle , 
\end{aligned} 
\label{eq:64TX} 
\end{align}%
and the {\em time-$s$-ordered cumulant\/}, 
\begin{align} 
\begin{aligned} 
 &\protect\big \langle 
\mathcal{T}_s{\protect{\hat{\mathcal A}}_{\nu}(x),\protect{\hat{\mathcal A}}_{\nu '}(x ')}
 \big \rangle 
\\ &\quad
= 2\Re\protect\big [ 
\mathcal{F}^{(s+)}_{x_0}\mathcal{F}^{(s+)}_{x_0'}\protect\big \langle 
T_{+}\protect{\hat{\mathcal A}}_{\nu}(x),\protect{\hat{\mathcal A}}_{\nu '}(x ')
 \big \rangle
+\mathcal{F}^{(s-)}_{x_0}\mathcal{F}^{(s+)}_{x_0'}\protect\big \langle 
\protect{\hat{\mathcal A}}_{\nu}(x),\protect{\hat{\mathcal A}}_{\nu '}(x ')
 \big \rangle
 \big ] . 
\end{aligned} 
\label{eq:16WC} 
\end{align}%
Calculation leading to Eq.\ (\protect\ref{eq:63TW}), which are typical for our approach, are presented in some detail in appendix \ref{ch:CL}. 

The term {\em time-$s$-ordered cumulant\/} is justified by the formula, 
\begin{align} 
\begin{aligned} 
\protect\big \langle 
\mathcal{T}_s{\protect{\hat{\mathcal A}}_{\nu}(x),\protect{\hat{\mathcal A}}_{\nu '}(x ')}
 \big \rangle 
 &= \protect\big \langle 
\mathcal{T}_s{\protect{\hat{\mathcal A}}_{\nu}(x)\protect{\hat{\mathcal A}}_{\nu '}(x ')}
 \big \rangle 
-\protect\big \langle \protect{\hat{\mathcal A}}_{\nu}(x)\big \rangle\protect\big \langle \protect{\hat{\mathcal A}}_{\nu '}(x ') \big \rangle , 
\end{aligned} 
\label{eq:18AN} 
\end{align}%
where the {\em time-$s$-ordered product\/} of two field operators reads, 
\begin{align} 
\begin{aligned} 
\mathcal{T}_s{\protect{\hat{\mathcal A}}_{\nu}(x)\protect{\hat{\mathcal A}}_{\nu '}(x ')} &= 
\mathcal{F}^{(s+)}_{x_0}\mathcal{F}^{(s+)}_{x_0'} 
T_{+}\protect{\hat{\mathcal A}}_{\nu}(x)\protect{\hat{\mathcal A}}_{\nu '}(x ')
+ 
\protect{\hat{\mathcal A}}^{(s-)}_{\nu}(x)\protect{\hat{\mathcal A}}^{(s+)}_{\nu '}(x ') + \textrm{H.c.}\, . 
\end{aligned} 
\label{eq:19AP} 
\end{align}%
Equation (\protect\ref{eq:18AN}) is readily verified using Eq.\ (\protect\ref{eq:21AR}). 
Calling the expression on the right-hand side (rhs) of (\ref{eq:19AP}) an ``ordered operator product'' agrees with the use of this term, e.g., in photodetection theory \cite{GlauberPhDet,KelleyKleiner,GlauberTN,MandelWolf}. For the general definition of the time-$s$-ordering of Heisenberg\ operators see appendix \ref{ch:T}. 

Equations (\protect\ref{eq:63TW})--(\ref{eq:19AP}) are exact relations applicable to arbitrary bosonic fields, whether free or interacting. Cancellation of the term quadratic in \mbox{$j_s$} and the explicitly causal nature of \mbox{$\mathcal{D}_{\text{R}}$} are a manifestation of general causality in the ``rotated'' theory, cf.\ \cite{APII,APIII,RelCaus,RelCausMadrid} and remarks at the end of \mbox{Sec.\ \ref{ch:TXC}}. 

\subsection{Formulae for closed-time-loop\ cumulants}%
\label{ch:TXR}
It is equally easy to obtain formulae for the closed-time-loop\ cumulants (\ref{eq:59TS}) in terms of the rotated ones. Applying substitutions (\ref{eq:28SJ}) to the rotated form (\ref{eq:63TW}) and comparing the result to the initial form (\ref{eq:60TT}) we find, 
\begin{align} 
\protect\big \langle 
T_+\protect{\hat{\mathcal A}}_{\nu }(x),\protect{\hat{\mathcal A}}_{\nu '}(x')
 \big \rangle 
 &
= \protect\big \langle 
\mathcal{T}_{s}{\protect{\hat{\mathcal A}}_{\nu }(x),\protect{\hat{\mathcal A}}_{\nu '}(x')}
 \big \rangle 
\nonumber\\ &\quad
+ i\hbar c\protect\big [ 
\mathcal{F}^{(s-)}_{x_0'}\mathcal{D}_{\mathrm{R}\nu \nu '}(x,x') 
+ 
\mathcal{F}^{(s-)}_{x_0}\mathcal{D}_{\mathrm{R}\nu' \nu }(x',x)
 \big ] , 
\label{eq:4LZ} 
\\ 
\protect\big \langle 
\protect{\hat{\mathcal A}}_{\nu }(x),\protect{\hat{\mathcal A}}_{\nu '}(x')
 \big \rangle 
 &
= \protect\big \langle 
\mathcal{T}_{s}{\protect{\hat{\mathcal A}}_{\nu }(x),\protect{\hat{\mathcal A}}_{\nu '}(x')}
 \big \rangle 
\nonumber\\ &\quad
+ i\hbar c\protect\big [ 
\mathcal{F}^{(s-)}_{x_0'}\mathcal{D}_{\mathrm{R}\nu \nu '}(x,x') 
- 
\mathcal{F}^{(s+)}_{x_0}\mathcal{D}_{\mathrm{R}\nu' \nu }(x',x)
 \big ] , 
\label{eq:99BE} 
\end{align}%
where 
\begin{align} 
\begin{aligned} 
\mathcal{D}_{\mathrm{R}\nu\nu '}(x,x') = 
g_{ \nu '\bar\nu}\mathcal{D}_{\mathrm{R}\nu}^{\bar\nu}(x,x') . 
\end{aligned} 
\label{eq:81UR} 
\end{align}%
Derivation of Eqs.\ (\protect\ref{eq:4LZ}), (\ref{eq:99BE}) relies on Eq.\ (\protect\ref{eq:11VX}). 
\section{``Quantum noise'' and operator ordering}%
\label{ch:QN}
\subsection{Classical connotations of generalised Keldysh rotations}%
\label{ch:GKD}
We temporarily return to the harmonic oscillator. For all definitions see \mbox{Sec.\ \ref{ch:A}}. 

As is shown in appendix \ref{ch:OS}, Eq.\ (\protect\ref{eq:65PR}) is a particular case of the general formula, relating the closed-time-loop ordering of the free-field operators to the \mbox{$O_s$}-ordering of the creation and annihilation operators, 
\begin{align} 
\begin{aligned} 
\Xi[\eta_+,\eta_-] 
\settoheight{\auxlv}{$|$}%
\raisebox{-0.3\auxlv}{$|_{\eta_{\pm}\to\eta ,j_s}$} 
 &= \exp\big(
-i\eta D_{\text{R}}j_s 
 \big) 
\protect\big \langle 
O_s\exp\big(
-i\eta \hat q
 \big) 
 \big \rangle 
, 
\end{aligned} 
\label{eq:19RY} 
\end{align}%
where
\begin{align} 
\begin{aligned} 
 &\Xi[\eta_+,\eta_-] 
= \protect\big \langle 
T_C\exp\big(
-i\eta_+\hat q_+
+i\eta_-\hat q_-
 \big) 
 \big \rangle 
\end{aligned} 
\label{eq:45XK} 
\end{align}%
is the generating functional of the closed-time-loop\ Green functions of the oscillator (for all definitions see \mbox{Secs \ref{ch:G}} and \ref{ch:AF}). We use notation (\ref{eq:47XM}). Equation (\protect\ref{eq:19RY}) holds irrespective of the quantum state (i.e., it is in fact an operator formula). 
For Gaussian states, Wick's theorem, 
\begin{align} 
\begin{aligned} 
\Xi[\eta_+,\eta_-] = \exp\Lambda[\eta_+,\eta_-] , 
\end{aligned} 
\label{eq:15RU} 
\end{align}%
reduces transformation of all closed-time-loop\ Green functions to transformation of the kernels (\ref{eq:8VU}). 

A remarkable feature of Eq.\ (\protect\ref{eq:19RY}) is the absence of Planck's constant. Any quantum relation without $\hbar $ survives the classical limit \mbox{$
\hbar \to 0
$} unchanged, and must therefore have a classical counterpart. Indeed, consider a driven classical oscillator. Its displacement is given by the formula, 
\begin{align} 
\begin{aligned} 
q(t) = q_{\mathrm{in}}(t) + \int dt'D_{\text{R}}(t-t')j(t) , 
\end{aligned} 
\label{eq:18MQ} 
\end{align}%
where \mbox{$
j(t)
$} is the driving force \cite{endF}, and the in-field \mbox{$q_{\mathrm{in}}(t)$} is given by Eq.\ (\protect\ref{eq:20RZ}). 
If the complex amplitude \mbox{$\alpha$} in (\ref{eq:20RZ}) is a random variable distributed with probability \mbox{$p(\alpha )$}, the generating functional of stohastic moments of \mbox{$q(t)$} reads, 
\begin{align} 
\begin{aligned} 
\protect\big \langle \exp\big(
-i\eta q
 \big) \big \rangle 
 &= \exp\big(
-i\eta D_{\text{R}}j_s 
 \big) 
\int d^2\alpha p(\alpha )\exp\big(
-i\eta q_{\mathrm{in}}
 \big) . 
\end{aligned} 
\label{eq:69PV} 
\end{align}%
We use the same notation (angle brackets) for the quantum and classical statistical averaging; what we have in mind is clear in the context. Equation (\protect\ref{eq:69PV}) is the classical limit of Eq.\ (\protect\ref{eq:19RY}), when the $s$-ordered quantum averages turn into classical averages. To see this clearly, express the quantum average on the rhs of (\ref{eq:19RY}) as a quasiaverage, 
\begin{align} 
\begin{aligned} 
 &\protect\big \langle 
O_s\exp\big(
-i\eta \hat q
 \big)
 \big \rangle 
= \int d^2\alpha p_s(\alpha )\exp\big(
-i\eta q_{\mathrm{in}}
 \big) , 
\end{aligned} 
\label{eq:21SA} 
\end{align}%
so that Eq.\ (\protect\ref{eq:19RY}) becomes, 
\begin{align} 
\begin{aligned} 
\Xi[\eta_+,\eta_-] 
\settoheight{\auxlv}{$|$}%
\raisebox{-0.3\auxlv}{$|_{\eta_{\pm}\to\eta ,j_s}$} 
 &= \exp\big(
-i\eta D_{\text{R}}j_s 
 \big) 
\int d^2\alpha p_s(\alpha )\exp\big(
-i\eta q_{\mathrm{in}}
 \big) . 
\end{aligned} 
\label{eq:81DF} 
\end{align}%
Equations (\protect\ref{eq:69PV}) and (\ref{eq:81DF}) coinside up to the replacement of quasiprobability by probability, \mbox{$p_s(\alpha )\to p(\alpha )$}, which is natural in the limit \mbox{$
\hbar \to 0
$}. Without this limit, each generalised Keldysh rotation defines a phase-space representation (mapping) of the quantum system. The conventional Keldysh rotation is a phase-space mapping based on the Weyl ordering. For an in-depth discussion see \mbox{Refs.\ \protect\cite{MandelWolf,Schleich,Bettina}}. 

\subsection{What is ``the best rotation''?}%
\label{ch:GKX}
For simplicity, consider the Gaussian case, when the state of the oscillator is fully described by the \mbox{$s$}-ordered quantum average \mbox{$\protect\big \langle 
O_s\hat q(t)\hat q(t')
 \big \rangle$}. According to the arguments of \mbox{Sec.\ \ref{ch:GKD}}, this average is a quantum counterpart of the classical statistical average \mbox{$\protect\big \langle 
q_{\mathrm{in}}(t)q_{\mathrm{in}}(t')
 \big \rangle$}. The latter characterises noise in a classical system. One may therefore say that the former represents {\em quantum noise\/}. This concept is by definition associated with operator ordering: ``quantum noise'' is ordering-specific and thus nonunique.
 
It is instructive to put these arguments in context with Eq.\ (\protect\ref{eq:30BA}). The latter stipulates that quantum noises according to different orderings differ in fact in a formal admixture of the linear response. Setting \mbox{$s'=1$} in (\ref{eq:30BA}) and remembering that \mbox{$O_1$} is the normal ordering we have, 
\begin{align} 
\begin{aligned} 
 &\protect\big \langle 
O_s\hat q(t)\hat q(t')
 \big \rangle = 
\protect\big \langle 
{\bf :}
\hat q(t)\hat q(t')
{\bf :} 
 \big \rangle + 2s_-\, Z(t-t') , 
\end{aligned} 
\label{eq:85DL} 
\end{align}%
where \mbox{$Z(t-t')$} is given by Eq.\ (\protect\ref{eq:86DM}). Recalling that, 
\begin{align} 
\begin{aligned} 
\protect\big \langle 0\big| 
{\bf :}
\hat q(t)\hat q(t')
{\bf :} 
 \big |0\big\rangle = 0, 
\end{aligned} 
\label{eq:82DH} 
\end{align}%
for the vacuum state of the oscillator 
we obtain, 
\begin{align} 
\begin{aligned} 
 &\protect\big \langle 0\big| 
O_s\hat q(t)\hat q(t')
 \big |0\big\rangle 
= 2s_- \, Z(t-t') . 
\end{aligned} 
\label{eq:35BF} 
\end{align}%
This way, Eq.\ (\protect\ref{eq:85DL}) represents quantum noise of the field according to the \mbox{$O_s$}-ordering as a sum of a normal contribution and zero-point fluctuations. The latter is nothing but a coded information about response properties of the oscillator. Indeed, it is easy to show the formula, 
\begin{align} 
\begin{aligned} 
D_{\text{R}}(\tau ) = (2/i\hbar )\theta(\tau )\protect\big [ 
Z^{(+)}(\tau )-Z^{(-)}(\tau )
 \big ] . 
\end{aligned} 
\label{eq:87DN} 
\end{align}%
The difference between a complete quantum characterisation of the oscillator and its characterisation in terms of response is contained in the normal average. It is therefore natural to talk about physical and response components of quantum noise. Under the normal ordering, the response component vanishes. It may therefore be seen as a formal ``contamination'' of quantum noise by linear response under orderings other than normal. 

We stress that, mathematically, all rotations are equal. The triad \mbox{$
s=0,\pm 1
$} have been extensively used in quantum optics \cite{MandelWolf,Schleich}. The conventional Keldysh rotation \mbox{($
s=0
$)} is the standard one in the real-time QFT (for references see the introduction). It also happens to be of help for practical calculations with few-mode nonlinear bosonic systems, see \cite{Bettina} and references therein. 

Things change if we consider spatial (in particular, relativistic) problems, characterised by infinite number of modes. Zero-point fluctuations make phase-space images of quantised energy, momentum, etc., badly defined, {\em except in the normally ordered representation\/}. 
For this reason standard quantisation of free fields employs normal ordering. 
Cancellation of zero-point fluctuations also makes normal ordering special when analysing the relation between quantum and classical mechanics. This ordering assures direct correspondence between the quantum and classical vacua: 
\begin{align} 
\begin{aligned} 
\protect\big \langle 0\big| {\bf :}
\hat q(t_1)\cdots \hat q(t_m)
{\bf :} \big |0\big\rangle=0 
\Longleftrightarrow 
\protect\big \langle q(t_1)\cdots q(t_m) \big \rangle_{\mathrm{vac}}=0 
. 
\end{aligned} 
\label{eq:1LW} 
\end{align}%
With other orderings, quantum vacuum turns out to be non-empty, and one is doomed to encounter a formal discrepancy between quantum and classical mechanics (except in the limit \mbox{$
\hbar \to 0
$}). 
With normal ordering, this kind of discrepancy is eliminated \cite{endElim}\nocite{Caves}. 
\subsection{``Quantum noise'' of interacting fields}%
\label{ch:QNA}
Apart from the presence of the average field, the ``Heisenberg'' formula (\ref{eq:63TW}) has the same structure as the ``free'' formula (\ref{eq:65PR}), and has the same classical connotations (cf.\ \mbox{Sec.\ \ref{ch:GKD}}). It isolates the ``quantum noise'' of the field given by Eq.\ (\protect\ref{eq:16WC}). The latter is the only quantity in (\ref{eq:63TW}) that is specific to the rotation. The average field and the linear response are shared by all rotations. For the conventional Keldysh rotation, the quantum noise is given by the symmetrised cumulant \cite{endTS}, 
\begin{align} 
\begin{aligned} 
 &\protect\big \langle 
\mathcal{W}{\protect{\hat{\mathcal A}}_{\nu}(x),\protect{\hat{\mathcal A}}_{\nu '}(x ')}
 \big \rangle 
= \frac{1}{2}\protect\big [ 
\protect\big \langle 
\protect{\hat{\mathcal A}}_{\nu}(x),\protect{\hat{\mathcal A}}_{\nu '}(x ')
 \big \rangle+\protect\big \langle 
\protect{\hat{\mathcal A}}_{\nu '}(x '),\protect{\hat{\mathcal A}}_{\nu}(x)
 \big \rangle
 \big ] 
 , 
\end{aligned} 
\label{eq:19WF} 
\end{align}%
while response transformation leads to the {\em time-normal\/} cumulant, 
\begin{align} 
\begin{aligned} 
 &\protect\big \langle 
{\mathcal T}{\bf :}\protect{\hat{\mathcal A}}_{\nu}(x),\protect{\hat{\mathcal A}}_{\nu '}(x '){\bf :}
 \big \rangle 
\\ &\quad
= 2\Re\protect\big [ 
\mathcal{F}^{(+)}_{x_0}\mathcal{F}^{(+)}_{x_0'}\protect\big \langle 
T_{+}\protect{\hat{\mathcal A}}_{\nu}(x),\protect{\hat{\mathcal A}}_{\nu '}(x ')
 \big \rangle
+\mathcal{F}^{(-)}_{x_0}\mathcal{F}^{(+)}_{x_0'}\protect\big \langle 
\protect{\hat{\mathcal A}}_{\nu}(x),\protect{\hat{\mathcal A}}_{\nu '}(x ')
 \big \rangle
 \big ] . 
\end{aligned} 
\label{eq:65TY} 
\end{align}%
Other rotations lead to other definitions of quantum noise. {\em For the Heisenberg\ as well as for free fields, the very concept of quantum noise is by definition associated with operator ordering.\/} 

\subsection{Reordering of Heisenberg\ operators}%
\label{ch:QNB}
It is straightforward to generalise Eq.\ (\protect\ref{eq:30BA}) connecting ``quantum noises'' under different orderings to the Heisenberg\ field. Namely, using either of Eqs.\ (\protect\ref{eq:4LZ}), (\ref{eq:99BE}) we find, 
\begin{align} 
\begin{aligned} 
 &\protect\big \langle 
\mathcal{T}_{s}{\protect{\hat{\mathcal A}}_{\nu }(x),\protect{\hat{\mathcal A}}_{\nu '}(x')}
 \big \rangle - \protect\big \langle 
\mathcal{T}_{\mathrm{s'}}{\protect{\hat{\mathcal A}}_{\nu }(x),\protect{\hat{\mathcal A}}_{\nu '}(x')}
 \big \rangle 
= (s'-s)\mathcal{Z}_{\nu\nu '}(x,x'), 
\end{aligned} 
\label{eq:26AW} 
\end{align}%
where, 
\begin{align} 
\begin{aligned} 
\mathcal{Z}_{\nu\nu '}(x,x') 
 &
= \frac{i\hbar c}{2}\protect\big [ 
\mathcal{F}^{(-)}_{x_0'}\mathcal{D}_{\mathrm{R}\nu \nu '}(x,x') 
+ 
\mathcal{F}^{(-)}_{x_0}\mathcal{D}_{\mathrm{R}\nu' \nu }(x',x)
\\ &\quad
- 
\mathcal{F}^{(+)}_{x_0'}\mathcal{D}_{\mathrm{R}\nu \nu '}(x,x') 
- 
\mathcal{F}^{(+)}_{x_0}\mathcal{D}_{\mathrm{R}\nu' \nu }(x',x)
 \big ] . 
\end{aligned} 
\label{eq:27AX} 
\end{align}%
Thus, for free as well as for interacting fields, quantum noises according to different orderings differ in a formal admixture of the linear response. Having calculated response and noise in one representation, it is straightforward to transform the latter to any other representation, e.g., 
\begin{align} 
\begin{aligned} 
\protect\big \langle 
\mathcal{T}_{s}{\protect{\hat{\mathcal A}}_{\nu }(x),\protect{\hat{\mathcal A}}_{\nu '}(x')}
 \big \rangle 
 &
= \protect\big \langle 
\mathcal{W}{\protect{\hat{\mathcal A}}_{\nu}(x),\protect{\hat{\mathcal A}}_{\nu '}(x ')}
 \big \rangle - s\mathcal{Z}_{\nu\nu '}(x,x')
\\ &
= \protect\big \langle 
{\mathcal T}{\bf :}\protect{\hat{\mathcal A}}_{\nu}(x),\protect{\hat{\mathcal A}}_{\nu '}(x '){\bf :}
 \big \rangle + 2s_-\mathcal{Z}_{\nu\nu '}(x,x') , 
\end{aligned} 
\label{eq:28AY} 
\end{align}%
etc. 

The second of Eqs.\ (\protect\ref{eq:28AY}) extends Eq.\ (\protect\ref{eq:85DL}) to interacting fields. This relation is general and does not depend on details of quantum dynamics. {\em Were we also able to generalise to interacting fields Eq.\ (\protect\ref{eq:82DH})\/}, 
\begin{align} 
\begin{aligned} 
\protect\big \langle 0\big| {\mathcal T}{\bf :}\protect{\hat{\mathcal A}}_{\nu }(x),\protect{\hat{\mathcal A}}_{\nu '}(x'){\bf :} \big |0\big\rangle = 0 , 
\end{aligned} 
\label{eq:40BM} 
\end{align}%
the whole philosophy of physical {\em versus\/} response components of quantum noise (cf.\ \mbox{Sec.\ \ref{ch:GKX}}) would become applicable to interacting fields. However, unlike Eq.\ (\protect\ref{eq:28AY}), Eq.\ (\protect\ref{eq:40BM}) does depend on details of quantum dynamics. It may be shown for arbitrary polynomial interactions that, if all fields are in a vacuum state and this vacuum is stable, all time-normal averages in the theory are zero. Verification of this result in its entirety requires advanced formal tools \cite{LPEAUnp} that we tend to avoid in this paper. Here we demonstrate it in spinor quantum electrodynamics\ in the first nonvanishing approximation in interaction of the electromagnetic\ field with Dirac vacuum (see \mbox{Sec.\ \ref{ch:ZP}} below). General analyses will be presented elsewhere. 

\begin{center}* * *\end{center}\nopagebreak 
Natural quantum-classical correspondence with well-defined quantum vacuum make response transformation the natural choice if we are interested in the classical limit and/or in relativistic problems. The price to pay is that response substitution is an integral rather than an algebraic transfomation, i.e., nonlocal in time. This raises nontrivial causality issues. As a structural concept, response transformation was discussed for free bosonic fields in \cite{API}, for interacting bosons in \cite{APII} and for interacting fermions in \cite{APIII}. Its association with Wick's theorem was subject of \cite{WickCaus}. Causality problems were taken care of in \cite{APII,APIII,RelCaus,RelCausMadrid}. 

In the rest of the paper we concern ourselves with two questions, 
\begin{itemize}
\item[(a)]
whether suppression of zero-point fluctuations (empty vacuum) indeed extends to quantum dynamics, and
\item[(b)]
whether the formal structures established in \mbox{Sec.\ \ref{ch:TX}} are consistent with renormalisations. 
\end{itemize}
Answering these questions in full implies generalisation of Eq.\ (\protect\ref{eq:63TW}) to higher-order cumulants and arbitrary interactions. This is a formidable task, which may take a cumulative effort of many researchers. In this paper we restrict ourselves to the quantised electromagnetic\ field in linear media, where ``media'' include such nontrivial QFT object as the Dirac (spinor) field. This problem is rich enough to illustrate all the important points without too much mathematics. 
\section{Classical stochastic, semiclassical and quantum approaches to electromagnetic\ field in a linear medium}%
\label{ch:M}
\subsection{Driven electromagnetic\ field}%
\label{ch:CR}
Throughout the rest of the paper we consider (or imply) a driven relativistic electromagnetic\ field governed by the Hamiltonian, (in the interaction picture)
\begin{align} 
\begin{aligned} 
\hat H(x_0) &= \hat H_{\mathrm{ff}}(x_0) + \hat H_{\mathrm{m}}(x_0) 
+ \int d^3\mbox{\bf x}\hat A_{\nu}(x)\protect\big [ 
J_{\mathrm{e}}^{\nu}(x)+\hat J^{\nu}(x)
 \big ] . 
\end{aligned} 
\label{eq:72PY} 
\end{align}%
Here, \mbox{$\hat A_{\nu}(x)$} and \mbox{$\hat J_{\nu}(x)$} are the quantised potential and current operators in the interaction picture. The corresponding Heisenberg\ operators are \mbox{$\protect{\hat{\mathcal A}}_{\nu}(x)$} and \mbox{$\protect{\hat{\mathcal J}}_{\nu}(x)$}. \mbox{$
\hat H_{\mathrm{ff}}(x_0)
$} is the free-field Hamiltonian and \mbox{$
\hat H_{\mathrm{m}}(x_0)
$} comprises Hamiltonians of quantised matter fields responsible for the quantum current \mbox{$\hat J^{\nu}(x)$} (ff stands for {\em free field\/} and m for {\em matter\/}). Unlike the free-field Hamiltonian, the matter Hamiltonians are not bound to be free (quadratic). The initial (Heisenberg) state of the field is vacuum, while that of the matter may be arbitrary. 
\subsection{C-number electromagnetic field in a linear medium}%
\label{ch:DQ}
\subsubsection{Susceptibility and random source}%
\label{ch:DQA}
In classical electrodynamics, the equation for the electromagnetic potential reads, (in SI units)
\begin{align} 
\begin{aligned} 
\mu_{\mathrm{vac}}^{-1}\Box \mathcal{A}_{\nu }(x) = J_{\nu }(x), 
\end{aligned} 
\label{eq:17PA} 
\end{align}%
where $\mu_{\mathrm{vac}}$ is the magnetic constant (permeability of vacuum), $\Box$ is the d'Alembertian, 
\begin{align} 
\Box = \partial_{\nu }\partial^{\nu } = \frac{1}{c^2}\,\frac{\partial^2 }{\partial t^2} - \frac{\partial^2}{\partial {\bf x}^2} , 
\label{eq:18PB} 
\end{align}%
and $\partial_{\nu },\partial^{\nu }$ 
are the co- and contravariant derivatives, 
\begin{align} 
\partial_{\nu } = \frac{\partial }{\partial x^{\nu }}, \quad
\partial^{\nu } = \frac{\partial }{\partial x_{\nu }}. 
\label{eq:19PC} 
\end{align}%
\mbox{$
\mathcal{A}_{\nu }(x)
$} is subject to the Lorentz condition, 
\begin{align} 
\partial^{\nu }\mathcal{A}_{\nu }(x) = 0, 
\label{eq:24TR} 
\end{align}%
consistent with conservation of current, 
\begin{align} 
\partial^{\nu }J_{\nu }(x) = 0. 
\label{eq:25TS} 
\end{align}%
Free propagation of the field corresponds to Eq.\ (\protect\ref{eq:17PA}) with \mbox{$
J_{\nu }(x) = J_{\mathrm{e}\nu }(x)
$}, where \mbox{$
J_{\mathrm{e}\nu }(x)
$} is an external source. 

Now the field emitted by the source propagates in a linear medium. Formally, the latter is characterised by two objects: the {\em microscopic linear susceptibility\/} \mbox{$\Pi_{\mathrm{R}\nu}^{\nu' }(x,x')$} and the random current (random source) in the medium \mbox{$J_{\mathrm{r}\nu }(x)$}. Linearity of the medium means that neither \mbox{$\Pi_{\text{R}}$} nor stochastic properties of \mbox{$J_{\mathrm{r}}$} depend on \mbox{$J_{\mathrm{e}}$}. 
We do not assume the medium to be homogeneous, nor stationary. 
Total current is the sum of the random and induced currents, 
\begin{align} 
\begin{aligned} 
J_{\nu}(x)= J_{\mathrm{r}\nu }(x) 
+ \int d^4x' \Pi_{\mathrm{R}\nu}^{\nu' }(x,x')\mathcal{A}_{\nu '}(x ') . 
\end{aligned} 
\label{eq:42SY} 
\end{align}%
The self-consistent equation for the field in the medium is then found to be, 
\begin{align} 
\begin{aligned} 
 &\mu_{\mathrm{vac}}^{-1}\Box \mathcal{A}_{\nu }(x) 
-\int d^4 x' \Pi_{\mathrm{R}\nu}^{\nu' }(x-x')\mathcal{A}_{\nu' }(x') 
= J_{\mathrm{e}\nu }(x) + J_{\mathrm{r}\nu }(x) . 
\end{aligned} 
\label{eq:21PE} 
\end{align}%
Solution to (\ref{eq:21PE}) is written in terms of the {\em macroscopic linear susceptibility\/} of the medium \mbox{$\mathcal{D}_{\mathrm{R}\nu }^{\nu '}(x,x')$}, 
\begin{align} 
\begin{aligned} 
\mathcal{A}_{\nu}(x)= \int d^4x'\mathcal{D}_{\mathrm{R}\nu }^{\nu '}(x,x')\protect\big [ 
J_{\mathrm{e}\nu '}(x') + J_{\mathrm{r}\nu '}(x')
 \big ] . 
\end{aligned} 
\label{eq:44TA} 
\end{align}%
where \mbox{$\mathcal{D}_{\mathrm{R}\nu }^{\nu '}(x,x')$} is defined as a retarded Green function of Eq.\ (\protect\ref{eq:21PE}), 
\begin{align} 
\begin{aligned} 
\mu_{\mathrm{vac}}^{-1}\Box \mathcal{D}_{\mathrm{R}\nu }^{\nu '}(x,x')
-\int d^4 x'' \Pi_{\mathrm{R}\nu}^{\nu'' }(x,x'')
\mathcal{D}_{\mathrm{R}\nu''}^{\nu '}(x'',x') 
= \delta^{(4)}(x-x') , 
\\ 
\mathcal{D}_{\mathrm{R}\nu }^{\nu '}(x,x') = 0 , \quad x_0<x_0' . 
\end{aligned} 
\label{eq:43SZ} 
\end{align}%
For linear media, \mbox{$\mathcal{D}_{\text{R}}$} is indepenedent of the external current \mbox{$J_{\mathrm{e}}$}. 
 
Equation (\protect\ref{eq:44TA}) calls for a remark. It contains no in-field, which means that all field sources (currents) are accounted for explicitly. In other words, our analyses apply to a closed system in the strict meaning of the term. For a field in thermal equilibrium, currents in a heatbath should formally be included in \mbox{$J_{\mathrm{r}}$}. Since no restriction is imposed on the latter, the absence of an in-field contribution in (\ref{eq:44TA}) is not a limitation. 

\subsubsection{Cumulants of the random field}%
\label{ch:DC}
We are interested in the average field 
\mbox{$\protect\big \langle \mathcal{A}_{\nu }(x) \big \rangle$} and in the stochastic cumulant, 
\begin{align} 
\begin{aligned} 
\protect\big \langle \mathcal{A}_{\nu }(x),\mathcal{A}_{\nu '}(x') \big \rangle 
 &= 
\protect\big \langle \mathcal{A}_{\nu }(x)\mathcal{A}_{\nu '}(x') \big \rangle 
- \protect\big \langle \mathcal{A}_{\nu }(x) \big \rangle
\protect\big \langle \mathcal{A}_{\nu }(x) \big \rangle 
. 
\end{aligned} 
\label{eq:49TF} 
\end{align}%
For the average field we find, 
\begin{align} 
\begin{aligned} 
\protect\big \langle \mathcal{A}_{\nu }(x) \big \rangle 
 &= \int d^4x'\mathcal{D}_{\mathrm{R}\nu }^{\nu '}(x,x')
\protect\big [ 
J_{\mathrm{e}\nu '}(x') + \protect\big \langle J_{\mathrm{r}\nu }(x) \big \rangle
 \big ] . 
\end{aligned} 
\label{eq:46TC} 
\end{align}%
We allowed for nonzero average current in the medium \cite{endAvC}, 
\begin{align} 
\begin{aligned} 
\protect\big \langle J_{\mathrm{r}\nu }(x) \big \rangle \neq 0 . 
\end{aligned} 
\label{eq:45TB} 
\end{align}%
Furthermore,
\begin{align} 
\begin{aligned} 
 &\protect\big \langle \mathcal{A}_{\nu }(x),\mathcal{A}_{\nu '}(x') \big \rangle 
= 
\int d^4\bar x d^4\bar x'
\mathcal{D}_{\mathrm{R}\nu }^{\bar \nu }(x,\bar x) 
\mathcal{D}_{\mathrm{R}\nu '}^{\bar \nu '}(x',\bar x')
\Pi _{\mathrm{N}\bar \nu \bar\nu '}(\bar x,\bar x'), 
\end{aligned} 
\label{eq:47TD} 
\end{align}%
where 
\begin{align} 
\begin{aligned} 
\Pi _{\mathrm{N}\nu\nu '}(x,x') 
 &= \protect\big \langle 
J_{\mathrm{r}\nu }(x),J_{\mathrm{r}\nu '}(x')
 \big \rangle 
= \protect\big \langle 
J_{\mathrm{r}\nu }(x)J_{\mathrm{r}\nu '}(x')
 \big \rangle - \protect\big \langle 
J_{\mathrm{r}\nu }(x) \big \rangle\protect\big \langle J_{\mathrm{r}\nu '}(x')
 \big \rangle. 
\end{aligned} 
\label{eq:48TE} 
\end{align}%
For linear media, cumulant (\ref{eq:47TD}) is independent of the external current. The latter only enters through Eq.\ (\protect\ref{eq:46TC}) for the average field. 
\subsection{Semiclassical theory}%
\label{ch:C}
In a semiclassical approach, one keeps the field classical (a c-number), while attempting to calculate microscopic quantities \mbox{$\Pi_{\text{R}}$}, \mbox{$\Pi_{\mathrm{N}}$} and \mbox{$\protect\big \langle 
J_{\mathrm{r}}
 \big \rangle$} in a suitable quantum model of the medium. 
Their general quantum definition is a nontrivial question \cite{endMicr} which we shall discuss in full elsewhere. 
One way to define these quantities consistently is to stick to the first nonvanishing approximation in the electromagnetic\ interaction \cite{Maxwell,QDynResp}. Formally, one considers a driven quantum medim interacting with a c-number field source, 
\begin{align} 
\begin{aligned} 
\hat H_{\mathrm{dm}}(x_0) &= \hat H_{\mathrm{m}}(x_0) 
+ \int d^3\mbox{\bf x}A_{\mathrm{e}\nu}(x)\hat J^{\nu}(x)
. 
\end{aligned} 
\label{eq:20WH} 
\end{align}%
\mbox{$\hat H_{\mathrm{m}}(x_0)$} and \mbox{$\hat J_{\nu}(x)$} here are the same as in (\ref{eq:72PY}). \mbox{$\Pi _{\text{R}}$} is identified with Kubo's linear response function, 
\begin{align} 
\Pi_{\mathrm{R}\nu }^{\nu '}(x,x') 
 &= \frac{\delta \protect\big \langle 
\hat J^{\mathrm{d}}_{\nu}(x)
 \big \rangle }{\delta A_{\mathrm{e}\nu '}(x')} 
\settoheight{\auxlv}{$\big|$}%
\raisebox{-0.3\auxlv}{$\big|_{A_{\mathrm{e}}=0}$}
= (i\hbar c)^{-1}\theta(x_0-x_0')\protect\big \langle 
\protect\big [ 
\hat J_{\nu}(x),\hat J^{\nu '}(x ')
 \big ] 
 \big \rangle , 
\label{eq:67UA} 
\end{align}%
and stochastic cumulants \mbox{$\protect\big \langle 
J_{\mathrm{r}}
 \big \rangle $} and \mbox{$\Pi _{\mathrm{N}}$}---with the corresponding time-normal current cumulants, 
\begin{align} 
\protect\big \langle 
J_{\mathrm{r}\nu }(x)
 \big \rangle &= \protect\big \langle 
\hat J_{\nu}(x)
 \big \rangle, 
\label{eq:22WK} 
\\
\Pi_{\mathrm{N}\nu\nu '}(x,x') 
 &= \protect\big \langle 
{\mathcal T}{\bf :}\hat J_{\nu}(x),\hat J^{\nu '}(x '){\bf :}
 \big \rangle , 
\label{eq:68UB} 
\end{align}%
where 
\begin{align} 
\begin{aligned} 
 &\protect\big \langle 
{\mathcal T}{\bf :}\hat J_{\nu}(x),\hat J^{\nu '}(x '){\bf :}
 \big \rangle 
= \protect\big \langle 
{\mathcal T}{\bf :}\hat J_{\nu}(x)\hat J^{\nu '}(x '){\bf :}
 \big \rangle 
- \protect\big \langle 
\hat J_{\nu}(x)\big \rangle\protect\big \langle \hat J^{\nu '}(x ')
 \big \rangle . 
\end{aligned} 
\label{eq:90VA} 
\end{align}%
Quantum averaging in Eqs.\ (\protect\ref{eq:67UA})--(\ref{eq:90VA}) is over the initial state of the medium which may be arbitrary. \mbox{$\hat J^{\mathrm{d}}$} in (\ref{eq:67UA}) is the Heisenberg\ current operator according to Hamiltonian (\ref{eq:20WH}). All other relations employ the ``free'' current operator \mbox{$\hat J$}. We put {\em free\/} in quotation marks because \mbox{$\hat H_{\mathrm{m}}$} may contain nonlinearities. {\em Caution:\/} generalisation of Eqs.\ (\protect\ref{eq:67UA})--(\ref{eq:68UB}) by replacing \mbox{$\hat J$} by the Heisenberg\ operator \mbox{$\protect{\hat{\mathcal J}}$} (defined in \mbox{Sec.\ \ref{ch:CR}}) is an error, cf.\ endnote \cite{endMicr}.

\subsection{Quantum theory}%
\label{ch:CQ}
The semiclassical model follows by redefining stochastic cumulants of the c-number current as time-normal cumulants of the q-number one, 
\begin{align} 
\begin{aligned} 
\protect\big \langle 
J_{\mathrm{r}\nu }(x)
 \big \rangle &= \protect\big \langle 
\hat J_{\nu}(x)
 \big \rangle, 
\quad 
\protect\big \langle 
J_{\mathrm{r}\nu '}(x'), 
J_{\mathrm{r}\nu }(x)
 \big \rangle = \protect\big \langle 
{\mathcal T}{\bf :}\hat J_{\nu}(x),\hat J_{\nu '}(x '){\bf :}
 \big \rangle . 
\end{aligned} 
\label{eq:52BZ} 
\end{align}%
There does not seem to be any limit to a classical interpretation of \mbox{$\protect\big \langle 
\hat J
 \big \rangle $} and hence of \mbox{$\Pi _{\text{R}}$} (see, however, the word of caution in \mbox{Sec.\ \ref{ch:CM}}). Such interpretation cannot be guarantied for the time-normal cumulant \mbox{$\protect\big \langle 
{\mathcal T}{\bf :}\hat J,\hat J{\bf :}
 \big \rangle $}. Whether this quantity may be expressed as a stochatic cumulant of a c-number current is a ``joint decision'' of the quantum dynamics (expressed by \mbox{$\hat H_{\mathrm{m}}$}) and quantum state (expressed by the averaging). If this interpretation fails, quantisation of the field becomes compulsory. 

A fully quantum theory emerges by replacing all stochastic cumulants by their quantum counterparts. This applies to the current, 
\begin{align} 
\begin{aligned} 
\protect\big \langle 
J_{\mathrm{r}\nu }(x)
 \big \rangle &\Longrightarrow \protect\big \langle 
\hat J_{\nu}(x)
 \big \rangle, 
\quad 
\protect\big \langle 
J_{\mathrm{r}\nu '}(x'), 
J_{\mathrm{r}\nu }(x)
 \big \rangle\Longrightarrow \protect\big \langle 
{\mathcal T}{\bf :}\hat J_{\nu}(x),\hat J_{\nu '}(x '){\bf :}
 \big \rangle . 
\end{aligned} 
\label{eq:23WL} 
\end{align}%
as well as to the field, 
\begin{align} 
\begin{aligned} 
\protect\big \langle 
\mathcal{A}_{\nu}(x)
 \big \rangle 
 &\Longrightarrow 
\protect\big \langle 
\protect{\hat{\mathcal A}}_{\nu}(x)
 \big \rangle , 
\quad 
\protect\big \langle 
\mathcal{A}_{\nu}(x),
\mathcal{A}_{\nu '}(x ')
 \big \rangle 
\Longrightarrow 
\protect\big \langle 
{\mathcal T}{\bf :}\protect{\hat{\mathcal A}}_{\nu}(x),
\protect{\hat{\mathcal A}}_{\nu '}(x ') 
{\bf :} 
 \big \rangle . 
\end{aligned} 
\label{eq:89UZ} 
\end{align}%
{\em There is no warranty whatsoever that the dynamical relations (\ref{eq:46TC}), (\ref{eq:47TD}) would survive such formal upgrade unchanged\/}. Nonetheless this is the case: 
quantised electromagnetic\ field in a linear medium 
is solved by the formulae, 
\begin{align} 
\begin{aligned} 
\protect\big \langle \protect{\hat{\mathcal A}}_{\nu }(x) \big \rangle 
 &= \int d^4x'\mathcal{D}_{\mathrm{R}\nu }^{\nu '}(x,x')
\protect\big [ 
J_{\mathrm{e}\nu '}(x') + \protect\big \langle \hat J_{\nu }(x) \big \rangle
 \big ] , 
\end{aligned} 
\label{eq:66TZ} 
\end{align}%
and
\begin{align} 
\begin{aligned} 
 &\protect\big \langle {\mathcal T}{\bf :}\protect{\hat{\mathcal A}}_{\nu }(x),\protect{\hat{\mathcal A}}_{\nu '}(x'){\bf :} \big \rangle 
= 
\int d^4\bar x d^4\bar x'
\mathcal{D}_{\mathrm{R}\nu }^{\bar \nu }(x,\bar x) 
\mathcal{D}_{\mathrm{R}\nu '}^{\bar \nu '}(x',\bar x')
\Pi _{\mathrm{N}\bar \nu \bar\nu '}(\bar x,\bar x'). 
\end{aligned} 
\label{eq:92VC} 
\end{align}%
The closed-time-loop\ cumulants (\ref{eq:59TS}), which are a more traditional form of a quantum solution, may be recovered from Eqs.\ (\protect\ref{eq:99BE}), (\ref{eq:81UR}). 
 
We stress that, while Eq.\ (\protect\ref{eq:66TZ}) is expectable, validity of Eq.\ (\protect\ref{eq:92VC}) is in no way automatic. The reason that Eq.\ (\protect\ref{eq:92VC}) emerges as a formal upgrade of the corresponding classical relation is that {\em it lacks contribution from zero-point fluctuations\/}. 
Absence of a physical contribution is evidently due to the initial vacuum state of the field (and to the assumption of closed system, cf.\ the remark at the end of \mbox{Sec.\ \ref{ch:DQA}}). Cancellation of the vacuum contribution (zero-point fluctuations in the true meaning of the term) is a property of the time-normal ordering. 

Equations (\protect\ref{eq:66TZ}), (\ref{eq:92VC}) are verified in appendix \ref{ch:ZZZ}. In particular, we show that \mbox{$\mathcal{D}_{\text{R}}$} occuring in these relations obeys the classical Eq.\ (\protect\ref{eq:43SZ}) with quantum \mbox{$\Pi _{\text{R}}$} given by (\ref{eq:67UA}). Consistency of Eq.\ (\protect\ref{eq:66TZ}) with Kubo's linear response theory \cite{KuboIrrevI,KuboTodaHashitsumeII} then warrants that \mbox{$\mathcal{D}_{\text{R}}$} is given by Eq.\ (\protect\ref{eq:64TX}). 

The macroscopic susceptibility \mbox{$\mathcal{D}_{\text{R}}$} is shared by the classical, semiclassical and quantum viewpoints. To find it from Eq.\ (\protect\ref{eq:43SZ}) is the only nontrivial part of the problem. The rest of the calculation reduces to quadratures [Eqs.\ (\protect\ref{eq:66TZ}), (\ref{eq:92VC})], and to separation of the frequency-positive and frequency-negative\ parts of known functions [Eqs.\ (\protect\ref{eq:99BE}), (\ref{eq:81UR})]. 

\section{Results and discussion}%
\label{ch:RD}
\subsection{What is known and what is not}%
\label{ch:CK}
We presume everything associated with Eq.\ (\protect\ref{eq:66TZ}) known---although we cannot think of a reference where the theory would be formulated in a similar way. Indeed, this equation and Eq.\ (\protect\ref{eq:43SZ}) do not depend on the type of Keldysh rotation used, and must appear, perhaps in disguise, in conventional real-time QFT. So, Aurenche and Becherrawy \cite{Therm1} derive a Dyson equation for the retarded propagator which is easily shown to be equivalent to (\ref{eq:43SZ}). {\em The result of this paper is Eq.\ (\protect\ref{eq:92VC})\/}. It expresses two remarkable features of the response picture: cancellation of zero-point fluctuations, and the resulting strict parallelism between quantum and classical stochastic electrodynamics. 
\subsection{The quantum and the classical in QED}%
\label{ch:CM}
Validity of the semiclassical approach of \mbox{Sec.\ \ref{ch:C}} may be seen as a generalisation of classical states of free electromagnetic\ field \cite{MandelWolf} to the interacting one. 
Classical states of free fields are defined by the condition, 
\begin{align} 
\begin{aligned} 
\protect\big \langle {\bf :}
\hat A_{\nu _1}(x _1)\cdots \hat A_{\nu _m}(x _m)
{\bf :} \big \rangle 
= 
\protect\big \langle A_{\nu _1}(x _1)\cdots A_{\nu _m}(x _m) \big \rangle 
, 
\end{aligned} 
\label{eq:93VD} 
\end{align}%
where \mbox{$A_{\nu}(x)$} is a random free classical electromagnetic\ potential. 
For the field interacting with a linear medium, classicality is defined postulating that upgrade (\ref{eq:89UZ}) reduces to a tautology, 
\begin{align} 
\begin{aligned} 
\protect\big \langle 
\mathcal{A}_{\nu}(x)
 \big \rangle 
 &= 
\protect\big \langle 
\protect{\hat{\mathcal A}}_{\nu}(x)
 \big \rangle , 
\quad
\protect\big \langle 
\mathcal{A}_{\nu}(x),
\mathcal{A}_{\nu '}(x ')
 \big \rangle 
= 
\protect\big \langle 
{\mathcal T}{\bf :}\protect{\hat{\mathcal A}}_{\nu}(x),
\protect{\hat{\mathcal A}}_{\nu '}(x ') 
{\bf :} 
 \big \rangle . 
\end{aligned} 
\label{eq:94VE} 
\end{align}%
Validity of these relations defines a linear medium which {\em appears\/} classical to a macroscopic observer \cite{endObs}\nocite{Sudar}, and which is {\em ipso facto\/} subject to the semiclassical theory of \mbox{Sec.\ \ref{ch:C}}. 
Since no immediate quantality may be observed in the average field, the crucial part is the interpretation of \mbox{$\protect\big \langle 
{\mathcal T}{\bf :}\protect{\hat{\mathcal A}}\protect{\hat{\mathcal A}}{\bf :}
 \big \rangle $} as a stochastic average. If this interpretation fails, a macroscopic observer detects a {\em quantum state\/} \cite{MandelWolf} of self-radiation of the linear medium. For a classical linear medium, Eqs.\ (\protect\ref{eq:4LZ}), (\ref{eq:99BE}) constitute in fact {\em quantisation relations\/} which turn a classical stochastic theory of the medium into a QED one. 

A word of extreme caution is in place here. A distinction should be maintained between {\em direct observation\/} and {\em inference\/} based on attempts at explanation (theoretical modelling). The best known example is black-body radiation. By itself, it is in a classical state and may be {\em described\/} in terms of classical statistics. Its quantum nature is inferred from the fact that classical dynamical models invariably fail to {\em explain\/} it, due to the equidistribution theorem. Similarly, we do not need quantum mechanics to {\em formulate\/} the results of observation of \mbox{$\mathcal{D}_{\text{R}}$}, but may require it to {\em explain\/} what we observe. For the time-normal cumulant, we already have three possibilities: 
\begin{itemize}
\item
this quantity cannot be interpreted classically (quantum manifestation by direct observation, e.g., violation of a Bell inequality); 
\item
it can be interpreted but not derived classically (quantum manifestation by inference, e.g., black-body radiation); 
\item
its observation agrees with some classical model (no quantum manifestation in observation). 
\end{itemize}
Saying that something is quantum, we always imply a quantum manifestation amenable to direct observation rather than following by inference. It is strictly according to this meaning that we say that the average field is always classical, while the time-normal cumulant may happen to be quantum. 
\subsection{Electromagnetic response of the Dirac sea (renormalisation for pedestrians)}%
\label{ch:D}
\subsubsection{The Dirac sea}%
\label{ch:DP}
As an example of a dynamical theory based on the quantum-classical correspondences formulated in the previous section, we consider the electromagnetic\ field in the Dirac vacuum. We assume that the reader is familiar with the Dirac equation, $\gamma $-matrices, 4-component spinor field \mbox{$
\hat \psi(x)
$}, the Dirac-adjoint \mbox{$
\hat{\bar\psi}(x)=\hat\psi^{\dag}(x)\gamma_0
$}, and other basic concepts. 
The quantized current is given by the standard formula, 
\begin{align} 
\hat J^{\nu }(x) = e c\,{\bf :}
\hat{\bar\psi}(x)\gamma^{\nu }\hat\psi(x)
{\bf :} . 
\label{eq:27PM} 
\end{align}%
The symbol \mbox{$
{\bf :}
\cdots
{\bf :} 
$} denotes the normal operator ordering \cite{Schweber,Bogol,MandelWolf}. 
The factor $c$ in (\ref{eq:27PM}) leads to the charge density defined naturally as, 
\begin{align} 
\hat\rho (x) = e\,{\bf :}
\hat{\bar\psi} (x)\gamma_0\hat{\psi} (x)
{\bf :} 
= e\,{\bf :}
\hat{\psi}^{\dag}(x)\hat{\psi} (x) 
{\bf :} , 
\label{eq:80WB} 
\end{align}%
so that dimension of the Dirac field is \mbox{m$^{-3/2}$}. 
With this reservation in mind, all formulae relevant to the Dirac field may be borrowed from the texts \cite{Itzykson,Bogol,Schweber}. The necessary minimum is summarised in appendix \ref{ch:DB}. 

Applied to the spinor field in a vacuum state, the general formulae (\ref{eq:65TY}), (\ref{eq:67UA}) and (\ref{eq:68UB}) yield, 
\begin{align} 
\Pi_{\mathrm{R}\nu }^{\nu '}(x-x') 
 &= - \frac{i}{\hbar c}\theta(x_0-x_0')\protect\big \langle 0\big| 
\protect\big [ 
\hat J_{\nu}(x),\hat J^{\nu '}(x ')
 \big ] 
 \big |0\big\rangle , 
\label{eq:72UF} 
\\ 
\Pi_{\mathrm{N}\nu\nu '}(x-x') 
 &= 2\Re\protect\big \langle 0\big| 
\hat J^{(-)}_{\nu}(x)\hat J^{(+)}_{\nu '}(x ')
 \big |0\big\rangle . 
\label{eq:73UH} 
\end{align}%
Simplifications to the second formula here compared to (\ref{eq:65TY}) are due to the fact that the averages, 
\begin{align} 
\begin{aligned} 
\protect\big \langle 0\big| 
T_+\hat J_{\nu}(x)\hat J_{\nu '}(x ')
 \big |0\big\rangle, \quad 
\protect\big \langle 0\big| 
\hat J_{\nu}(x)\hat J_{\nu '}(x ')
 \big |0\big\rangle, 
\end{aligned} 
\label{eq:74UJ} 
\end{align}%
depend on the argument difference. Recalling Eqs.\ (\protect\ref{eq:39BL}) and (\ref{eq:59XZ}), for any function of time difference we have, 
\begin{align} 
\begin{aligned} 
\mathcal{F}^{(\pm)}_{t} \mathcal{F}^{(\pm)}_{t'} f(t-t') = \mathcal{F}^{(\pm)}_{t} \mathcal{F}^{(\mp)}_{t} f(t-t') = 0. 
\end{aligned} 
\label{eq:76UL} 
\end{align}%
Hence the first term in (\ref{eq:65TY}) does not contribute to \mbox{$\protect\big \langle 0\big| 
{\mathcal T}{\bf :}\hat J\hat J{\bf :}
 \big |0\big\rangle $}. In the second term we moved the \mbox{${^{(\pm)}}$} operators inside the average. 

Both Eq.\ (\protect\ref{eq:72UF}) and Eq.\ (\protect\ref{eq:73UH}) require regularisation, the former explicitly and the latter implicitly. Indeed, when deriving the latter, we ignored the divergent nature of \mbox{$\protect\big \langle 0\big| 
T_+\hat J\hat J
 \big |0\big\rangle $}, i.e., assumed it regularised. Such regularization is carried out in any standard text \cite{Itzykson,Schweber,Bogol}. For an approach best suited to our purposes see \mbox{Sec.\ \ref{ch:L}} below.

\subsubsection{The commutator of spinor currents}%
\label{ch:DS}
Calculation of \mbox{$\Pi _{\text{R}}$} starts from calculation of the vacuum average of the commutator of currents in (\ref{eq:72UF}). The latter is a well-defined (convergent) quantity. Its calculation is a textbook exercise. It reduces to a large extent to recognizing implications of conservation of current (4-transversality) and relativistic covariance. In view of these properties we look for the average current commutator in the form, 
\begin{align} 
 &\protect\big \langle 0\big| 
\protect\big [ 
\hat J_{\mu }(x),\hat J_{\nu }(x')
 \big ] 
 \big |0\big\rangle 
= e^2c^2\big(
g_{\mu \nu }\Box - \partial_{\mu }\partial_{\nu }
 \big) \int \frac{d^4k}{(2\pi )^4}\text{e}^{-ik(x-x')}\varepsilon (k_0)
K\big(
k^2
 \big) , 
\label{eq:30PQ} 
\end{align}%
where $K(y)$ is a scalar function of scalar argument. As is shown in appendix \ref{ch:OE}, 
\begin{align} 
K\big(
k^2
 \big) = \frac{1}{6\pi } F\Big(
\frac{k^2}{4\mu_0^2}
 \Big), 
\label{eq:33PT} 
\end{align}%
where 
\begin{align} 
F(y) = \theta(y-1)\Big(
1 + \frac{1}{2y}
 \Big) \sqrt{1 - \frac{1}{y}}, 
\label{eq:28PN} 
\end{align}%
and 
$\mu_0$ is the mass of the electron in units of inverse length, 
\begin{align} 
\mu_0 = \frac{c\hspace{1pt}m_{\mathrm{electron}}}{\hbar } . 
\label{eq:81WC} 
\end{align}%
For details of the calculation see the appendix. 

\subsubsection{Regularized linear susceptibility of the Dirac sea}%
\label{ch:DD}
Commutator (\ref{eq:30PQ}) is a singular (generalized) function. Multiplying it by the step-function as per Eq.\ (\protect\ref{eq:72UF}) is not defined; ignoring this leads to divergences. Rather than proceeding formally with the multiplication and then sorting out the mess, we apply the Pauli-Villars regularization \cite{PauliVillars} directly to the average commutator, replacing, 
\begin{align} 
K(k^2) \to K^{\mathrm{reg}}(k^2) = \frac{1}{6\pi }\sum_{l=0}^N (-1)^l d_l F\Big(
\frac{k^2}{4\mu_l^2}
 \Big) , 
\label{eq:32PS} 
\end{align}%
where \mbox{$
d_0=1
$}, and \mbox{$
\mu_l\gg \mu_0,\ l=1,\cdots,N
$}, are regularization masses. The necessary number of these and the coefficients $d_l$ for \mbox{$
l\geq 1
$} are specified stipulating that the function, 
\begin{align} 
K^{\mathrm{reg}}(x) = \int \frac{d^4k}{(2\pi )^4}\text{e}^{-ikx}\varepsilon (k_0)
K^{\mathrm{reg}}(k^2), 
\label{eq:34PU} 
\end{align}%
be analytically benign, with a given number of continuous derivatives in the whole 4D space. Such regularization scheme is constructed in appendix \ref{ch:AR}. 

Calculation of the linear susceptibility with the regularized commutator is uneventful (appendix \ref{ch:OA}). The result reads, (with ``obs'' meaning {\em observable\/}) 
\begin{align} 
 &-\mu_{\mathrm{vac}}\Pi^{\mathrm{reg}}_{\mathrm{R}\mu \nu }(x-x')
=\big(
g_{\mu \nu }\Box - \partial_{\mu }\partial_{\nu }
 \big)\protect\big [ 
 R_0 \delta^{(4)}(x-x') 
+ R^{\mathrm{obs}}(x-x') 
 \big ] 
. 
\label{eq:54QR} 
\end{align}%
In this relation, \mbox{$
R^{\mathrm{obs}}(x-x')
$} stands for the quantity, 
\begin{align} 
 &R^{\mathrm{obs}}(x-x') = \int \frac{d^4k}{(2\pi )^4}\text{e}^{-ik(x-x')}
R^{\mathrm{obs}}\big(
k
 \big) 
, 
\label{eq:26TT} 
\end{align}%
where \cite{endSobs} 
\begin{align} 
R^{\mathrm{obs}}(k) 
 &= \frac{\alpha k^2}{3\pi }
\int_{4\mu_0^2}^{\infty} 
\frac{d\mu^2F\big(
{\mu^2}/{4\mu_0^2}
 \big)}{\mu^2\big(
\mu^2-k^2-i 0^+ \operatorname{sign} k_0 \big)
} 
, 
\label{eq:53QQ} 
\end{align}%
$\alpha $ is the fine structure constant \cite{endAlpha}, 
\begin{align} 
\alpha =\frac{e^2 c\mu_{\mathrm{vac}}}{4\pi \hbar } =\frac{e^2}{4\pi\varepsilon_{\mathrm{vac}}\hbar c}, 
\label{eq:83WE} 
\end{align}%
and 
\mbox{$
R_0
$} is a logarithmically divergent constant, 
\begin{align} 
 R_0 = -\frac{\alpha }{3\pi}\sum_{l=1}^N(-1)^l d_l 
 \ln \frac{\mu_l^2}{\mu_0^2} . 
\label{eq:59QW} 
\end{align}%
{\em Divergence\/} in this context refers to the way \mbox{$
 R_0
$} depends on regularization masses; thanks to early regularization, all quantities we work with are finite. The infinitesimal imaginary shift in Eq.\ (\protect\ref{eq:53QQ}) assures retardation of \mbox{$
R^{\mathrm{obs}}(x-x')
$}. Regularization of the commutator has dramatically reduced the degree of divergence of the linear susceptibility (from 
quadratic to logarithmic). Elimination of the remaining divergence is a matter of physics rather than mathematics. 
\subsubsection{The long-wavelength limit and renormalization}%
\label{ch:DY}
When substituting (\ref{eq:54QR}) in Eq.\ (\protect\ref{eq:21PE}), the terms proportional to $\partial_{\mu }\partial_{\nu }$ vanish due to the Lorentz condition (\ref{eq:24TR}). For Fourier-components of the field and current we then find, 
\begin{align} 
-k^2 \protect\big [ 
1+ R_0 + R^{\mathrm{obs}}(k)
 \big ] 
A_{\mu }(k) 
= \mu_{\mathrm{vac}}J_{\mathrm{e}\mu }(k) . 
\label{eq:85WH} 
\end{align}%
If the source current changes slowly in space-time on the scale of \mbox{$
\mu_0^{-1}
$}, the relevant range of $k$ is limited to \mbox{$
|k^2|\ll \mu_0^2
$}. In this limit, the integral in Eq.\ (\protect\ref{eq:53QQ}) is a constant. By direct integration, 
\begin{align} 
R^{\mathrm{obs}}(k) &= \frac{\alpha k^2}{15\pi\mu_0^2},\quad |k^2|\ll \mu_0^2 , 
\label{eq:28TV} 
\end{align}%
and we find, 
\begin{align} 
-k^2\Big(
1+ R_0+\frac{\alpha k^2}{15\pi \mu_0^2}
 \Big)A_{\mu }(k) = \mu_{\mathrm{vac}}J_{\mathrm{e}\mu }(k),\quad |k^2|\ll \mu_0^2. 
\label{eq:29TW} 
\end{align}%
The standard renormalization condition is that corrections due to vacuum polarization must disappear for macroscopic distances and low frequencies. Thus the renormalization condition is, 
\begin{align} 
\begin{aligned} 
R_0=0 . 
\end{aligned} 
\label{eq:81QJ} 
\end{align}%
By amending the regularization scheme one can make \mbox{$
 R_0
$} equal anything. One simply treats Eq.\ (\protect\ref{eq:59QW}) {\em with given\/} \mbox{$
 R_0
$} as an additional condition for the $d_l$'s. \mbox{$
 R_0
$} may be made logarithmically divergent, finite, or, indeed, zero. {\em The renormalization condition \mbox{$
 R_0=0
$} may hence be imposed directly on the regularization scheme, making our semiclassical approach self-contained.\/} For details see appendix \ref{ch:AR}. Observable linear susceptibility of the Dirac vacuum is given by (\ref{eq:54QR}) with \mbox{$R_0=0$}. 

\subsection{Simplest current-related kernels in spinor QED}%
\label{ch:L} 
\subsubsection{C-number kernels associated with the current operator}%
\label{ch:LK} 
In terms of the questions (a) and (b) formulated at the end of \mbox{Sec.\ \ref{ch:TX}}, the example in \mbox{Sec.\ \ref{ch:D}} showed, in particular, that renormalisation may be naturally included in the response viewpoint. The obvious question is how the results of \mbox{Sec.\ \ref{ch:D}} correspond to more traditional approaches. Here, we show that the ``pedestrian'' approach of \mbox{Sec.\ \ref{ch:D}} may be seamlessly integrated into conventional techniques of QFT. Certain texbook results directly follow from our approach. 

Susceptibility \mbox{$\Pi _{\text{R}}$} calculated in \mbox{Sec.\ \ref{ch:D}} is part of the set of c-number kernels associated with the free current operator, 
\begin{align} 
\Pi_{\nu \nu '}(x-x') &= (i\hbar c )^{-1}\protect\big \langle 0\big| 
\protect\big [ 
\hat J_{\nu }(x),\hat J_{\nu' }(x')
 \big ] 
 \big |0\big\rangle 
= \Pi^{(+)}_{\nu \nu '}(x-x')+\Pi^{(-)}_{\nu \nu '}(x-x'), 
\label{eq:44JL} 
\\ 
\Pi^{(+)}_{\nu \nu '}(x-x') &= (i\hbar c )^{-1}\protect\big \langle 0\big| 
\hat J_{\nu }(x)\hat J_{\nu' }(x')
 \big |0\big\rangle 
= - \Pi^{(-)}_{\nu' \nu }(x'-x) , 
\label{eq:45JM} 
\\ 
\Pi_{\mathrm{F}\nu \nu '}(x-x') &= (i\hbar c )^{-1}\protect\big \langle 0\big| 
T_+ \hat J_{\nu }(x)\hat J_{\nu' }(x') \big |0\big\rangle 
\nonumber\\ &
= \theta(x_0-x_0')\Pi^{(+)}_{\nu \nu '}(x-x')
- \theta(x_0'-x_0)\Pi^{(-)}_{\nu \nu '}(x-x')
, 
\label{eq:34YK} 
\\
\Pi_{\mathrm{R}\mu \mu '}(x-x') 
 &= (i\hbar c )^{-1}\theta(x_0-x_0')\protect\big \langle 0\big| 
\protect\big [ 
\hat J_{\nu }(x),\hat J_{\nu' }(x')
 \big ] 
 \big |0\big\rangle
\nonumber\\ &
= \theta(x_0-x_0')\Pi_{\mu \mu '}(x-x') . 
\label{eq:39YQ} 
\end{align}%
Unlike for free bosonic fields, the commutator in (\ref{eq:44JL}) is not a c-number, so that the vacuum averaging is essential. Otherwise Eqs.\ (\protect\ref{eq:44JL})--(\ref{eq:39YQ}) follow the pattern of the ``algebra of Green functions'' of a free bosonic field discussed in \cite{API,WickCaus}. 
\subsubsection{Response transformation of the current-related kernels}%
\label{ch:LRX} 
Similarities with free fields include the critical point: frequency-positiveness of \mbox{$\Pi^{(+)}$} (as the notation suggests). Basically, this quantity is defined by Eq.\ (\protect\ref{eq:45JM}) and not as the frequency-positive\ part of \mbox{$\Pi $}. 
However, owing to the normal ordering of the current operator (\ref{eq:27PM}), we have, 
\begin{gather} 
\begin{gathered} 
\hat J_{\nu }(x)\left|0\right\rangle
=\hat I_{\nu }^{(-)}(x)\left|0\right\rangle, 
\quad 
\protect \langle 
0
 | \hat J_{\nu }(x) 
= \protect \langle 
0
 |\hat I_{\nu }^{(+)}(x) , 
\end{gathered} 
\label{eq:25XZ} 
\end{gather}%
where \mbox{$\hat I_{\nu }^{(+)}(x)$} is a {\em frequency-positive\/} bosonic operator, 
\begin{align} 
\begin{aligned} 
\hat I_{\nu }^{(+)}(x) = \protect\big [ 
\hat I_{\nu }^{(-)}(x) \big ]^{\dag}= \hat{\bar\psi}^{(+)}(x)\gamma_{\nu } \hat\psi^{(+)}(x) . 
\end{aligned} 
\label{eq:46JN} 
\end{align}%
This way, 
\begin{align} 
\begin{aligned} 
 &\protect\big \langle 0\big| 
\hat J^{\nu }(x)\hat J^{\nu' }(x')
 \big |0\big\rangle 
= \protect\big \langle 0\big| 
\hat I^{\nu(+)}(x)
\hat I^{\nu'(-)}(x')
 \big |0\big\rangle , 
\end{aligned} 
\label{eq:77UM} 
\end{align}%
making frequency-positiveness of \mbox{$\Pi ^{(+)}$} evident. Response transformation of the kernels \mbox{$
\Pi_{\text{F}}
$} and \mbox{$
\Pi^{(+)}
$}, namely, 
\begin{align} 
\Pi_{\mathrm{F}\nu \nu '}(x-x') 
= \Pi^{(+)}_{\mathrm{R}\nu \nu '}(x-x') 
+ \Pi^{(+)}_{\mathrm{R}\nu' \nu}(x'-x) 
, 
\label{eq:37YN} 
\\ 
\Pi^{(+)}_{\nu \nu '}(x-x') 
= \Pi^{(+)}_{\mathrm{R}\nu \nu '}(x-x') 
- \Pi^{(-)}_{\mathrm{R}\nu' \nu}(x'-x) . 
\label{eq:47JP} 
\end{align}%
may then be shown exactly as for free quantized bosonic fields \cite{API,APII} (cf.\ also appendix \ref{ch:B}). 

\subsubsection{Regularisation of the divergent kernels}%
\label{ch:LRY} 
The kernels \mbox{$\Pi _{\text{F}}$} and \mbox{$\Pi _{\text{R}}$} are divergent, so that Eqs.\ (\ref{eq:34YK}), (\ref{eq:39YQ}), (\ref{eq:37YN}), and (\ref{eq:47JP}) are only symbolic. A consistent way of simultaneous regularisation of both divergent kernels is to replace the unregularised current commutator (\ref{eq:44JL}) by its regularised version as in \mbox{Sec.\ \ref{ch:DD}}. By definition, Eqs.\ (\protect\ref{eq:34YK}), (\ref{eq:39YQ}) then specify regularised versions of \mbox{$\Pi _{\text{F}}$} and \mbox{$\Pi _{\text{R}}$}, denoted \mbox{$\Pi _{\text{F}}^{\mathrm{reg}}$} and \mbox{$\Pi _{\text{R}}^{\mathrm{reg}}$}. For the latter, we rediscover Eq.\ (\protect\ref{eq:54QR}). \mbox{$\Pi _{\text{F}}^{\mathrm{reg}}$} may be obtained by similar means. However, with \mbox{$\Pi _{\text{R}}^{\mathrm{reg}}$} known, the easiest way to obtain \mbox{$
\Pi _{\text{F}}^{\mathrm{reg}}
$} is to use Eq.\ (\protect\ref{eq:37YN}). 
Indeed, frequency-positiveness of \mbox{$
\Pi ^{(+)}
$} extends to the corresponding regularized quantity, 
so that response transformations (\ref{eq:37YN}), (\ref{eq:47JP}) also survive regularisation. In fact, \mbox{$
\Pi ^{(+)}
$} is not divergent; it follows by replacing \mbox{$
\varepsilon (k_0)\to \theta(k_0)
$} in Eq.\ (\protect\ref{eq:30PQ}) for the unregularized commutator. As a demonstration of consistency, we also calculate it from Eq.\ (\protect\ref{eq:47JP}). 

\subsubsection{Explicit formula for regularised \mbox{$\Pi _{\text{F}}$}}%
\label{ch:LS} 
We change to momentum representation, so as to have a more direct connection to conventional QFT. In momentum space, Eq.\ (\protect\ref{eq:53QQ}) becomes, 
\begin{align} 
\begin{aligned} 
-\mu_{\mathrm{vac}}\Pi_{\mathrm{R}\nu \nu '}^{\mathrm{reg}}(k) &= \big(
k_{\nu }k_{\nu '}-k^2 g_{\nu \nu '}
 \big) 
\protect\bigg [ 
R_0 
+ \frac{\alpha k^2}{3\pi }
\int_{4\mu_0^2}^{\infty} 
\frac{d\mu^2F\big(
{\mu^2}/{4\mu_0^2}
 \big)}{\mu^2\big(
\mu^2-k^2-i \operatorname{sign}k_0 \,0^+ \big)
} 
 \bigg ] , 
\end{aligned} 
\label{eq:49JR} 
\end{align}%
where \mbox{$
F(y)
$} is given by Eq.\ (\protect\ref{eq:28PN}). We use the same notation for a function and its Fourier-image; what we have in mind is clear from the notation for the argument. 

Transformations (\ref{eq:37YN}), (\ref{eq:47JP}) in momentum representation become, 
\begin{align} 
\begin{aligned} 
\Pi_{\mathrm{F}\nu \nu '}^{\mathrm{reg}}(k) 
 &= \theta(k_0)\Pi_{\mathrm{R}\nu \nu '}^{\mathrm{reg}}(k)
+ \theta(-k_0)\Pi_{\mathrm{R}\nu'\nu}^{\mathrm{reg}}(-k), 
\quad 
\Pi_{\nu \nu '}^{(+)}(k) 
= 2i\theta(k_0)\Im \Pi_{\mathrm{R}\nu \nu '}^{\mathrm{reg}}(k). 
\end{aligned} 
\label{eq:48JQ} 
\end{align}%
In obtaining these formulae, we employed the ``dictionary'' relating operations on kernels in space and in momentum space, 
\begin{align} 
\begin{aligned} 
\Pi (x) &\rightarrow \Pi (k), &
\Pi^{(\pm)}(x) &\rightarrow \theta(\pm k_0)\Pi (k), \\
\Pi (-x) &\rightarrow \Pi (-k), &
\Pi^* (-x) &\rightarrow \Pi^* (k) . 
\end{aligned} 
\label{eq:50JS} 
\end{align}%
Use was also made of the formula, 
\begin{align} 
\begin{aligned} 
\Pi ^{(\pm)}(x-x') = \protect\big [ 
\Pi ^{(\mp)}(x-x')
 \big ] ^*, 
\end{aligned} 
\label{eq:51JT} 
\end{align}%
valid for any real kernel. Taking notice of the symmetry of \mbox{$
\Pi_{\mathrm{R}\nu \nu '}^{\mathrm{reg}}
$} as a 4-tensor we recover the textbook result \cite{Itzykson,Schweber,Bogol}, 
\begin{align} 
\begin{aligned} 
-\mu_{\mathrm{vac}}\Pi_{\mathrm{F}\nu \nu '}^{\mathrm{reg}}(k) 
 &= \big(
k_{\nu }k_{\nu '}-k^2 g_{\nu \nu '}
 \big) 
\protect\bigg [ 
R_0 
+ \frac{\alpha k^2}{3\pi }
\int_{4\mu_0^2}^{\infty} 
\frac{d\mu^2F\big(
{\mu^2}/{4\mu_0^2}
 \big)}{\mu^2\big(
\mu^2-k^2-i 0^+ \big)
} 
 \bigg ] . 
\end{aligned} 
\label{eq:53JV} 
\end{align}%
The renormalized expression follows with \mbox{$
R_0 =0
$}. For \mbox{$
\Pi ^{(+)}
$} we use the formula 
\begin{align} 
\begin{aligned} 
\Im \frac{1}{\mu ^2-k^2-i0^+} = \pi \delta(\mu ^2-k^2) . 
\end{aligned} 
\label{eq:54JW} 
\end{align}%
By direct integration we then obtain, 
\begin{align} 
\begin{aligned} 
-\mu_{\mathrm{vac}}\Pi_{\nu \nu '}^{(+)}(k)
 &= \frac{2i\alpha }{3} \theta(k_0)\theta(k^2-4\nu_0^2) 
(k_{\nu }k_{\nu '}-k^2 g_{\nu \nu '}) 
F\Big(
\frac{k^2}{4\nu_0^2}
 \Big) . 
\end{aligned} 
\label{eq:55JX} 
\end{align}%
This coincides with the formula one finds directly from Eq.\ (\protect\ref{eq:30PQ}), subject to Eq.\ (\protect\ref{eq:83WE}). The renormalization parameter \mbox{$
R_0
$} has canceled as expected: since Eq.\ (\protect\ref{eq:30PQ}) does not ``know'' about \mbox{$
R_0
$}, any alternative derivation must also somehow ``forget'' about it. 
\subsection{Zero-point fluctuations in the Dirac vacuum}%
\label{ch:ZP}
\subsubsection{No zero-point fluctuations in the Dirac vacuum in the time-normally-ordered representation}%
\label{ch:JS}
It is instructive to consider zero-point fluctuations of the electromagnetic field. We start from showing that \mbox{$\Pi _{\mathrm{N}}$} given by Eq.\ (\protect\ref{eq:73UH}) vanishes. Indeed, remembering Eq.\ (\protect\ref{eq:77UM}), 
\begin{align} 
 &\protect\big \langle 0\big| 
{\mathcal T}{\bf :}\hat J^{\nu }(x)\hat J^{\nu' }(x'){\bf :}
 \big |0\big\rangle 
= 2\Re\protect\big \langle 0\big| 
\protect\big [ 
\hat I^{\nu(+)}(x)
 \big ]^{(-)}\protect\big [ 
\hat I^{\nu'(-)}(x')
 \big ]^{(+)}
 \big |0\big\rangle = 0 . 
\label{eq:75VW} 
\end{align}%
This quantity iz zero, because the frequency-negative\ part of any frequency-positive\ quantity is zero, and {\em vice versa\/}. 

This way, irrespective of what \mbox{$\mathcal{D}_{\text{R}}$} in the Dirac vacuum is, Eqs.\ (\protect\ref{eq:92VC}) and (\ref{eq:75VW}) predict cancellation of zero-point fluctuations of the electromagnetic\ field in the Dirac vacuum in the time-normally ordered representation. We have indeed recovered Eq.\ (\protect\ref{eq:40BM}), introduced as a conjecture in \mbox{Sec.\ \ref{ch:QNB}}. 
\subsubsection{Zero-point fluctuations under orderings other than time-normal}%
\label{ch:DA}
Combining Eqs.\ (\protect\ref{eq:28AY}) and (\ref{eq:40BM}) we find, 
\begin{align} 
\begin{aligned} 
\protect\big \langle 0\big| \mathcal{T}_s{\protect{\hat{\mathcal A}}_{\nu }(x),\protect{\hat{\mathcal A}}_{\nu '}(x')} \big |0\big\rangle = 2 s_-\mathcal{Z}_{\nu\nu '}(x,x'), 
\end{aligned} 
\label{eq:41BN} 
\end{align}%
where \mbox{$\mathcal{Z}$} is given by Eq.\ (\protect\ref{eq:27AX}). For the Dirac sea, this quantity depends on \mbox{$x-x'$}. We calculate it in momentum representation, 
\begin{align} 
\begin{aligned} 
\mathcal{Z}_{\nu\nu '}(x,x') = -g_{\nu \nu '}\int \frac{d^4k}{(2\pi )^4} 
\text{e}^{-ik(x-x')}\mathcal{Z}(k). 
\end{aligned} 
\label{eq:42BP} 
\end{align}%
This formula implies the Feynman gauge for \mbox{$\mathcal{D}_{\text{R}}$}, 
\begin{align} 
\begin{aligned} 
\mathcal{D}_{\mathrm{R}\nu\nu '}(x,x') = -g_{\nu \nu '}\int \frac{d^4k}{(2\pi )^4} 
\text{e}^{-ik(x-x')}\mathcal{D}_{\mathrm{R}}(k). 
\end{aligned} 
\label{eq:48BV} 
\end{align}%
From Eq.\ (\protect\ref{eq:27AX}) we have, 
\begin{align} 
\begin{aligned} 
\mathcal{Z}(k) = \frac{i\hbar c}{2}\protect\big [ 
\mathcal{D}_{\mathrm{R}}(k)-\mathcal{D}_{\mathrm{R}}(-k)
 \big ] \operatorname{sign}k_0 . 
\end{aligned} 
\label{eq:44BR} 
\end{align}%
For \mbox{$\mathcal{D}_{\mathrm{R}}(k)$} we find from Eq.\ (\protect\ref{eq:85WH}), 
\begin{align} 
\begin{aligned} 
\mathcal{D}_{\mathrm{R}}(k) = \frac{\mu_{\mathrm{vac}}}{\big(
k^2 + i0^+\operatorname{sign}k_0
 \big) \protect\big [ 
1 + \mathcal{R}^{\mathrm{obs}}(k)
 \big ] } . 
\end{aligned} 
\label{eq:43BQ} 
\end{align}%
The infininitesimal imaginary shift is also present in the definition of \mbox{$\mathcal{R}^{\mathrm{obs}}(k)$} by Eq.\ (\protect\ref{eq:53QQ}). Inversion of the argument of \mbox{$\mathcal{D}_{\text{R}}$} in (\ref{eq:44BR}) manifests itself only through inversion of this shift, hence, 
\begin{align} 
\begin{aligned} 
\mathcal{D}_{\mathrm{R}}(k)-\mathcal{D}_{\mathrm{R}}(-k) = 2 i \Im\mathcal{D}_{\text{R}}(k). 
\end{aligned} 
\label{eq:45BS} 
\end{align}%
The two factors in (\ref{eq:43BQ}) never happen to be complex simultaneously. The first factor is purely imaginary for \mbox{$k^2=0$}, when \mbox{$\mathcal{R}^{\mathrm{obs}}(k)=0$}, and real otherwise. The second factor acquires an imaginary part only for \mbox{$k^2>4\mu _0^2$}. With these observations it is straightforward to obtain, 
\begin{align} 
\begin{aligned} 
\mathcal{Z}(k) &= \hbar c\mu_{\mathrm{vac}}
\protect\Bigg [ 
\pi \delta(k^2) + 
\frac{}{}\theta(k^2-4\mu _0^2) 
\frac{\alpha F\big(
k^2/4\mu _0^2
 \big) }{3k^2\big|1+\mathcal{R}(k)\big|^2}
 \Bigg ] . 
\end{aligned} 
\label{eq:46BT} 
\end{align}%
Unlike \mbox{$\mathcal{D}_{\mathrm{R}}(k)$}, \mbox{$\mathcal{Z}(k)$} is a full relativistic scalar; it depends only on \mbox{$k^2$}. The first term in (\ref{eq:46BT}) is a free-field contribution. The second one comes from interactions (virtual pair creation). 

\subsubsection{Unphysical nature of electromagnetic\ zero-point fluctuations in the Dirac vacuum}%
\label{ch:UPN}
Our actual motivation for deriving Eqs.\ (\protect\ref{eq:41BN}), (\ref{eq:42BP}) and (\ref{eq:46BT}) was to show that, {\em while impeccable mathematically, physically they are inconsistent\/}. Indeed, according to Eq.\ (\protect\ref{eq:42BP}), the zero-point fluctuations do not obey the Lorentz condition, nor Maxwell's equations. Their interpretation as fluctuations of the electromagnetic\ field, whether quantum or classical, appears to be too much of a stretch. 
 
From the first glance, the problem originates in the Feynman gauge in (\ref{eq:48BV}). Physically, \mbox{$\mathcal{D}_{\text{R}}$} is defined up to the transformation, 
\begin{align} 
\begin{aligned} 
\mathcal{D}_{\mathrm{R}\nu\nu '}(x,x') \to 
\mathcal{D}_{\mathrm{R}\nu\nu '}(x,x') 
+ \partial_{\nu}\partial_{\nu '}\mathcal{D}_{\parallel}(x-x'), 
\end{aligned} 
\label{eq:83DJ} 
\end{align}%
where \mbox{$\mathcal{D}_{\parallel}$} is to a large extent arbitrary. When deriving Eqs.\ (160), (165), \mbox{$\mathcal{D}_{\parallel}$} was chosen so as to ensure the Feynman gauge for \mbox{$\mathcal{D}_{\mathrm{R}}$}. The result is thus correct only up to a gauge transformation. One may try to choose it so as to assure the replacement, 
\begin{align} 
\begin{aligned} 
g_{\nu \nu '}\mathcal{Z}(k)\to\Big(
g_{\nu \nu '} - \frac{k_{\nu }k_{\nu '}}{k^2}
 \Big) \mathcal{Z}(k). 
\end{aligned} 
\label{eq:84DK} 
\end{align}%
However, this leads to emergence of a mathematically meaningless term \mbox{$\propto\delta (k^2)/k^2$}. A physical inconsistency is substituted by a mathematical one. 

In the Feynman gauge, the expression for the zero-point fluctuations of the electromagnetic\ field may be obtained, but it happens to be physically unsatisfactory. An attempt to rescure the situation by using the 4-transverse (Lorentz) gauge fails on mathematical grounds---not to mention that non-gauge-invariant zero-point fluctuations of the electromagnetic\ field are a serious problem by themselves. There does not seem to be a way of defining the zero-point fluctuations of the electromagnetic field\ in a consistent way---except in response representation, where they vanish. 

Obviously, we have not discovered anything fundamentally new. The Gupta-Bleuler potential operator does not obey the Lorentz condition, nor Maxwell's equations, nor do its {\em arbitrary\/} matrix elements. In the standard texts, this problem is dealt with imposing the Lorentz condition on the states of the electromagnetic\ field. This suffices to ensure that the average field obeys this condition. However, as can be seen from Eq.\ (\protect\ref{eq:46BT}), this is not enough to ensure that fluctuations of the electromagnetic\ field obey the Lorentz condition and Maxwell's equations (formally, because in averages of operator products, one cannot impose conditions on intermediate quantum states). As soon as we become interested in the Hanbury Brown-Twiss kind of measurements \cite{BrownTwissI,BrownTwissII}, the problem reemerges. The exception is the response representation. This is yet another argument in favour of the special role of this representation and of the related time-normal ordering, at least in QED. We return to this question elsewhere. 

\section{Conclusion and outlook}%
\label{ch:W}
In conclusion, it is shown that the Keldysh rotation in the real-time QFT and the response transformation \cite{API,APII,APIII} are particular cases of generalised Keldysh rotations. The latter are defined as phase-space mappings of interacting quantum fields, each based on a particular type of operator ordering. General structural properties of response transformation shown in \cite{APII,APIII} hold in fact for arbitrary generalised rotation. This includes causality properties. The characteristic feature of response transformation is cancellation of zero-point fluctuations in dynamics, shown in this paper for a relativistic electromagnetic\ field interacting with the Dirac vacuum in the first nonvanishing order of perturbation theory. 

While results of this paper are encouraging, the general consistency between response transformations and renormalizations remains an open problem. Cancellation of zero-point fluctuations has an interesting side-effect: the free quantised electromagnetic\ field which in the standard Gupta-Bleuler theory does not obey Maxwell's equations is eliminated from the theory. One should therefore expect that, under response transformation, the electromagnetic\ field should obey the Lorentz condition and the Maxwell equations. In fact things are more subtle, because of the so-called Schwinger terms \cite{SchwingerTerms,WeinbergI}. This question also remains open for discussion. 

An interesting physical question we have not even touched upon is the interplay between response and quantum noise under Lorentz transformations. Separation of the frequency-positive and frequency-negative\ parts is not an invariant operation, except for free fields. Observable response and noise properties of a macroscopic device (say) should therefore depend on the reference frame. Understanding this is subject to further work. 

\section*{Acknowledgements}
Support of SFB/TRR 21 and of the Humboldt Foundation is acknowledged. 

\appendix


\section{Generalised Keldysh rotations for free fields beyond Gaussian states}%
\label{ch:OS}
\subsection{Wick's theorem\ldots}%
\label{ch:OSW}
The goal of this appendix is derivation of Eq.\ (\protect\ref{eq:19RY}). We do not posess an equivalent of Eq.\ (\protect\ref{eq:17MP}) for operators, and therefore make our starting point the normal rather that the symmetric ordering. The relation between the closed-time-loop\ and normally-ordered operator products is well known: it is given by Wick's theorem \cite{Keldysh} (not to be confused with Wick's theorem for Gaussian systems (\ref{eq:15RU})). 

As was shown in \cite{API}, Wick's theorem for the closed-time-loop\ ordering may be written as a closed formula, 
\begin{align} 
\begin{aligned} 
\hat\Xi[\eta_+,\eta_-] = \protect\big \langle 0\big| 
\hat\Xi[\eta_+,\eta_-]
 \big |0\big\rangle\ {\bf :}
\exp\protect\big [ 
-i(\eta_+-\eta_-)\hat q
 \big ] 
{\bf :}, 
\end{aligned} 
\label{eq:46XL} 
\end{align}%
where \mbox{$\hat\Xi[\eta_+,\eta_-]$} is the operator-valued generating functional of the closed-time-loop\-ordered products of the displacement operator, 
\begin{align} 
\begin{aligned} 
 &\hat\Xi[\eta_+,\eta_-] 
= 
T_C\exp\big(
-i\eta_+\hat q_+
+i\eta_-\hat q_-
 \big) . 
\end{aligned} 
\label{eq:13MK} 
\end{align}%
For definitions see \mbox{Sec.\ \ref{ch:A}}. 
We use condensed notation (\ref{eq:47XM}) where possible. Furthermore, 
\begin{align} 
\begin{aligned} 
 &\protect\big \langle 0\big| 
\hat\Xi[\eta_+,\eta_-]
 \big |0\big\rangle 
= \exp\protect\Big [ 
-\frac{i\hbar }{2}\eta_+D_{\text{F}}\eta_+
+\frac{i\hbar }{2}\eta_-D_{\text{F}}^*\eta_-
+i\hbar\eta_-D^{(+)}\eta_+
 \Big ] , 
\end{aligned} 
\label{eq:48XN} 
\end{align}%
where \mbox{$D_{\text{F}},D^{(+)}$} are the Keldysh contractions, 
\begin{align} 
\begin{aligned} 
i\hbar D_{\text{F}}(t-t') &= \protect\big \langle 0\big| 
T_+\hat q(t)\hat q(t')
 \big |0\big\rangle, 
\quad 
i\hbar D^{(+)}(t-t') = \protect\big \langle 0\big| 
\hat q(t)\hat q(t')
 \big |0\big\rangle . 
\end{aligned} 
\label{eq:49XP} 
\end{align}%
Wick's theorem as such follows by expanding (\ref{eq:46XL}) in a functional Taylor series. 
\subsection{\ldots\ and its response transformation}%
\label{ch:OSZ}
It was also shown in \cite{API} that response transformation reduces the bilinear form in the exponent to the single kernel \mbox{$D_{\text{R}}$}, 
\begin{align} 
\begin{aligned} 
\protect\big \langle 0\big| 
\hat\Xi[\eta_+,\eta_-]
 \big |0\big\rangle\settoheight{\auxlv}{$|$}%
\raisebox{-0.3\auxlv}{$|_{\eta_{\pm}\to\eta,j_{\mathrm{e}}}$} = \exp\big(
-i\eta D_{\text{R}}j_{\mathrm{e}}
 \big) , 
\end{aligned} 
\label{eq:50XQ} 
\end{align}%
where 
\begin{align} 
\begin{aligned} 
j_{\mathrm{e}}(t) &= \hbar \protect\big [ 
\eta^{(+)}_+(t) + \eta^{(-)}_-(t) 
 \big ] , 
\end{aligned} 
\label{eq:13VZ} 
\end{align}%
and \mbox{$\eta (t)$} is given by Eq.\ (\protect\ref{eq:51BY}). 
Combining Eqs.\ (\protect\ref{eq:46XL}) and (\ref{eq:50XQ}) 
we find the formula, 
\begin{align} 
\begin{aligned} 
 &\hat\Xi[\eta_+,\eta_-]\settoheight{\auxlv}{$|$}%
\raisebox{-0.3\auxlv}{$|_{\eta_{\pm}\to\eta,j_{\mathrm{e}}}$} 
= \exp\big(
-i\eta D_{\text{R}}j_{\mathrm{e}}
 \big)\ {\bf :}
\exp\big(
-i\eta\hat q
 \big) 
{\bf :} . 
\end{aligned} 
\label{eq:51XR} 
\end{align}%
This relation is the starting point of analyses in this appendix. 
\subsection{Functional reordering formula for free fields}%
\label{ch:OSF}
To extend Eq.\ (\protect\ref{eq:51XR}) to other orderings, we construct a formula connecting \mbox{$O_s$}-ordered products of displacement operators for two different values of \mbox{$s$}. From Eq.\ (\protect\ref{eq:23MV}) we have, 
\begin{align} 
\begin{aligned} 
O_{s'} \exp\big(
\beta \hat a^{\dag}- \beta ^*\hat a
 \big) 
 &= 
O_{s} \exp\big(
\beta \hat a^{\dag}- \beta ^*\hat a
 \big) 
\exp\frac{(s'-s)|\beta |^2}{2} . 
\end{aligned} 
\label{eq:41XE} 
\end{align}%
Now, let 
\begin{align} 
\begin{aligned} 
\beta = - i \sqrt{\frac{\hbar }{2}} 
\int dt \eta (t) \text{e}^{i\omega_0t} . 
\end{aligned} 
\label{eq:40NP} 
\end{align}%
Equation (\ref{eq:41XE}) then turns into a reordering formula for the displacement operators, 
\begin{align} 
\begin{aligned} 
 &O_{s'} \exp\big(
-i\eta\hat q
 \big) 
= 
O_{s} \exp\big(
-i\eta\hat q
 \big) 
\exp\int dt dt'\protect\bigg [ 
\frac{(s'-s)\hbar }{4}\eta (t)\eta (t')\text{e}^{i\omega _0(t-t')}
 \bigg ] . 
\end{aligned} 
\label{eq:41NQ} 
\end{align}%
Because of symmetrisation imposed by the integration, the exponent here may be replaced by the cosine, 
\begin{align} 
\begin{aligned} 
\text{e}^{i\omega _0(t-t')}\to \cos\omega _0(t-t'). 
\end{aligned} 
\label{eq:42XF} 
\end{align}%
Recalling (\ref{eq:30BA}) we arrive at the relation sought, 
\begin{align} 
\begin{aligned} 
 &O_{s'} \exp\big(
-i\eta\hat q
 \big) 
= 
O_{s} \exp\big(
-i\eta\hat q
 \big) 
\exp\protect\Big \{ 
\frac{i(s'-s)\hbar }{2} 
\eta D_{\text{R}}\protect\big [ 
\eta^{(+)}-\eta^{(-)}
 \big ]
 \Big \} . 
\end{aligned} 
\label{eq:43XH} 
\end{align}%
This formula generalises Eq.\ (\protect\ref{eq:30BA}) beyond Gaussian states of the oscillator. 
\subsection{Derivation of Eq.\ (\protect\ref{eq:19RY})}%
\label{ch:GKC}
We now recall that response transformation is the generalised Keldysh rotation for \mbox{$s=1$}, and that the related ordering coincides with the normal ordering, \mbox{$O_1\cdots = {\bf :}
\cdots
{\bf :} $}. Using the reordering formula (\ref{eq:43XH}) with \mbox{$s'=1$} we find, 
\begin{align} 
\begin{aligned} 
{\bf :}
\exp\big(
-i\eta\hat q
 \big){\bf :} &= O_s\exp\big(
-i\eta\hat q
 \big) 
\exp\protect\Big \{ 
i\hbar s_- 
\eta D_{\text{R}}\protect\big [ 
\eta^{(+)}-\eta^{(-)}
 \big ]
 \Big \} , 
\end{aligned} 
\label{eq:52XS} 
\end{align}%
where \mbox{$s_-$} is given by Eq.\ (\protect\ref{eq:82QK}). Combining Eqs.\ (\protect\ref{eq:51XR}) and (\ref{eq:52XS}) yields, 
\begin{align} 
\begin{aligned} 
\hat\Xi[\eta_+,\eta_-]\settoheight{\auxlv}{$|$}%
\raisebox{-0.3\auxlv}{$|_{\eta_{\pm}\to\eta,j_s}$} = \exp\big(
-i\eta D_{\text{R}}j_s
 \big)O_s
\exp\big(
-i\eta\hat q
 \big) , 
\end{aligned} 
\label{eq:53XT} 
\end{align}%
where 
\begin{align} 
\begin{aligned} 
j_s(t) = j_{\mathrm{e}}(t) - \hbar s_- \protect\big [ 
\eta^{(+)}(t)-\eta^{(-)}(t)
 \big ] . 
\end{aligned} 
\label{eq:54XU} 
\end{align}%
Recalling Eqs.\ (\protect\ref{eq:51BY}) and (\ref{eq:13VZ}) for \mbox{$\eta(t)$} and \mbox{$j_{\mathrm{e}}(t)$} we recover Eq.\ (\protect\ref{eq:25MX}) for \mbox{$j_s(t)$} (as expected). Equation (\protect\ref{eq:19RY}) follows by applying quantum averaging to Eq.\ (\protect\ref{eq:53XT}). 

\section{Operator orderings, generalised Keldysh rotations and the nonlinear response problem for interacting fields}%
\label{ch:T}
\subsection{The Kubo and Schwinger currents}%
\label{ch:KSC}
In this appendix we briefly reiterate results of \mbox{Ref.\ \protect\cite{APII}}, taking this opportunity to generalise them to an arbitrary generalised Keldysh rotation. 
The assemblage of closed-time-loop Green functions of the electromagnetic\ field governed by Hamiltonian (\ref{eq:72PY}) are conveniently accessed through their generating functional, 
\begin{align} 
\begin{aligned} 
 &\Xi \protect{\big[ \eta _+,\eta _- \big| J_{\mathrm{e}} \big]} 
= \protect\big \langle 
T_C\exp\big(
-i\protect{\hat{\mathcal A}}_{+}\eta _+
+i\protect{\hat{\mathcal A}}_{-}\eta _-
 \big) 
 \big \rangle , 
\end{aligned} 
\label{eq:73PZ} 
\end{align}%
where \mbox{$
\eta _{\pm}^{\nu}(x)
$} is a pair of auxiliary c-number 4-vector functions. The \mbox{$T_C$}-ordering is defined in \mbox{Sec.\ \ref{ch:G}}. We use condensed notation (\ref{eq:62TV}). The averaging in (\ref{eq:73PZ}) is over the initial (Heisenberg) state of the field, which may be arbitrary. The potential operator is by definition dependent (conditional) on the source; in \mbox{$
\Xi 
$}, this dependence is made explicit. 

Schwinger \cite{SchwingerC} introduced functional (\ref{eq:73PZ}) following the idea of evolution forward and backward in time. Indeed, if we rescale the arguments of the functional, 
\begin{align} 
\begin{aligned} 
\eta _{\pm}^{\nu}(x)= (\hbar c)^{-1}j _{\pm}^{\nu}(x), 
\end{aligned} 
\label{eq:23SC} 
\end{align}%
it may be seen as an average of the product of the forward and backward S-matrices, 
\begin{align} 
\begin{aligned} 
 &\Xi \protect{\big[ \eta _+,\eta _- \big| J_{\mathrm{e}} \big]}
= \protect\big \langle 
\protect{\hat{\mathcal S}}_- \protect{\hat{\mathcal S}}_+
 \big \rangle , 
\quad
\protect{\hat{\mathcal S}}_{\pm} = T_{\pm}\exp\protect\big [ 
\pm (i\hbar c)^{-1}\protect{\hat{\mathcal A}}_{\nu}j_{\pm}
 \big ] . 
\end{aligned} 
\label{eq:24SD} 
\end{align}%
The natural question is whether there is any relation between the Kubo current \mbox{$J_{\mathrm{e}}^{\nu}(x)$} and the Schwinger ones \mbox{$j_{\pm}^{\nu}(x)$}. As is shown in our paper \cite{APII}, functional (\ref{eq:73PZ}) may be reduced to itself with Kubo's current put to zero, 
\begin{align} 
\begin{aligned} 
\Xi \protect{\big[ \eta _+,\eta _- \big| J_{\mathrm{e}} \big]} 
= \Xi \protect{\big[ \eta _++(\hbar c)^{-1}J_{\mathrm{e}}, 
\eta _-+(\hbar c)^{-1}J_{\mathrm{e}} \big| 0 \big]} . 
\end{aligned} 
\label{eq:25SE} 
\end{align}%
Functional (\ref{eq:73PZ}) thus depends only on the linear combinations of the Kubo and Schwinger currents, 
\begin{align} 
\begin{aligned} 
\eta_{\pm}^{\nu}(x)+ (\hbar c)^{-1}J_{\mathrm{e}}^{\nu}(x)= 
(\hbar c)^{-1}\protect\big [ 
j_{\pm}^{\nu}(x)+ J_{\mathrm{e}}^{\nu}(x)
 \big ] , 
\end{aligned} 
\label{eq:26SF} 
\end{align}%
and not on all three quantities separately. 
This formal redundancy is at the heart of our approach. 
\subsection{Generalised Keldysh rotations and nonlinear response problem}%
\label{ch:TR}
Consider now the generalised Keldysh rotation of functional (\ref{eq:73PZ}), defined as the change of functional variables according to Eqs.\ (\protect\ref{eq:28SJ}), (\ref{eq:27SH}). It is instructive to consider the interplay of redundancy (\ref{eq:25SE}) and substitution (\ref{eq:28SJ}). Functional (\ref{eq:73PZ}) may be defined with \mbox{$J_{\mathrm{e}}=0$}, and then extended to the response problem by replacing, 
\begin{align} 
\begin{aligned} 
\eta _{\pm}^{\nu}(x)\to \eta _{\pm}^{\nu}(x)+ (\hbar c)^{-1}J_{\mathrm{e}}^{\nu}(x). 
\end{aligned} 
\label{eq:29SK} 
\end{align}%
In variables \mbox{$\eta ^{\nu}(x)$}, \mbox{$j_{s}^{\nu}(x)$}, this replacement becomes, 
\begin{align} 
\begin{aligned} 
\eta ^{\nu}(x)\to \eta ^{\nu}(x), 
\quad 
j_{s}^{\nu}(x)\to j_{s}^{\nu}(x)+ J_{\mathrm{e}}^{\nu}(x). 
\end{aligned} 
\label{eq:30SL} 
\end{align}%
Hence the ``rotated'' functional (\ref{eq:73PZ}) depends on the sum \mbox{$j_{s} + J_{\mathrm{e}}$}, 
\begin{align} 
\begin{aligned} 
\Xi \protect{\big[ \eta _+,\eta _- \big| J_{\mathrm{e}} \big]} 
\settoheight{\auxlv}{$|$}%
\raisebox{-0.3\auxlv}{$|_{\eta _{\pm}\to\eta,j_{s}}$} = \Phi_s\protect{\big[ \eta \big| j_{s}+J_{\mathrm{e}} \big]} . 
\end{aligned} 
\label{eq:31SM} 
\end{align}%
This property of generalised Keldysh rotations is independent of the parameter \mbox{$s$} (while the rotated functional \mbox{$\Phi_s$} certainly depends on it). Functional \mbox{$\Phi _s$} is thus ideally suited for discussion of the quantum nonlinear response problem and {\em ipso facto\/} of the real-time QFT. 
\subsection{Quantum response functions and time-$s$-ordered products of Heisenberg\ operators}%
\label{ch:TN}
Interpretation of functional \mbox{$\Phi _s$} follows an analogy with the classical statitical response problem. Let \mbox{$\mathcal{A}_{\nu}(x)$} be a classical random field dependent (conditional) on the external source current \mbox{$J_{\mathrm{e}\nu }(x)$}. Full formal characterisation of such system is given by the {\em stochastic response functions\/}, 
\begin{align} 
\begin{aligned} 
 &R_{\nu _1\cdots\nu _m}^{\nu '_1\cdots\nu '_n} 
\big(x_1,\cdots x_m;x_1',\cdots x_n' \big) 
= 
\frac{\delta^n\protect\big \langle 
\mathcal{A}_{\nu _1}(x _1)\cdots \mathcal{A}_{\nu _m}(x _m)
 \big \rangle }{\delta J_{\mathrm{e}\nu _1'}(x_1')\cdots
\delta J_{\mathrm{e}\nu _n'}(x_n')} 
\settoheight{\auxlv}{$\big|$}%
\raisebox{-0.3\auxlv}{$\big|_{J_{\mathrm{e}}=0}$} . 
\end{aligned} 
\label{eq:97VJ} 
\end{align}%
Stochastic averages \mbox{$\protect\big \langle 
\mathcal{A}\cdots \mathcal{A}
 \big \rangle$} may be ``stored'' in the generating functional, 
\begin{align} 
\begin{aligned} 
\Phi_{\mathrm{cl}} \protect{\big[ \eta \big| J_{\mathrm{e}} \big]} &= 
\protect\big \langle 
\exp\big(
-i\eta \mathcal{A}
 \big) 
 \big \rangle , 
\\ 
\protect\big \langle 
\mathcal{A}_{\nu _1}(x _1)\cdots \mathcal{A}_{\nu _m}(x _m)
 \big \rangle &= \frac{\delta^m \Phi_{\mathrm{cl}} 
\protect{\big[ \eta \big| J_{\mathrm{e}} \big]}}
{\delta \eta^{\nu _1}(x _1)\cdots\delta \eta^{\nu _m}(x _m)} 
\settoheight{\auxlv}{$\big|$}%
\raisebox{-0.3\auxlv}{$\big|_{\eta=0}$} . 
\end{aligned} 
\label{eq:98VK} 
\end{align}%
All quantum-classical correspondences follow the analogy, 
\begin{align} 
\begin{aligned} 
\Phi_{\mathrm{cl}} \protect{\big[ \eta \big| J_{\mathrm{e}} \big]}
\Longleftrightarrow 
\Phi_{s} \protect{\big[ \eta \big| J_{\mathrm{e}} \big]} . 
\end{aligned} 
\label{eq:99VL} 
\end{align}%
To start with, note that, if \mbox{$\protect{\hat{\mathcal A}}_{\nu}(x)$} commutes with itself for different \mbox{$x$}, the \mbox{$T_C$} ordering in (\ref{eq:73PZ}) may be neglected, and 
\begin{align} 
\begin{aligned} 
\Phi_{s} \protect{\big[ \eta \big| J_{\mathrm{e}} \big]}
\Longrightarrow
\protect\big \langle 
\exp\big(
-i\eta \protect{\hat{\mathcal A}}
 \big) 
 \big \rangle . 
\end{aligned} 
\label{eq:1VM} 
\end{align}%
Dependence on the parameter $s$ is gone together with the dependence on \mbox{$j_s$}. Both thus belong to the quantum realm of noncommuting objects. 

In general, we define the {\em time-$s$-ordering\/} of operators \mbox{$\mathcal{T}_s$} by declaring \mbox{$\Phi _s$} the generating functional of the corresponding averages, 
\begin{align} 
\begin{aligned} 
\Phi_{s} \protect{\big[ \eta \big| J_{\mathrm{e}} \big]}
 &\equiv
\protect\big \langle 
\mathcal{T}_s\exp\big(
-i\eta \protect{\hat{\mathcal A}}
 \big) 
 \big \rangle , 
\\ 
\protect\big \langle 
\mathcal{T}_s\protect{\hat{\mathcal A}}_{\nu _1}(x _1)\cdots \protect{\hat{\mathcal A}}_{\nu _m}(x _m)
 \big \rangle &\equiv \frac{\delta^m \Phi_s \protect{\big[ \eta \big| J_{\mathrm{e}} \big]}}
{\delta \eta^{\nu _1}(x _1)\cdots\delta \eta^{\nu _m}(x _m)} 
\settoheight{\auxlv}{$\big|$}%
\raisebox{-0.3\auxlv}{$\big|_{\eta=0}$} , 
\end{aligned} 
\label{eq:2VN} 
\end{align}%
cf.\ the remark on terminology after Eq.\ (\protect\ref{eq:18AN}). 
Quantum response functions are defined following the pattern of Eq.\ (\protect\ref{eq:97VJ}), 
\begin{align} 
\begin{aligned} 
 &\mathcal{R}_{s\,\nu _1\cdots\nu _m}^{\nu '_1\cdots\nu '_n} 
\big(x_1,\cdots x_m;x_1',\cdots x_n' \big) 
= 
\frac{\delta^n\protect\big \langle 
\mathcal{T}_s\protect{\hat{\mathcal A}}_{\nu _1}(x _1)\cdots \protect{\hat{\mathcal A}}_{\nu _m}(x _m)
 \big \rangle }{\delta J_{\mathrm{e}\nu _1'}(x_1')\cdots
\delta J_{\mathrm{e}\nu _n'}(x_n')} 
\settoheight{\auxlv}{$\big|$}%
\raisebox{-0.3\auxlv}{$\big|_{J_{\mathrm{e}}=0}$} . 
\end{aligned} 
\label{eq:3VP} 
\end{align}%
Response functions without inputs, 
\begin{align} 
\begin{aligned} 
\mathcal{R}_{s\,\nu _1\cdots\nu _m} = \protect\big \langle 
\mathcal{T}_s\protect{\hat{\mathcal A}}_{\nu _1}(x _1)\cdots \protect{\hat{\mathcal A}}_{\nu _m}(x _m)
 \big \rangle
\settoheight{\auxlv}{$|$}%
\raisebox{-0.3\auxlv}{$|_{J_{\mathrm{e}}=0}$} , 
\end{aligned} 
\label{eq:88DP} 
\end{align}%
express self-radiation of the system, and the rest---its dependence on the source. Since \mbox{$\mathcal{T}_s\protect{\hat{\mathcal A}}= \protect{\hat{\mathcal A}}$}, the nonlinear response functions in the true meaning of the term \mbox{$\mathcal{R}_{s\,\nu}^{\nu '_1\cdots\nu '_n}$} do not depend on $s$. The crucial property of the quantum response functions is their explicit causality, 
\begin{align} 
\begin{aligned} 
\mathcal{R}_{s}^{\nu '_1\cdots\nu '_n} 
\big(;x_1',\cdots x_n' \big) &= 0 
, 
\\ 
\mathcal{R}_{s\,\nu _1\cdots\nu _m}^{\nu '_1\cdots\nu '_n} 
\big(x_1,\cdots x_m;x_1',\cdots x_n' \big) &= 0, 
\quad 
\operatorname{max}\big(
x_1^0,\cdots,x_m^0
 \big) < \operatorname{max}\big(
x_1^{\prime 0},\cdots,x_n^{\prime 0}
 \big) 
. 
\end{aligned} 
\label{eq:4VQ} 
\end{align}%
Proof of these relations is a straightforward generalisation of that in \cite{APII,APIII} from \mbox{$s=1$} to arbitrary \mbox{$s$}. Indeed, apart from general properties of closed-time-loop\ ordered products, proof in \cite{APII,APIII} depends only on the first of Eqs.\ (\protect\ref{eq:28SJ}). We note also that, due to Eq.\ (\protect\ref{eq:31SM}), quantum response functions may be defined in terms of the Heisenberg\ operator at zero source, 
\begin{align} 
\begin{aligned} 
 &\mathcal{R}_{s\,\nu _1\cdots\nu _m}^{\nu '_1\cdots\nu '_n} 
\big(x_1,\cdots x_m;x_1',\cdots x_n' \big) 
\\ &\quad
= 
\frac{\delta^{m+n}\Phi_s \protect{\big[ \eta \big| j_s \big]}}
{\delta \eta^{\nu _1}(x _1)\cdots\delta \eta^{\nu _m}(x _m)
\delta j_{s\nu _1'}(x_1')\cdots
\delta j_{s\nu _n'}(x_n')} 
\settoheight{\auxlv}{$\big|$}%
\raisebox{-0.3\auxlv}{$\big|_{\eta=j_s=0}$} . 
\end{aligned} 
\label{eq:5VR} 
\end{align}%
This relation constitutes a general formal solution to the nonlinear quantum response problem. Explicit formulae following from Eqs.\ (\protect\ref{eq:2VN}) and (\ref{eq:5VR}) are rather tangled. Examples for \mbox{$s=1$} (time-normal ordering) may be found in \cite{APII,APIII,RelCaus}. 
\subsection{Time-normal operator ordering}%
\label{ch:TO}
As in \cite{API,APII,APIII}, of actual interest to us are the properties of the generalised Keldysh rotation based on the normal ordering, termed in the quoted papers {\em response transformation\/}. Formally, it emerges by setting \mbox{$s=s_+=1$}, \mbox{$s_-=0$} in Eqs.\ (\protect\ref{eq:28SJ}), (\ref{eq:27SH}) and (\ref{eq:31SM}): 
\begin{align} 
\begin{aligned} 
\Xi \protect{\big[ \eta _+,\eta _- \big| J_{\mathrm{e}} \big]} 
\settoheight{\auxlv}{$|$}%
\raisebox{-0.3\auxlv}{$|_{\eta _{\pm}\to\eta,j_{\mathrm{e}}}$} = \Phi\protect{\big[ \eta \big| j_{\mathrm{e}}+J_{\mathrm{e}} \big]} . 
\end{aligned} 
\label{eq:40SW} 
\end{align}%
Note the change of notation, \mbox{$j_1\to j_{\mathrm{e}}$} and \mbox{$\Phi_1\to\Phi$}. 

We postulate functional \mbox{$\Phi[\eta|J_{\mathrm{e}}]$} to be the generating one of {\em time-normal averages\/} \cite{KelleyKleiner,GlauberTN,MandelWolf,APII,APIII} of the Heisenberg\ operator \mbox{$\protect{\hat{\mathcal A}}_{\nu}(x)$}, 
\begin{align} 
\begin{aligned} 
\Phi \protect{\big[ \eta \big| J_{\mathrm{e}} \big]} 
 &\equiv 
\protect\big \langle 
{\mathcal T}{\bf :}\exp\big(
-i\protect{\hat{\mathcal A}}\eta 
 \big)
{\bf :}
 \big \rangle 
= \protect\big \langle 
T_C\exp\protect\big [ 
-i\protect{\hat{\mathcal A}}_{+}\eta^{(-)}
+i\protect{\hat{\mathcal A}}_{-}\eta^{(+)}
 \big ] 
 \big \rangle . 
\end{aligned} 
\label{eq:78QE} 
\end{align}%
This formula applies with an arbitrary Heisenberg\ state, so that the averaging is in fact irrelevant.
Using definitions (\ref{eq:86UW}) of the \mbox{$^{(\pm)}$} operations and their properties (\ref{eq:11VX}), 
we find the explicit operator formula, 
\begin{align} 
\begin{aligned} 
{\mathcal T}{\bf :}\protect{\hat{\mathcal A}}_{\nu _1}(x _1)\cdots\protect{\hat{\mathcal A}}_{\nu _n}(x _n){\bf :} 
 &= \int dx_{01}'\cdots dx_{0n}' 
T_C\prod_{m=1}^n\protect\big [ 
\protect{\hat{\mathcal A}}_{\nu_m +}(x'_m)\delta ^{(+)}(x_{0m}'-x_{0m})
\\ &\quad
+ 
\protect{\hat{\mathcal A}}_{\nu_m -}(x'_m)\delta ^{(-)}(x_{0m}'-x_{0m})
 \big ] , 
\end{aligned} 
\label{eq:79QF} 
\end{align}%
where \mbox{$x_m=\protect \{ 
x_{0m},{\bf r}_m
 \} $} and \mbox{$x_m'=\protect \{ 
x_{0m}',{\bf r}_m
 \} $}. 

The reader familiar with Glauber-Kelley-Kleiner's photodetection theory \cite{GlauberPhDet,KelleyKleiner,GlauberTN,MandelWolf} should have noticed that Eq.\ (\protect\ref{eq:79QF}) deviates from Kelley-Kleiner's, 
\begin{align} 
\begin{aligned} 
 &{\mathcal T}{\bf :}\protect{\hat{\mathcal A}}_{\nu _1}(x _1)\cdots\protect{\hat{\mathcal A}}_{\nu _n}(x _n){\bf :} 
= 
T_C\prod_{m=1}^n\protect\big [ 
\protect{\hat{\mathcal A}}_{\nu_m +}^{(+)}(x'_m)
+ 
\protect{\hat{\mathcal A}}_{\nu_m -}^{(-)}(x'_m)
 \big ] . 
\end{aligned} 
\label{eq:80QH} 
\end{align}%
In (\ref{eq:79QF}), operator ordering comes first and the \mbox{$^{(\pm)}$} operations second, while in (\ref{eq:80QH}) the \mbox{$^{(\pm)}$} operations are first and the ordering second. This implies that Eqs.\ (\protect\ref{eq:79QF}), (\ref{eq:80QH}) also differ in the definition of the \mbox{$T_C$}-ordering: in (\ref{eq:79QF}), it applies to ``entire'' operators, while in (\ref{eq:80QH})---to their frequency-positive and frequency-negative\ parts. Definitions (\ref{eq:79QF}), (\ref{eq:80QH}) coincide in the resonance approximation, but only the exact time-normal products (\ref{eq:79QF}) obey strict causality laws \cite{RelCaus,RelCausMadrid}. The Kelley-Kleiner\ products (\ref{eq:80QH}) are causal only in the resonance approximation. For details see \cite{deHaan,BykTat,Tat,RelCaus,RelCausMadrid} and references therein. 

\subsection{Generalised Keldysh rotation of the form \mbox{$\Lambda _2$}}%
\label{ch:CL}
Here we outline calculations leading to Eq.\ (\protect\ref{eq:63TW}). On applying substitution (\ref{eq:27SH}) to 
Eq.\ (\protect\ref{eq:60TT}) we obtain, 
\begin{gather} 
\begin{gathered} 
\Lambda_2[\eta_+,\eta_-] 
\settoheight{\auxlv}{$|$}%
\raisebox{-0.3\auxlv}{$|_{\eta_{\pm}\to\eta,j_{s}}$} 
= \Lambda_2^{a}[\eta] 
+ \Lambda_2^{b}[\eta,j_{s}] 
+ \Lambda_2^{c}[\eta] 
+ \Lambda_2^{d}[j_s], 
\end{gathered} 
\label{eq:14AJ} 
\end{gather}%
where 
\begin{align} 
\begin{aligned} 
\Lambda_2^{a}[\eta] &= 
- i\eta \protect \langle 
\protect{\hat{\mathcal A}}
 \rangle , 
\\ 
\Lambda_2^{b}[\eta,j_{s}] &= 
(\hbar c)^{-1}\protect\big [ 
j_{s}\big(
\protect \langle 
\protect{\hat{\mathcal A}},\protect{\hat{\mathcal A}}
 \rangle - 
\protect \langle 
T_+\protect{\hat{\mathcal A}},\protect{\hat{\mathcal A}}
 \rangle 
 \big)\eta^{(s-)} 
\\ &\quad
+ \eta^{(s+)}\big(
\protect \langle 
T_-\protect{\hat{\mathcal A}},\protect{\hat{\mathcal A}}
 \rangle - 
\protect \langle 
\protect{\hat{\mathcal A}},\protect{\hat{\mathcal A}}
 \rangle 
 \big)j_{s} 
 \big ] , 
\\ 
\Lambda_2^{c}[\eta] &= 
\eta^{(s-)}\protect \langle T_+\protect{\hat{\mathcal A}},\protect{\hat{\mathcal A}}\rangle\eta^{(s-)} 
+ \eta^{(s+)}\protect \langle \protect{\hat{\mathcal A}},\protect{\hat{\mathcal A}}\rangle\eta^{(s-)} 
\\ &\quad
+ \eta^{(s-)}\protect \langle \protect{\hat{\mathcal A}},\protect{\hat{\mathcal A}}\rangle\eta^{(s+)} 
+ \eta^{(s+)}\protect \langle T_-\protect{\hat{\mathcal A}},\protect{\hat{\mathcal A}}\rangle\eta^{(s+)} , 
\\ 
\Lambda_2^{d}[j_s] &= 
(2\hbar c)^{-2}j_{s}\big(
2\protect \langle 
\protect{\hat{\mathcal A}},\protect{\hat{\mathcal A}}
 \rangle 
-
\protect \langle 
T_+\protect{\hat{\mathcal A}},\protect{\hat{\mathcal A}}
 \rangle
-
\protect \langle 
T_-\protect{\hat{\mathcal A}},\protect{\hat{\mathcal A}}
 \rangle
 \big) j_{s} . 
\end{aligned} 
\label{eq:89DQ} 
\end{align}%
We have tidied the raw formula up, taking notice of the first of Eqs.\ (\protect\ref{eq:28SJ}) and of the symmetry of \mbox{$\protect\big \langle 
T_{\pm}\protect{\hat{\mathcal A}}_{\nu}(x),\protect{\hat{\mathcal A}}_{\nu '}(x ')
 \big \rangle$}, and grouped contributions according to their dependence on \mbox{$\eta,j_{s}$}: linear in \mbox{$\eta$} (\mbox{$\Lambda _2^a$}), bilinear in \mbox{$\eta,j_{s}$} (\mbox{$\Lambda _2^b$}), quadratic in \mbox{$\eta$} (\mbox{$\Lambda _2^c$}), and quadratic in \mbox{$j_{s}$} (\mbox{$\Lambda _2^d$}). The last contribution cancels, due to the obvios relation, 
\begin{align} 
\begin{aligned} 
 &\protect\big \langle 
\protect{\hat{\mathcal A}}_{\nu}(x),\protect{\hat{\mathcal A}}_{\nu '}(x ')
 \big \rangle + \protect\big \langle 
\protect{\hat{\mathcal A}}_{\nu '}(x '),\protect{\hat{\mathcal A}}_{\nu}(x)
 \big \rangle \\ &\quad
- \protect\big \langle 
T_+\protect{\hat{\mathcal A}}_{\nu}(x),\protect{\hat{\mathcal A}}_{\nu '}(x ')
 \big \rangle - \protect\big \langle 
T_-\protect{\hat{\mathcal A}}_{\nu}(x),\protect{\hat{\mathcal A}}_{\nu '}(x ')
 \big \rangle = 0, 
\end{aligned} 
\label{eq:17AM} 
\end{align}%
and to the symmetrisation of \mbox{$\protect\big \langle 
\protect{\hat{\mathcal A}},\protect{\hat{\mathcal A}}
 \big \rangle$} imposed by the quadratic form. The contribution quadratic in \mbox{$\eta$} (\mbox{$\Lambda _2^c$}) is ``processed'' by making use of Eq.\ (\protect\ref{eq:11VX}), 
\begin{align} 
\begin{aligned} 
\eta^{(s-)}\protect\big \langle T_+\protect{\hat{\mathcal A}},\protect{\hat{\mathcal A}}\big \rangle\eta^{(s-)} = 
\eta\protect\big [ 
\mathcal{F}^{(s+)}_{x_0}\mathcal{F}^{(s+)}_{x_0'}\protect\big \langle T_+\protect{\hat{\mathcal A}},\protect{\hat{\mathcal A}}\big \rangle
 \big ]\eta , 
\end{aligned} 
\label{eq:24AU} 
\end{align}%
etc. The final formula (\ref{eq:16WC}) relies on Eq.\ (\protect\ref{eq:22AS}). For the contribution bilinear in \mbox{$\eta,j_{s}$} (\mbox{$\Lambda _2^b$}) we use the relations, (verified by direct calculation) 
\begin{align} 
\begin{aligned} 
 &\protect\big \langle 
\protect{\hat{\mathcal A}}^{\nu '}(x '),\protect{\hat{\mathcal A}}_{\nu}(x)
 \big \rangle - 
\protect\big \langle 
T_+\protect{\hat{\mathcal A}}^{\nu '}(x '),\protect{\hat{\mathcal A}}_{\nu}(x)
 \big \rangle 
\\ &\quad
= \protect\big \langle 
T_-\protect{\hat{\mathcal A}}_{\nu}(x),\protect{\hat{\mathcal A}}^{\nu '}(x ')
 \big \rangle 
- 
\protect\big \langle 
\protect{\hat{\mathcal A}}_{\nu}(x),\protect{\hat{\mathcal A}}^{\nu '}(x ')
 \big \rangle 
= -i\hbar c\mathcal{D}_{\mathrm{R}\nu}^{\nu '}(x,x') , 
\end{aligned} 
\label{eq:15AK} 
\end{align}%
whence we find, 
\begin{align} 
\begin{aligned} 
 &\Lambda_2^b[\eta ,j_s]
= -i \protect\big [ 
\eta^{(s-)} + \eta^{(s+)} 
 \big ] \mathcal{D}_{\mathrm{R}}j_{s} 
= -i\eta\mathcal{D}_{\mathrm{R}}j_{s} . 
\end{aligned} 
\label{eq:16AL} 
\end{align}%
In obtaining the final formula use was made of Eq.\ (\protect\ref{eq:21AR}).

\section{Green functions of quantised fields}%
\label{ch:FAB}
\subsection{Signs and dimensional factors}%
\label{ch:FA}
In this appendix, we summarise definitions of all kernels (propagators) of the electromagnetic\ and spinor fields used in the paper. We also develop a simple way of deriving relations between them, including response transformations \cite{API,APII,APIII}, by traditional QFT means. 

Sign and factor conventions of a theory depend on the interaction Hamiltonian. The relativistic electromagnetic\ interaction (\ref{eq:72PY}) is ``positive,'' which agrees with sign conventions of \mbox{Refs.\ \protect\cite{API,APII}}. Notation for electromagnetic\ propagators here follows those papers ($D$'s as opposed to $G$'s in \mbox{Refs.\ \protect\cite{APIII,WickCaus}}). 

The electromagnetic\ field enters the theory through three c-number kernels. The most important one is the linear response function, also known as the retarded Green function, or retarded propagator. It is shared by quantum electrodynamics\ and classical stochastic electrodynamics, and serves as a bridge between the two worlds, the quantum and the classical one. In QED, it is given by the formula, 
\begin{align} 
\begin{aligned} 
D_{\mathrm{R}\nu \nu' }(x-x') 
 &= \frac{\delta \ensuremath{\big\langle 
{\hat{\mathcal A}}_{\mathrm{e}\nu }(x)
\big\rangle} }{\delta J_{\mathrm{e}}^{\nu' }(x')}\settoheight{\auxlv}{$\big|$}%
\raisebox{-0.3\auxlv}{$\big|_{J_{\mathrm{e}}=0}$}
= (i\hbar c)^{-1}\theta(t-t')\protect\big [ 
\hat A_{\nu }(x),\hat A_{\nu' }(x')
 \big ] 
\, , 
\end{aligned} 
\label{eq:70ZY} 
\end{align}%
where \mbox{$
{\hat{\mathcal A}}_{\mathrm{e}\nu }(x)
$} is the Heisenberg potential operator with respect to Hamiltonian (\ref{eq:72PY}) with \mbox{$
\hat J_{\nu }(x)=0
$}. 
The intermediate expression in (\ref{eq:70ZY}) is the definition of $ D_{\text{R}}$, and the last one is Kubo's formula for it. Kubo's proper expression contains quantum averaging of the commutator, which we omitted because the free-field commutator is a c-number anyway. An explicit expressions for \mbox{$D_{\text{R}}$} may be found in appendix \ref{ch:B}. 

The reader familiar with the linear response theory should have noticed that the coefficient in (\ref{eq:70ZY}) differs from Kubo's \mbox{$
i/\hbar 
$}. In fact, in that theory, the interaction is defined to be ``negative.'' This convention may be traced down to the minus in the relation between the force and potential gradient, making the particle accelerate parallel and not antiparallel to the force. However, the covariant form of the electromagnetic\ interaction is ``positive.'' This accounts for the sign change. As to the factor of $c$ in the denominator, Kubo's formula is derived regarding the source as a function of $t$ and not of \mbox{$
x_0=ct
$}. Rescaling the argument implies rescaling the functional derivative, 
\begin{align} 
\frac{\delta }{\delta A_{\mathrm{e}}^{\nu' }(x_0,{\bf x})} = 
\frac{1}{c}\,
\frac{\delta }{\delta A_{\mathrm{e}}^{\nu' }(t,{\bf x})} 
, 
\label{eq:82WD} 
\end{align}%
hence the spare $c$. 

In QFT, the field is characterised 
by the Feynman and Keldysh propagators, 
\begin{align} 
\begin{aligned} 
D_{\mathrm{F}\nu \nu' }(x-x') &= (i\hbar c)^{-1}\protect\big \langle 0\big| 
T_+ \hat A_{\nu }(x)\hat A_{\nu' }(x')
 \big |0\big\rangle , \\ 
D^{(+)}_{\nu \nu' }(x-x') &= (i\hbar c)^{-1}\protect\big \langle 0\big| 
\hat A_{\nu }(x)\hat A_{\nu' }(x')
 \big |0\big\rangle . 
\end{aligned} 
\label{eq:68RF} 
\end{align}%
The factor in Eq.\ (\protect\ref{eq:70ZY}) has been extended to these kernels by definition. Explicit expressions for the kernels may be found in appendix \ref{ch:B}. 

Up to the dimensional coefficient, Eqs.\ (\protect\ref{eq:70ZY}), (\ref{eq:68RF}) agree with conventions of Itzykson and Zuber (IZ), while Bogoliubov and Shirkov define these kernels with the opposite sign. IZ do not seem to introduce the kernel \mbox{$
D_{\text{R}}
$} explicitly, but what we say is consistent with their definition of the Feynman propagator. Both \mbox{$
D_{\text{R}}
$} and \mbox{$
D_{\text{F}}
$} are Green's functions of the inhomogeneous wave equation (\ref{eq:17PA}), 
\begin{align} 
\Box D_{\mathrm{R,F}\nu \nu' }(x-x') 
= \mu_{\mathrm{vac}}g_{\nu \nu' }\delta^{(4)}(x-x'). 
\label{eq:21TN} 
\end{align}%
This equation is evident from the explicit formulae in appendix \ref{ch:B}. In turn, those formulae imply that Eq.\ (3-107) for the potential operator in IZ is supplemented by the dimensional factor,
\begin{align} 
\sqrt{\hbar c\mu_{\mathrm{vac}}} = 
\sqrt{\frac{\hbar }{c\varepsilon_{\mathrm{vac}}}} \, , 
\label{eq:11XK} 
\end{align}%
where $\varepsilon_{\mathrm{vac}}$ and $\mu_{\mathrm{vac}}$ are the electric and magnetic constants (permittivity and permeability of vacuum). This assigns the potential operator --- which, as written by IZ, has the dimension of inverse length --- the right dimension of 
\mbox{V\hspace{1pt}s\hspace{1pt}m$^{-1}$}, cf.\ Eq.\ (\protect\ref{eq:79WA}). 

\subsection{Response transformation of propagators by manipulating integration contours in the complex plane of energy}%
\label{ch:B} 
Here we put response transformations \cite{API,APII,APIII} in the context with representation of free-field Green functions as integrals in the complex plane of energy, common in QFT texts \cite{Itzykson,Schweber,Bogol}. For simplicity we consider the scalar (Klein-Gordon) field $\hat\phi(x)$ with mass \mbox{$
\mu_0
$}. For all definitions we refer the reader to the texts. We follow phase conventions of Itzykson and Zuber's, and use units where \mbox{$
\hbar =c=1
$}. All quantities are measured in powers of length. This reservation matters for Eqs.\ (\protect\ref{eq:13XM}) and (\ref{eq:14XN}), which imply SI units. 

Of general interest are the following kernels, 
\begin{align} 
\begin{aligned} 
D_{\text{F}}(x-x') &= -i\protect\big \langle 0\big| 
T_+ \hat\phi(x)\hat\phi(x')
 \big |0\big\rangle , 
\\ 
D^{(+)}(x-x') &= -D^{(-)}(x'-x) 
= -i\protect\big \langle 0\big| 
\hat\phi(x)\hat\phi(x')
 \big |0\big\rangle , 
\\ 
D(x-x') &= D^{(+)}(x-x') + D^{(-)}(x-x') 
= -i\protect\big \langle 0\big| 
\protect\big [ 
\hat\phi(x),\hat\phi(x')
 \big ] 
 \big |0\big\rangle , 
\\ 
D_{\text{R}}(x-x') &= -i\theta(t-t')\protect\big \langle 0\big| 
\protect\big [ 
\hat\phi(x),\hat\phi(x')
 \big ] 
 \big |0\big\rangle , 
\\ 
D_{\text{A}}(x-x') &= -i\theta(t'-t)\protect\big \langle 0\big| 
\protect\big [ 
\hat\phi(x'),\hat\phi(x)
 \big ] 
 \big |0\big\rangle . 
\end{aligned} 
\label{eq:12XL} 
\end{align}%
In Eqs.\ (\protect\ref{eq:12XL}), $D$ is the Pauli-Jordan function, $D^{(\pm)}$ are its frequency-positive and frequency-negative\ parts, and $D_{\text{R}}$, $D_{\text{A}}$, and $D_{\text{F}}$ are, correspondingly, the retarded, advanced and causal (Feynman) Green functions. The corresponding electromagnetic quantities follow with \mbox{$
\mu_0=0
$} and a factor, 
\begin{align} 
D_{{X}\mu \mu '}(x-x')= -\mu_{\mathrm{vac}}g_{\mu \mu '}D_{{X}}(x-x')\settoheight{\auxlv}{$|$}%
\raisebox{-0.3\auxlv}{$|_{\mu_0=0}$}, 
\label{eq:13XM} 
\end{align}%
where $X$ enumerates the kernels (\ref{eq:12XL}). 
For the Dirac field with mass $\mu_0$, 
\begin{align} 
\Delta_{{X}}(x-x') = -\big(
\mu_0 + i \gamma ^{\mu }\partial_{\mu }
 \big) D_{{X}}(x-x'), 
\label{eq:14XN} 
\end{align}%
etc. 

\begin{figure}
\begin{center}
\includegraphics[width=350pt]{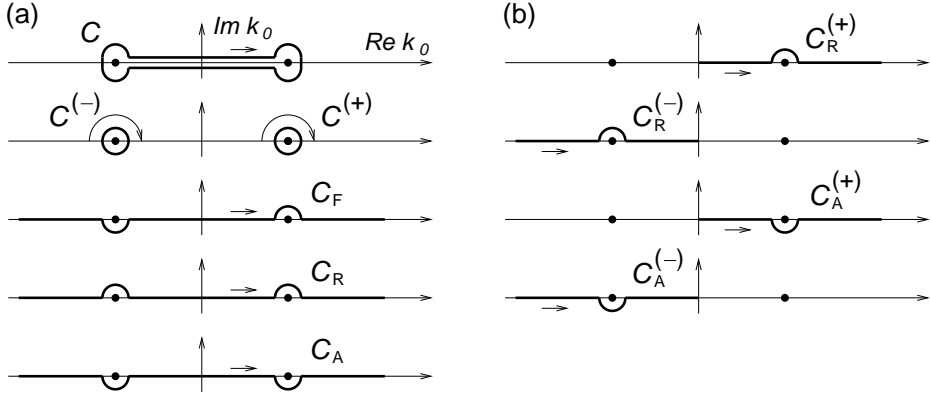}
\end{center}
\caption{%
Integration paths in the complex plane of $k_0$: (a) those used in definitions of Green's functions by Eq.\ (\protect\ref{eq:DlC}), and (b) those related to the frequency-positive and frequency-negative\ parts of the retarded Green function. Bold dots show schematically two poles of the integrand, positioned at \mbox{$
\pm\sqrt{{\bf k}^2+\mu_0^2}
$}. Arrows indicate travelling directions. 
}
\label{fig:Paths}
\end{figure}%
All kernels introduced by (\ref{eq:12XL}) 
afford a universal representation, 
\begin{align} 
\begin{aligned} 
D_X(x-x') &= \int \frac{d^3 {\bf k} }{(2\pi )^3}\oint_{C_X} \frac{dk_0}{2\pi }\, 
\frac{\text{e}^{-ik(x-x')}}{k^2-\mu_0^2}
\, , 
\end{aligned} 
\label{eq:DlC} 
\end{align}%
where $C_X$ is some contour in the complex plane of $k_0$. For brevity, we shall write this relation symbolically as, 
\begin{align} 
D_X \sim C_X. 
\label{eq:15XP} 
\end{align}%
For the kernels (\ref{eq:12XL}), 
\begin{align} 
\begin{aligned} 
D \sim C, \quad 
D^{(\pm)} &\sim C^{(\pm)}, \quad
D_{\text{R}}\sim C_{\text{R}}, \quad 
D_{\text{A}}\sim C_{\text{A}}, \quad 
D_{\text{F}}\sim C_{\text{F}}. 
\end{aligned} 
\label{eq:16XQ} 
\end{align}%
The contours mentioned here are depicted in Fig.\ \protect\ref{fig:Paths}a. For details we refer the reader to the texts \cite{Schweber,Bogol}. 

Representation (\ref{eq:DlC}) is very useful for deriving relations between the kernels. Indeed, the contours in Fig.\ \protect\ref{fig:Paths}a are not independent, 
\begin{align} 
\begin{aligned} 
C &= C^{(+)}+ C^{(-)}= C_{\text{R}}- C_{\text{A}}, \quad 
C^{(+)}= C_{\text{F}}- C_{\text{A}}, \quad 
C^{(-)}= C_{\text{R}}- C_{\text{F}}, 
\end{aligned} 
\label{eq:29YD} 
\end{align}%
etc. A contour occuring with a minus means that its travelling direction shown in Fig.\ \protect\ref{fig:Paths}a is changed to the opposite. Symbolic relations (\ref{eq:29YD}) amount to the analytical formulae, 
\begin{align} 
\begin{aligned} 
D &= D^{(+)}+ D^{(-)}= D_{\text{R}}- D_{\text{A}}, 
\quad
D^{(+)}= D_{\text{F}}- D_{\text{A}}, \quad 
D^{(-)}= D_{\text{R}}- D_{\text{F}}. 
\end{aligned} 
\label{eq:30YE} 
\end{align}%
Such relations were called in \mbox{Refs.\ \protect\cite{API,APII,APIII,WickCaus}} the {\em algebra of quantum Green functions\/}. The first two trivially follow from the definitions, while the other two already take a minor effort to obtain directly. 

The algebra of Green functions may be extended to conjugate kernels defined by the formula, 
\begin{align} 
D^{\dag}_X (x-x') = D^*_X(x'-x) . 
\label{eq:17XR} 
\end{align}%
For any kernel \mbox{$B$}, 
\begin{align} 
\begin{aligned} 
B(x-x') &= \int \frac{d^4x}{(2\pi )^4}\text{e}^{-ik(x-x')}B(k), 
\\ 
B^{\dag}(x-x') &= \int \frac{d^4x}{(2\pi )^4}\text{e}^{-ik(x-x')}B^*(k). 
\end{aligned} 
\label{eq:47YY} 
\end{align}%
Comparing this to Eq.\ (\protect\ref{eq:DlC}), we see that conjugation of a kernel corresponds to complex conjugation of the corresponding contour, 
\begin{align} 
D^{\dag}_X \sim C^*_X. 
\label{eq:18XS} 
\end{align}%
Visually, this means reflection of the contour with respect to the real axis. So, in Fig.\ \protect\ref{fig:Paths}a, $C_{\text{R}}$ and $C_{\text{A}}$ are conjugate, while conjugating \mbox{$
C
$}, \mbox{$
C^{(+)}
$} and \mbox{$
C^{(-)}
$} results in the same contour but with the opposite travelling direction. This amounts to the analytical properties, 
\begin{align} 
\begin{aligned} 
D_{\text{R}}^{\dag} &= D_{\text{A}}, &
D^{\dag} &= - D , &
D^{(\pm)\dagger} &= - D^{(\pm)}. 
\end{aligned} 
\label{eq:19XT} 
\end{align}%

In \mbox{Ref.\ \protect\cite{API}}, response transformation of propagators was derived from the algebra of Green's functions. This transformation also readily follows by manipulating the contours. Indeed, all contours in Fig.\ \protect\ref{fig:Paths}a may be constructed as combinations of the four legs shown in Fig.\ \protect\ref{fig:Paths}b. In particular 
\begin{align} 
\begin{aligned} 
C_{\text{F}} &=C_{\text{R}}^{(+)}+C_{\text{A}}^{(-)}, &
C^{(+)} &= C_{\text{R}}^{(+)}-C_{\text{A}}^{(+)}.
\end{aligned} 
\label{eq:20XU} 
\end{align}%
Owing to Eq.\ (\protect\ref{eq:18XS}), all four legs in Fig.\ \protect\ref{fig:Paths}b are related to the frequency-positive and frequency-negative\ parts of the retarded Green function and their conjugates, 
\begin{align} 
\begin{aligned} 
D_{\text{R}}^{(+)} &\sim C_{\text{R}}^{(+)}, \quad 
D_{\text{R}}^{(-)}\sim C_{\text{R}}^{(-)}, \quad 
D_{\text{R}}^{(+)\dagger} \sim C_{\text{A}}^{(+)}, \quad 
D_{\text{R}}^{(-)\dagger} \sim C_{\text{A}}^{(-)}. 
\end{aligned} 
\label{eq:21XV} 
\end{align}%
Combining Eqs.\ (\protect\ref{eq:16XQ}), (\ref{eq:20XU}) and (\ref{eq:21XV}) we find the analytical relations, 
\begin{align} 
\begin{aligned} 
D_{\text{F}}(x-x') &= D_{\text{R}}^{(+)}(x-x') + D_{\text{R}}^{(+)}(x'-x), \\ 
D^{(+)}(x-x') &= D_{\text{R}}^{(+)}(x-x') - D_{\text{R}}^{(-)}(x'-x) , 
\end{aligned} 
\label{eq:22XW} 
\end{align}%
where we also used that 
\begin{align} 
D_{\text{R}}^{(\pm)\dagger}(x-x') = D_{\text{R}}^{(\mp)}(x'-x) . 
\label{eq:23XX} 
\end{align}%
Eqs.\ (\protect\ref{eq:90SE}) used in appendix \ref{ch:ZZZ} below differ from Eqs.\ (\protect\ref{eq:22XW}) by the overall factor \mbox{$
-\mu_{\mathrm{vac}}g_{\mu \mu '}
$}. The differential operator in (\ref{eq:14XN}) does not interfere with rearranging of contours, so that Eqs.\ (\protect\ref{eq:22XW}) may also be extended to fermions, 
\begin{align} 
\begin{aligned} 
\Delta _{\text{F}}(x-x') &= \Delta_{\text{R}}^{(+)}(x-x') + \Delta_{\text{R}}^{(+)}(x'-x), \\ 
\Delta ^{(+)}(x-x') &= \Delta_{\text{R}}^{(+)}(x-x') - \Delta_{\text{R}}^{(-)}(x'-x) . 
\end{aligned} 
\label{eq:24XY} 
\end{align}%
These formulae are a particular case of those obtained in our paper \cite{WickCaus} for a {\em linear quantum channel\/}. 
\section{Diagram techniques for electromagnetic\ field in linear media}%
\label{ch:ZZZ}
\subsection{Diagrammatic solution to the classical field in linear media}%
\label{ch:Z}
\subsubsection{The Wyld diagram techniques}%
\label{ch:DW}
In this appendix we prove the conjectures of \mbox{Sec.\ \ref{ch:M}}. We start from embedding the classical theory of \mbox{Sec.\ \ref{ch:DQ}} in a Wyld-style diagram techniques \cite{Wyld}. We then show that this techniques and the conventional Perel-Keldysh series for the linear media in QED are connected by a one-to-one transformation. For a summary of the formal argument in this appendix see \mbox{Sec.\ \ref{ch:SumX}}. 

Let 
\mbox{$D_{\mathrm{R}\nu }^{\nu '}(x-x')$} be the {\em bare retarded propagator\/}, which is a solution to Eq.\ (\protect\ref{eq:43SZ}) with \mbox{$\Pi _{\text{R}}=0$}, 
\begin{align} 
\begin{aligned} 
\mu_{\mathrm{vac}}^{-1}\Box D_{\mathrm{R}\nu }^{\nu '}(x,x')
= \delta^{(4)}(x-x') , 
\qquad 
D_{\mathrm{R}\nu }^{\nu '}(x,x') = 0 , \quad x_0<x_0' . 
\end{aligned} 
\label{eq:54TM} 
\end{align}%
This quantity is known in QFT as the retarded Green function of the free quantised electromagnetic\ field; for an explicit formula see appendix \ref{ch:B}. Consider a diagram technique with the propagator, 
\begin{align} 
\begin{aligned}\raisebox{-0.6ex}{\includegraphics{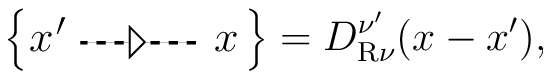}} 
\end{aligned} 
\label{eq:77KW} 
\end{align}%
and three generalised vertices, the {\em susceptibility vertex\/}, 
\begin{align} 
\begin{aligned}\raisebox{-0.6ex}{\includegraphics{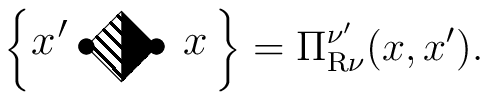}} 
\end{aligned} 
\label{eq:3XA} 
\end{align}%
the {\em regular-sorce\/} vertex, 
\begin{align} 
\begin{aligned}\raisebox{-0.6ex}{\includegraphics{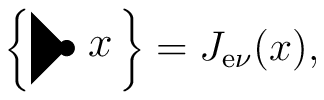}} \end{aligned} 
\label{eq:76KV} 
\end{align}%
and the {\em noise-source\/} vertex
\begin{align} 
\begin{aligned}\raisebox{-0.6ex}{\includegraphics{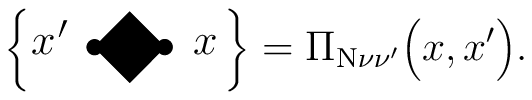}} 
\end{aligned} 
\label{eq:3ST} 
\end{align}%
The arguments of the regular-source and of the noise-souce vertices are regarded outputs, while the propagator and the susceptibility vertex have one output \mbox{($x$)} and one input \mbox{($x'$)}. The diagram rule is, match inputs of propagators to outputs of vertices, and {\em vice versa\/}, sum over matched 4-vector indices and integrate over matched space-time variables. All connected diagrams are linear chains with coefficient one, e.g., 
\begin{align} 
\begin{aligned}\raisebox{-0.6ex}{\includegraphics{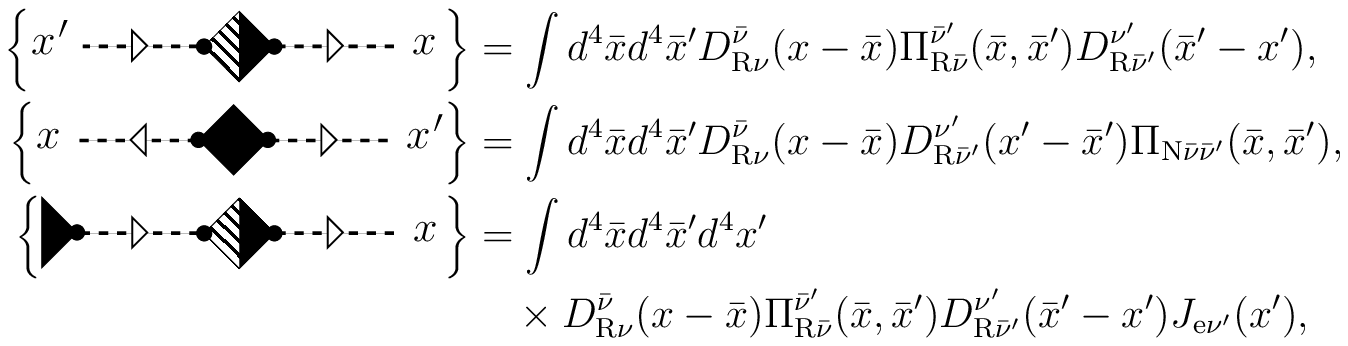}} 
\end{aligned} 
\label{eq:53TL} 
\end{align}%
etc. It is easy to see that no more than one source vertex---either (\ref{eq:76KV}) or (\ref{eq:3ST})---may occur in a chain. The number of susceptibility vertices is not limited. 
\subsubsection{Dyson equation for the retarded propagator}%
\label{ch:DE}
The only nontrivial class of connected diagrams are those containing only susceptibility vertices. Their sum defines the {\em dressed retarded propagator\/}, 
\begin{align} 
\begin{aligned}\raisebox{-0.6ex}{\includegraphics{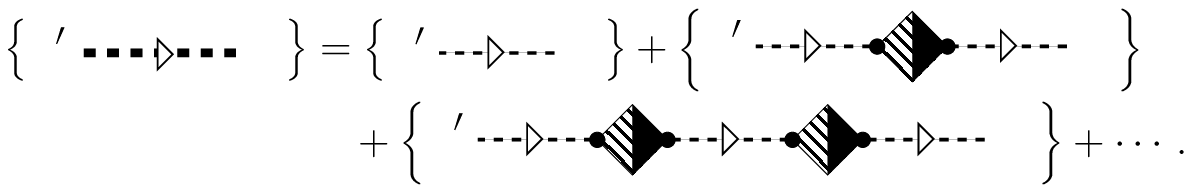}} 
\end{aligned} 
\label{eq:57VB} 
\end{align}%
It coincides with the retarded Green function defined by (\ref{eq:43SZ}), 
\begin{align} 
\begin{aligned}\raisebox{-0.6ex}{\includegraphics{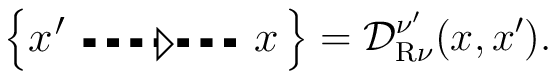}} 
\end{aligned} 
\label{eq:58VC} 
\end{align}%
Indeed, sum of series (\ref{eq:57VB}) obeys the Dyson equation, 
\begin{align} 
\begin{aligned}\raisebox{-0.6ex}{\includegraphics{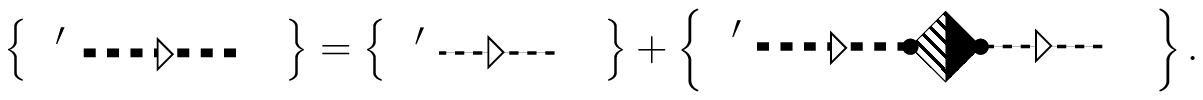}} 
\end{aligned} 
\label{eq:59VD} 
\end{align}%
Analytically, this amounts to the integral equation, 
\begin{align} 
\begin{aligned} 
\mathcal{D}_{\mathrm{R}\nu}^{\nu '}(x,x') &= D_{\mathrm{R}\nu}^{\nu '}(x-x')
+ \int d^4\bar xd^4\bar x'
D_{\mathrm{R}\nu}^{\bar\nu}(x-\bar x)
\Pi_{\mathrm{R}\bar\nu}^{\bar\nu '}(\bar x-\bar x') 
\mathcal{D}_{\mathrm{R}\bar\nu'}^{\nu '}(\bar x',x'). 
\end{aligned} 
\label{eq:65VL} 
\end{align}%
Acting on this equation by \mbox{$\mu_{\mathrm{vac}}^{-1}\Box$} and using (\ref{eq:54TM}) we arrive at Eq.\ (\protect\ref{eq:43SZ}). 

\subsubsection{Diagrammatic expressions for stochastic cumulants}%
\label{ch:DF}
Two other classes of connected diagrams are those with one regular-source vertex, and those with one noise-source vertex. 
These classes of diagrams sum up to two cumulants, 
\begin{align} 
\begin{aligned}\raisebox{-0.6ex}{\includegraphics{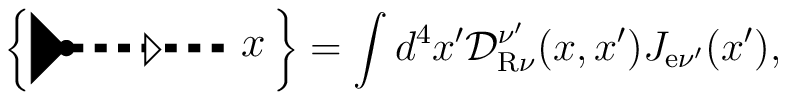}} 
\end{aligned} 
\label{eq:55TN} 
\end{align}%
and, 
\begin{align} 
\begin{aligned}\raisebox{-0.6ex}{\includegraphics{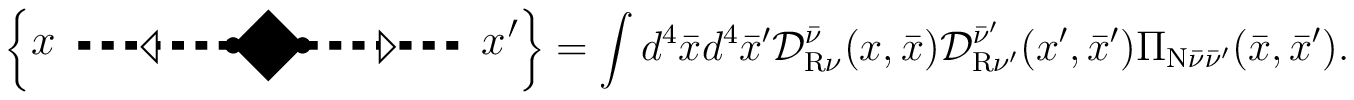}} 
\end{aligned} 
\label{eq:56TP} 
\end{align}%
These expressions coincide with Eqs.\ (\protect\ref{eq:46TC}), (\ref{eq:47TD}). This way, in diagram terms, 
\begin{align} 
&\raisebox{-0.6ex}{\includegraphics{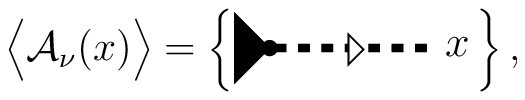}} 
\label{eq:57TQ} 
\\
&\raisebox{-0.6ex}{\includegraphics{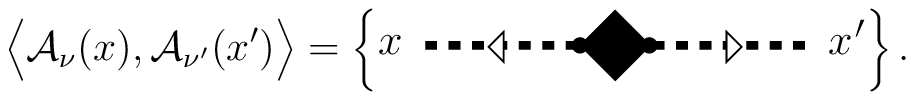}} 
\label{eq:58TR} 
\end{align}%
The diagram series built out of the elements (\ref{eq:77KW})--(\ref{eq:3ST}) thus indeed produce the macroscopic linear susceptibility and the stochastic cumulants as dressed connected cumulants. 

\subsection{The closed-time-loop formalism for electromagnetic\ field in a linear medium}%
\label{ch:F}
\subsubsection{Perel-Keldysh series in spinor electrodynamics}%
\label{ch:RS}
We now construct a consistent linearised quantum dynamical approach to the electromagnetic\ field in the Dirac vacuum within the Keldysh diagram techniques \cite{Keldysh}. Later it will be shown to be equivalent to the Wyld series of appendix \ref{ch:Z} and thus to the semiclassical approach of \mbox{Sec.\ \ref{ch:D}}. We assume familiarity of the reader with the concept of closed-time-loop ordering and its visualisation as an ordering on the so-called C-contour \cite{Perel,Keldysh,KamenevLevchenko,LandsmanVanWeert,RammerSmith}. We draw the C-contour with time increasing from left to right (unlike, e.g., Kamenev and Levchenko \cite{KamenevLevchenko}). For definitions of the \mbox{$T_{\pm}$} and \mbox{$T_C$} orderings see \mbox{Sec.\ \ref{ch:G}}. 

We introduce the graphical notation, for the electromagnetic\ propagators, 
\begin{align} 
\begin{aligned}\raisebox{-0.6ex}{\includegraphics{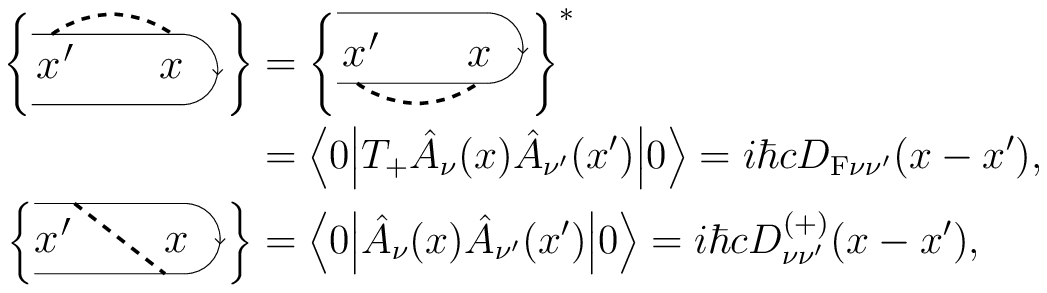}} 
\end{aligned} 
\label{eq:89WM} 
\end{align}%
for the fermionic ones, 
\begin{align} 
\begin{aligned}\raisebox{-0.6ex}{\includegraphics{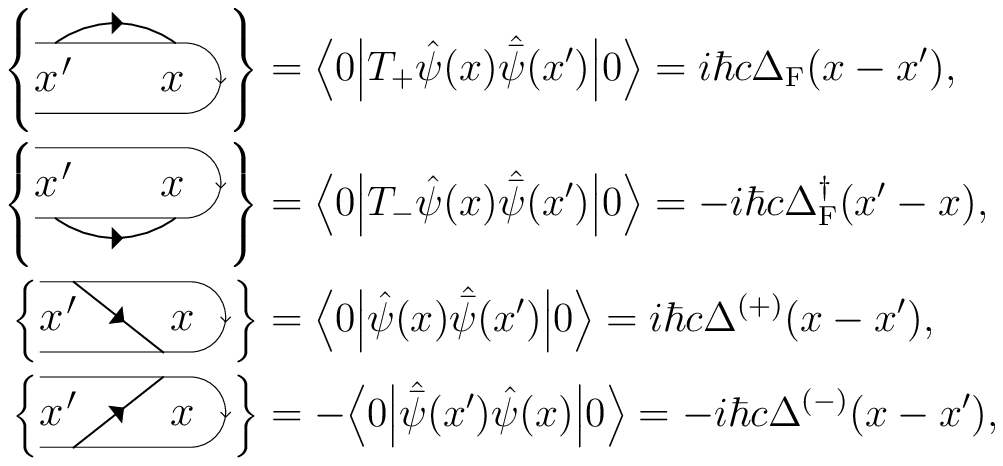}} 
\end{aligned} 
\label{eq:92WQ} 
\end{align}%
for the electromagnetic-interaction vertices, 
\begin{align} 
\begin{aligned}\raisebox{-0.6ex}{\includegraphics{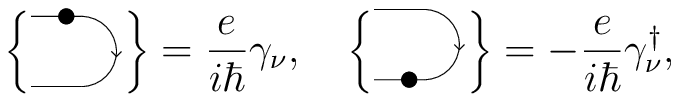}} 
\end{aligned} 
\label{eq:93WR} 
\end{align}%
and for the generalised vertices representing the external source, 
\begin{align} 
\begin{aligned}\raisebox{-0.6ex}{\includegraphics{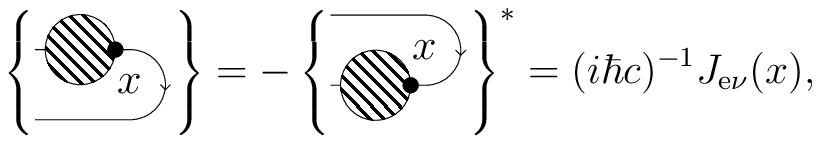}} 
\end{aligned} 
\label{eq:63KF} 
\end{align}%
In the graphical notation, the 4-vector and spinor indices are omitted.
The thin line symbolises the C-contour, with its forward branch on top. In Eqs.\ (\protect\ref{eq:92WQ})), (\ref{eq:93WR}), Hermitian conjugation applies to the matrix (spinor) structure of the propagators and vertices. The choice of signs and dimensional factors is discussed in appendix \ref{ch:FA}. Explicit formulae for all propagators may be found in appendix \ref{ch:B}, cf.\ also Eqs.\ (\protect\ref{eq:24PJ}). 
For general diagram rules we refer the reader to the literature \cite{Keldysh}.
Expressions for all diagrams of interest will be written explicitly.

\subsubsection{The linearised one-loop approximation}%
\label{ch:L1L}
In the first nonvanishing linearised approximation for the quantized electromagnetic\ field, information about the spinor field enters through the simplest fermionic loops, 
\begin{align}
&\raisebox{-0.6ex}{\includegraphics{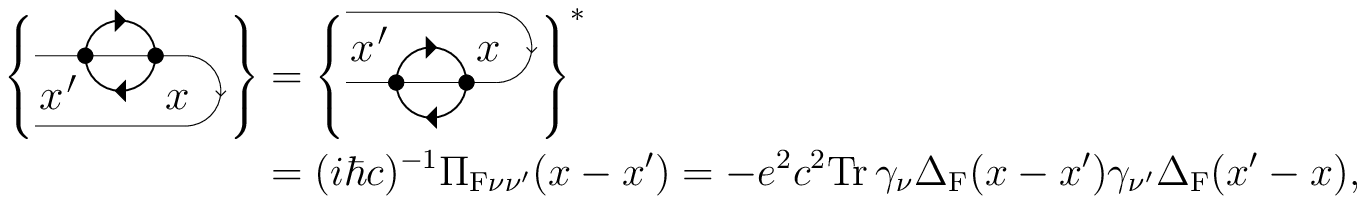}} 
\label{eq:57JZ} 
\\ 
&\raisebox{-0.6ex}{\includegraphics{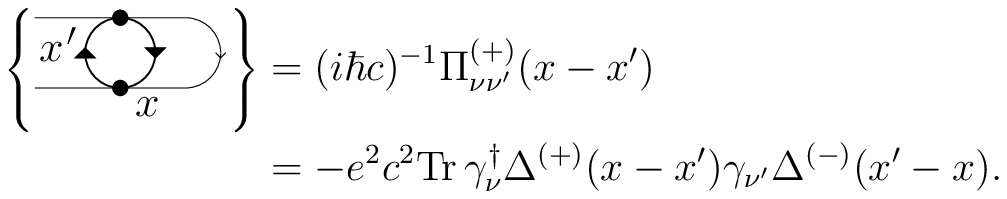}} 
\label{eq:59KB} 
\end{align}%
Loop (\ref{eq:59KB}) is convergent, while loops (\ref{eq:57JZ}) are not. 

\subsubsection{Generalised Perel-Keldysh series for one- and two-pole Green functions}%
\label{ch:RK}
Combining loops (\ref{eq:57JZ}), (\ref{eq:59KB}) together with propagators (\ref{eq:89WM}) and the source vertices (\ref{eq:63KF}) results in a series of chain diagrams for the one- and two-pole Green functions of the electromagnetic\ field, given by Eqs.\ (\protect\ref{eq:61TU}) and (\ref{eq:59TS}). This series is a particular case of {\em generalised linear Perel-Keldysh series\/}, comprising propagators (\ref{eq:89WM}), one-pole generalised vertices (\ref{eq:63KF}) and the two-pole generalised vertices, 
\begin{align} 
\begin{aligned}
\raisebox{-0.6ex}{\includegraphics{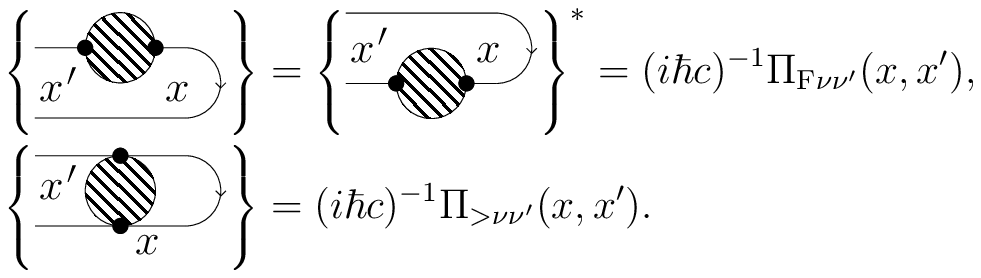}} 
\end{aligned} 
\label{eq:70KP} 
\end{align}%
For the Dirac sea, this may be regarded a graphical notation for regularised loops, 
\begin{align} 
\begin{aligned}
\raisebox{-0.6ex}{\includegraphics{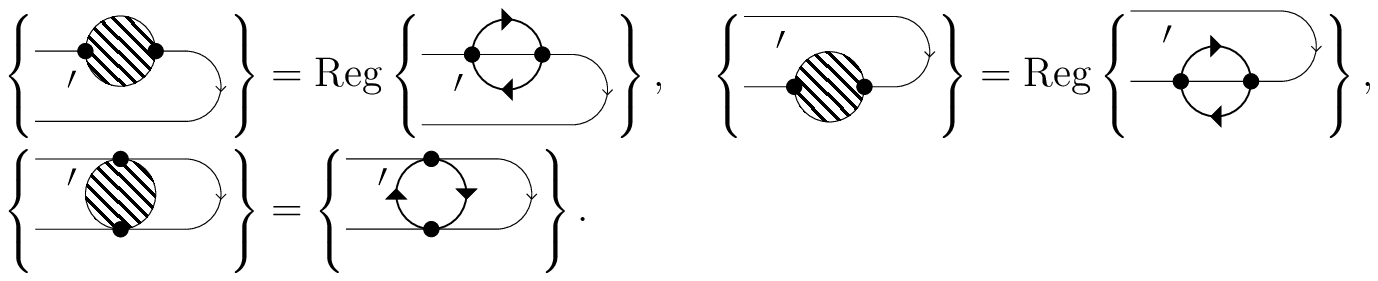}} 
\end{aligned} 
\label{eq:69KN} 
\end{align}%
For any medium, in the first nonvanishing linearised approximation we obtain the same linear chains, with the loops defined as 2-pole current cumulants, 
\begin{align} 
\begin{aligned}\raisebox{-0.6ex}{\includegraphics{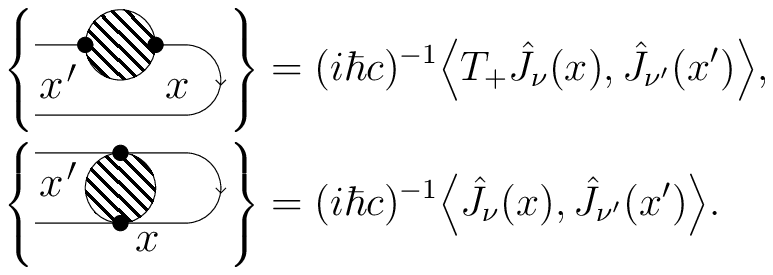}} 
\end{aligned} 
\label{eq:88SC} 
\end{align}%
Both the free current operator and the Heisenberg\ (initial) state of the medium here may in fact be arbitrary. Defining generalised vertices (\ref{eq:88SC}) as cumulants implies that the medium may be polarised \cite{endAvC}, 
in which case vertices (\ref{eq:63KF}) should be redefined replacing, 
\begin{align} 
\begin{aligned} 
J_{\mathrm{e}\nu }(x)\to J_{\mathrm{e}\nu }(x)+\protect\big \langle 
\hat J_{\nu}(x)
 \big \rangle . 
\end{aligned} 
\label{eq:96VH} 
\end{align}%
Furthermore, in spinor QED, the loops 
(\ref{eq:69KN}) are the first nonvanishing terms in series of {\em truncated 2-pole 1-particle irreducible\/} diagrams \cite{VasF,Itzykson,tHooft}, 
\begin{align} 
\begin{aligned}\raisebox{-0.6ex}{\includegraphics{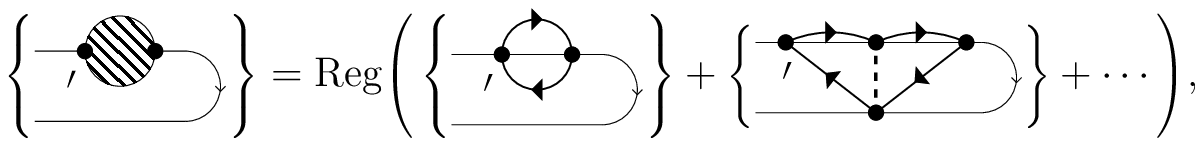}} 
\end{aligned} 
\label{eq:56JY} 
\end{align}%
and similarly for \mbox{$
\Pi_{>}^{\nu \nu '}(x,x')
$}. One may also generalise expansion (\ref{eq:56JY}) to non-vacuum states of the Dirac field, etc.

In Eqs.\ (\protect\ref{eq:88SC}), the media is not assumed to be homogeneous, nor stationary. Consequently, unlike \mbox{$
\Pi_{\text{F}}^{\nu\nu'}(x-x')
$} and \mbox{$
\Pi^{\nu\nu'(+)}(x-x')
$}, \mbox{$
\Pi_{\text{F}}^{\nu\nu'}(x,x')
$} and \mbox{$
\Pi_{>}^{\nu\nu'}(x,x')
$} are not regarded functions of the argument difference. As a useful side-effect, this makes the general case notationally distinct from the case of the Dirac sea. 

With few exception, all results below are subject to the conditions, 
\begin{align} 
\begin{aligned} 
\Pi_{\text{F}}^{\nu\nu'}(x,x') 
 &= \theta(x_0-x_0')\Pi_{>}^{\nu\nu'}(x,x') 
+\theta(x_0'-x_0)\Pi_{>}^{\nu'\nu}(x',x), \\ 
\protect\big [ 
\Pi_{>}^{\nu\nu'}(x,x')
 \big ]^* 
 &= -\Pi_{>}^{\nu'\nu}(x',x). 
\end{aligned} 
\label{eq:95SL} 
\end{align}%
These conditions are met by the perturbative definitions (\ref{eq:44JL}), (\ref{eq:34YK}) and (\ref{eq:88SC}). As was shown in \mbox{Sec.\ \ref{ch:LS}}, they survive through regularizations and hence hold for renormalized loops. In general, Eqs.\ (\protect\ref{eq:95SL}) are consistency requirements for the classes of diagrams that may contribute to \mbox{$
\Pi_{\mathrm{F}}
$} and \mbox{$
\Pi_{>}
$}. 

To summarize, our analyses apply at three levels of detalization. The {\em general, or structural, case\/} depends only on definitions (\ref{eq:70KP}) and conditions (\ref{eq:95SL}). The {\em perturbalive case\/} adds Eqs.\ (\protect\ref{eq:88SC}), where both the current operator and the quantum state of the media may be arbitrary. In the {\em case of the Dirac sea\/}, the medium is specified as the spinor field and its state as vacuum. 

As a word of caution, we remark that direct generalization of Eqs.\ (\protect\ref{eq:88SC}) beyond the first nonvanishing approximation by replacing the free current by the Heisenberg\ one \mbox{$
\protect{\hat{\mathcal J}}_{\nu}(x)
$} is incorrect. Those familiar with diagrammatics know that correct formulae are, 
\begin{align} 
\begin{aligned} 
\Pi_{\text{F}}^{\nu \nu' }(x,x') 
 &= (i\hbar c)^{-1}\mbox{\rm 1PI}\protect\big \langle 
T_+ \protect{\hat{\mathcal J}}^{\nu }(x)\protect{\hat{\mathcal J}}^{\nu' }(x')
 \big \rangle, \\ 
\Pi_{>}^{\nu \nu' }(x,x') 
 &= (i\hbar c)^{-1}\mbox{\rm 1PI}\protect\big \langle 
\protect{\hat{\mathcal J}}^{\nu }(x)\protect{\hat{\mathcal J}}^{\nu' }(x')
 \big \rangle , 
\end{aligned} 
\label{eq:71KQ} 
\end{align}%
where 1PI stands for separating the {\em 1-particle irreducible\/} part of the corresponding series 
\cite{Itzykson,tHooft}, cf.\ Eq.\ (\protect\ref{eq:56JY}). With the generalised vertices defined by Eq.\ (\protect\ref{eq:71KQ}), the linear series sum up to {\em exact\/} one- and two-pole Green functions of the electromagnetic\ field. 
\subsubsection{Formal summation of the linear series}%
\label{ch:RL}
The linear series comprises five types of connected diagrams, giving rise to five cumulants, 
\begin{align} 
\begin{aligned} 
 &\protect\big \langle 
T_+\protect{\hat{\mathcal A}}_{\nu}(x)
 \big \rangle, \quad 
\protect\big \langle 
T_-\protect{\hat{\mathcal A}}_{\nu}(x)
 \big \rangle, \\ 
 &\protect\big \langle 
\protect{\hat{\mathcal A}}_{\nu}(x),\protect{\hat{\mathcal A}}_{\nu '}(x ')
 \big \rangle, \quad 
\protect\big \langle 
T_+\protect{\hat{\mathcal A}}_{\nu}(x),\protect{\hat{\mathcal A}}_{\nu '}(x ')
 \big \rangle, \quad 
\protect\big \langle 
T_-\protect{\hat{\mathcal A}}_{\nu}(x),\protect{\hat{\mathcal A}}_{\nu '}(x ')
 \big \rangle . 
\end{aligned} 
\label{eq:80UQ} 
\end{align}%
Of these, only two are independent, 
\begin{align} 
\begin{aligned} 
\protect\big \langle 
T_+\protect{\hat{\mathcal A}}_{\nu}(x)
 \big \rangle &= 
\protect\big \langle 
T_-\protect{\hat{\mathcal A}}_{\nu}(x)
 \big \rangle = 
\protect\big \langle 
\protect{\hat{\mathcal A}}_{\nu}(x)
 \big \rangle , \\ 
\protect\big \langle 
T_+\protect{\hat{\mathcal A}}_{\nu}(x),\protect{\hat{\mathcal A}}_{\nu '}(x ')
 \big \rangle &= 
\protect\big \langle 
T_-\protect{\hat{\mathcal A}}_{\nu}(x),\protect{\hat{\mathcal A}}_{\nu '}(x ')
 \big \rangle^* 
\\ &= \theta(x_0-x_0')\protect\big \langle 
\protect{\hat{\mathcal A}}_{\nu}(x),\protect{\hat{\mathcal A}}_{\nu '}(x ')
 \big \rangle 
\\ &\quad
+ \theta(x_0'-x_0)\protect\big \langle 
\protect{\hat{\mathcal A}}_{\nu '}(x '),\protect{\hat{\mathcal A}}_{\nu}(x)
 \big \rangle . 
\end{aligned} 
\label{eq:79UP} 
\end{align}%
The one- and two-pole cumulants may be characterised as sums of diagrams, respectively, with and without the source vertices (\ref{eq:63KF}). 

The two-pole cumulants are given by the series, 
\begin{align} 
\begin{aligned}\raisebox{-0.6ex}{\includegraphics{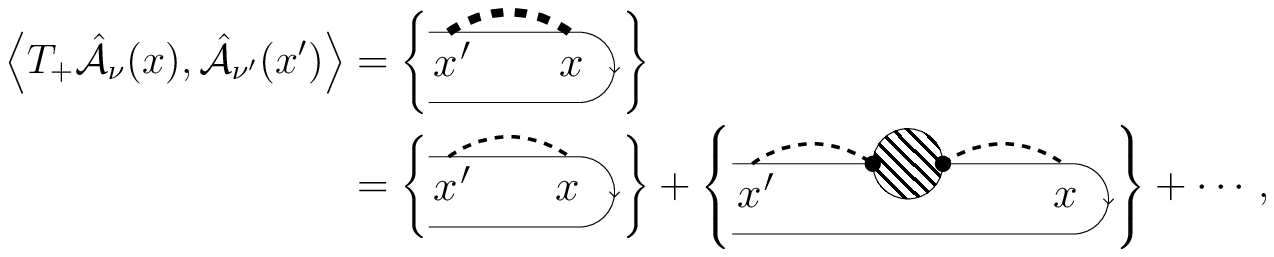}} 
\end{aligned} 
\label{eq:94WS} 
\end{align}%
\begin{align} 
\begin{aligned}\raisebox{-0.6ex}{\includegraphics{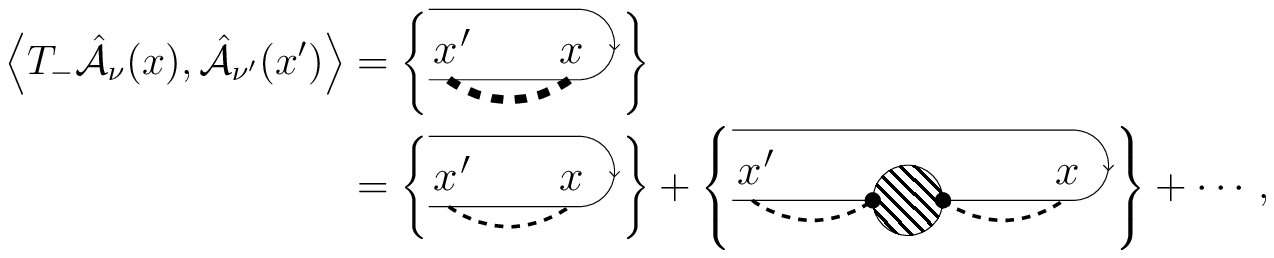}} 
\end{aligned} 
\label{eq:95WT} 
\end{align}%
\begin{align} 
\begin{aligned}\raisebox{-0.6ex}{\includegraphics{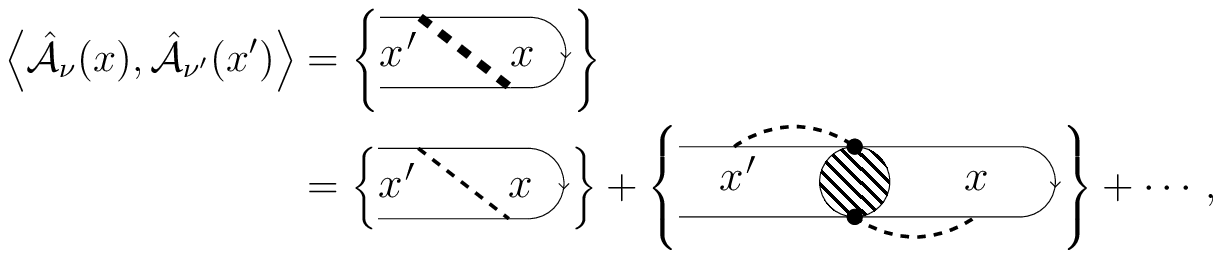}} 
\end{aligned} 
\label{eq:96WU} 
\end{align}%
where 
\begin{align} 
\begin{aligned}\raisebox{-0.6ex}{\includegraphics{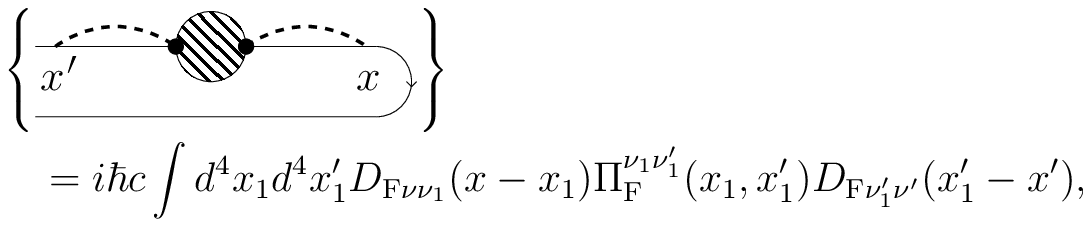}} 
\end{aligned} 
\label{eq:5XC} 
\end{align}%
\begin{align} 
\begin{aligned} \raisebox{-0.6ex}{\includegraphics{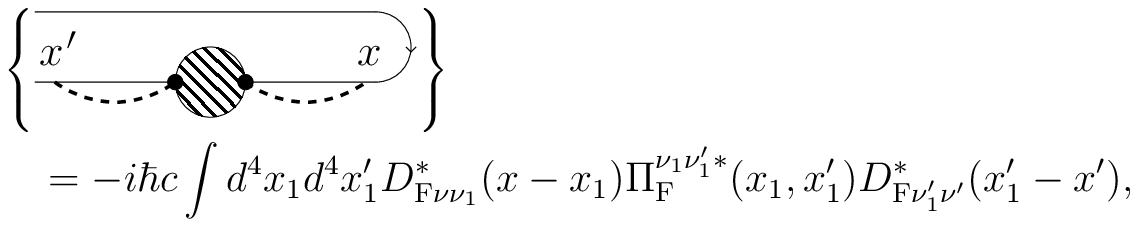}} 
\end{aligned} 
\label{eq:6XD} 
\end{align}%
\begin{align} 
\begin{aligned}\raisebox{-0.6ex}{\includegraphics{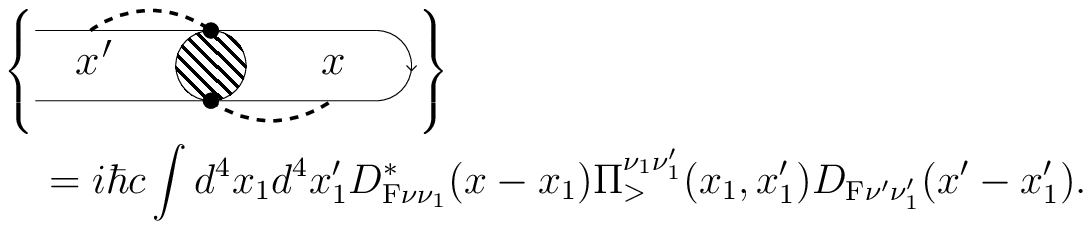}} 
\end{aligned} 
\label{eq:7XE} 
\end{align}%
All phase and dimensional factors are included into the definitions of the graphical elements, while diagrams as such occur with coefficient 1. 

The one-pole cumulants reduce to the two-pole ones: they follow by ``fastening'' the source vertices (\ref{eq:63KF}) to ends of chains (\ref{eq:94WS})--(\ref{eq:96WU}). If we assume that, 
\begin{align} 
\begin{aligned} 
J_{\mathrm{e}}^{\nu}(x)= 0 , 
\end{aligned} 
\label{eq:25AV} 
\end{align}%
the one-pole cumulants vanish. 
This assumption applies till appendix \ref{ch:KST}, where the source and the average field will be restored. 

Formally, we also encounter ``vacuum bubbles,'' such as, for instance,
\begin{align} 
\begin{aligned}\raisebox{-0.6ex}{\includegraphics{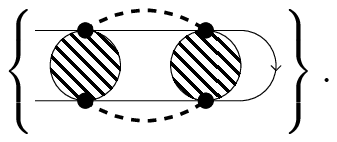}} 
\end{aligned} 
\label{eq:86SA} 
\end{align}%
In Perel-Keldysh's techniques, the sum of vacuum bubbles is zero, and we ignore them. 

\subsection{Response transformation of the linear Perel-Keldysh series}%
\label{ch:R}
\subsubsection{Response transformation of photon propagators}%
\label{ch:RP}
In this section we prove that the dressed graphical cumulants obey Eqs.\ (\protect\ref{eq:4LZ}), (\ref{eq:99BE}), where the time-normal cumulant is given by Eq.\ (\protect\ref{eq:92VC}). Our approach hinges on the formulae relating the Keldysh contractions to the retaded Green function \cite{API,WickCaus}: 
\begin{align} 
\begin{aligned} 
D_{\mathrm{F}\nu \nu '}(x-x') 
 &= D^{(+)}_{\mathrm{R}\nu \nu '}(x-x')
+ D^{(+)}_{\mathrm{R}\nu' \nu }(x'-x), \\ 
D^{(+)}_{\nu \nu '}(x-x') 
 &= D^{(+)}_{\mathrm{R}\nu \nu '}(x-x')
- D^{(-)}_{\mathrm{R}\nu' \nu }(x'-x). 
\end{aligned} 
\label{eq:90SE} 
\end{align}%
They are rederived by traditional QFT means in appendix \ref{ch:B}. 

It is instructive to write Eqs.\ (\protect\ref{eq:90SE}) in graphical terms. Notation for $D_{\text{R}}$ is given by Eq.\ (\protect\ref{eq:77KW}). To express graphically the \mbox{$^{(\pm)}$} operations, we introduce two auxiliary graphical elements, the line and the dummy vertex, 
\begin{align} 
\begin{aligned}\raisebox{-0.6ex}{\includegraphics{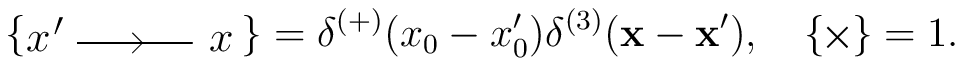}} 
\end{aligned} 
\label{eq:66KK} 
\end{align}%
Then, 
\begin{align} 
\begin{aligned}\raisebox{-0.6ex}{\includegraphics{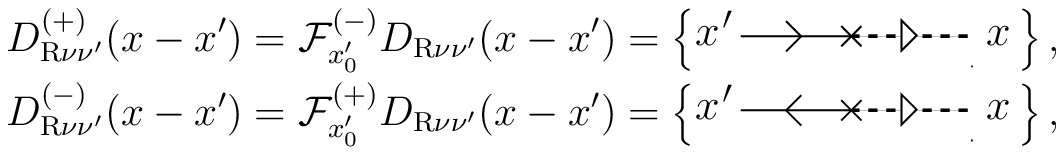}} 
\end{aligned} 
\label{eq:67KL} 
\end{align}%
cf.\ Eq.\ (\protect\ref{eq:36BH}). In this notation, Eqs.\ (\protect\ref{eq:90SE}) become, 
\begin{align} 
\begin{aligned}\raisebox{-0.6ex}{\includegraphics{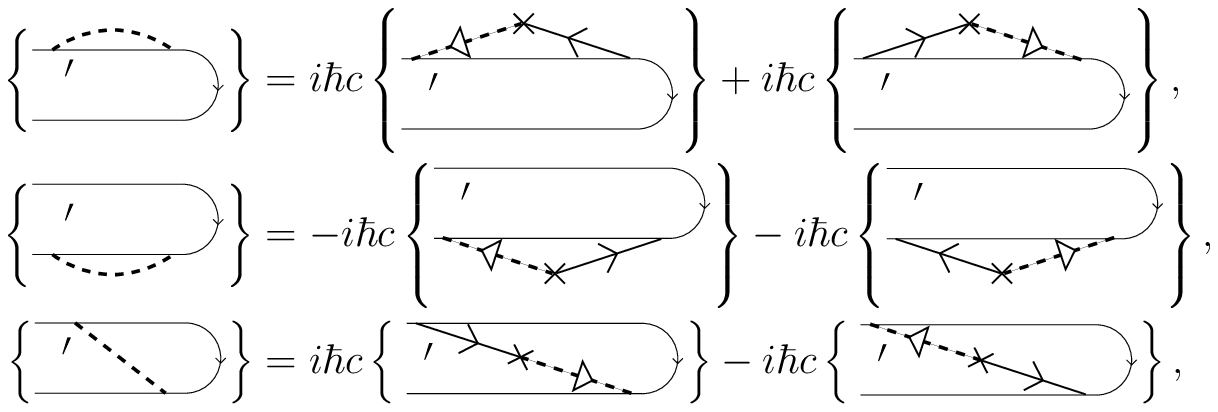}} 
\end{aligned} 
\label{eq:93SJ} 
\end{align}%
C-contours on the rhs indicates the way retarded and auxiliary propagators are ``fastened'' to other elements (vertices). By themselves, they are independent of the C-contour. Coefficients at the propagators are shown explicily, unlike for the Keldysh elements where they are part of the graphical notation. 
\subsubsection{Causal vertices}%
\label{ch:RV}
Equations (\protect\ref{eq:93SJ}) exhibit two crucial regularities: 
\begin{itemize}
\item 
Auxiliary propagators always ``start'' from the forward branch of the C-contour, and ``end'' at the reverse branch. 
\item 
The former occur with coefficient \mbox{$i\hbar c$} and the latter with coefficient \mbox{$-i\hbar c$}. 
\end{itemize}
In particular, combinations (\ref{eq:93SJ}) are indepenedent of the way the ``later'' ends of retarded propagators are fastened to the C-contour. 

These regularities allow for summation of the vertex-auxiliary line combinations, according to the way they are fastened to the retarded propagators (for the time being, ignore auxiliary propagators not fastened to vertices). This
transforms the Perel-Keldysh series into a causal (Wyld) diagram series \cite{BWO,Wyld}, with propagator \mbox{$
D_{\text{R}}
$} and three vertices: 
\begin{itemize}
\item
the {\em acausal vertex\/} where two retarded propagators end, 
\item
the {\em susceptibility vertex\/} where one retarded propagator ends and another starts, and 
\item
the {\em noise-source vertex\/} from which two retarded propagators start. 
\end{itemize}
This series differs from the classical series of appendix \ref{ch:Z} by the presence of the acausal vertex and absence of the regular-source vertex (\ref{eq:76KV}). The latter is natural with the external source put to zero. Dropping this assumption recovers the regular-source vertex, see \mbox{Sec.\ \ref{ch:RadE}} below. The acausal vertex has no meaningful physical interpretation and must vanish. 
Indeed, it equals the sum of four vertices, 
\begin{align} 
\begin{aligned}\raisebox{-0.6ex}{\includegraphics{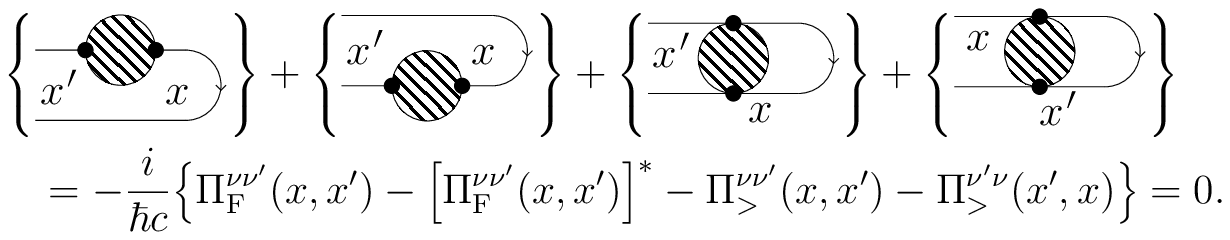}} 
\end{aligned} 
\label{eq:94SK} 
\end{align}%
The sum vanishes due to conditions (\ref{eq:95SL}). 

This way, restructuring of the Perel-Keldysh series initiated by transformations (\ref{eq:90SE}) yields the Wyld series of appendix \ref{ch:Z}, with vertices (\ref{eq:3XA}), (\ref{eq:3ST}) redefined in quantum terms, 
\begin{align}
\raisebox{-0.6ex}{\includegraphics{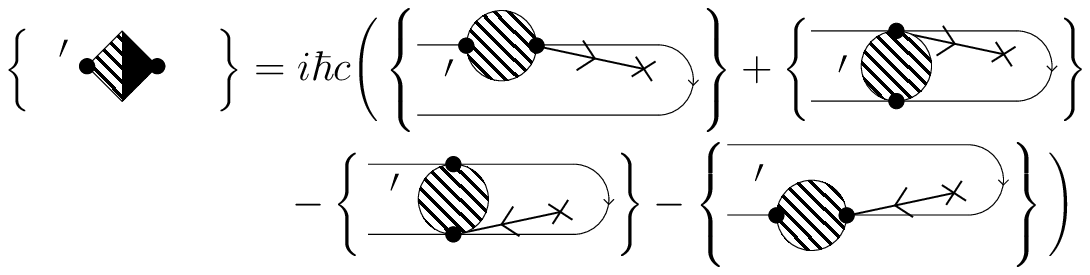}} , 
\label{eq:68KM} 
\end{align}%
and 
\begin{align} 
\begin{aligned}\raisebox{-0.6ex}{\includegraphics{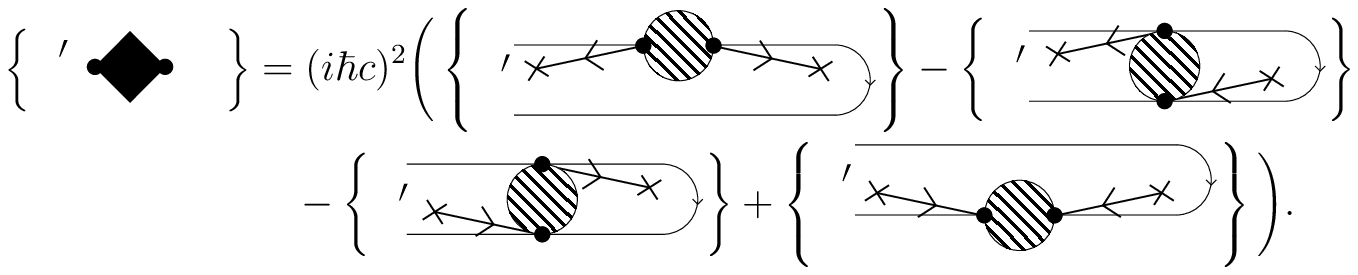}} 
\end{aligned} 
\label{eq:52TK} 
\end{align}%
To better orient the reader, in Eqs.\ (\protect\ref{eq:68KM}), (\ref{eq:52TK}) we retain the dummy vertices (crosses). Analytically, 
\begin{align} 
\begin{aligned} 
\Pi_{\text{R}}^{\nu \nu '}(x,x') 
= \theta(x_0-x_0')\protect\big [ 
\Pi_{>}^{\nu \nu '}(x,x') - \Pi_{>}^{\nu' \nu }(x',x)
 \big ] , 
\end{aligned} 
\label{eq:72KR} 
\end{align}%
and
\begin{align} 
\begin{aligned} 
 &\Pi_{\mathrm{N}}^{\nu \nu '} 
\big(
x,x'
 \big)
= -2\hbar c\Im
\protect\big [ 
\mathcal{F}^{(+)}_{x_0} 
\mathcal{F}^{(+)}_{x_0'} 
\Pi_{\text{F}}^{\nu \nu '}(x,x')
+\mathcal{F}^{(-)}_{x_0} 
\mathcal{F}^{(+)}_{x_0'} 
\Pi_{>}^{\nu \nu '}(x,x')
 \big ] 
. 
\end{aligned} 
\label{eq:4SU} 
\end{align}%
We preserve the analytical and graphical notation introduced by Eqs.\ (\protect\ref{eq:77KW})--(\ref{eq:3ST}). Equations (\protect\ref{eq:72KR}), (\ref{eq:4SU}) are subject to conditions (\ref{eq:95SL}). Equation (\protect\ref{eq:4SU}) directly follows from (\ref{eq:52TK}), while manipulations leading to (\ref{eq:72KR}) in essence repeat Eqs.\ (\protect\ref{eq:15AK}), (\ref{eq:16AL}) in appendix \ref{ch:CL}. 

In the perturbative case, including that of the Dirac sea, vertices \mbox{$\Pi _{\text{R}}$} and \mbox{$\Pi_{\mathrm{N}}$} coincide with the quantities given by Eqs.\ (\protect\ref{eq:67UA}), (\ref{eq:68UB}). The Wyld series we have recovered thus coincides with that considered in appendix \ref{ch:Z}, up to the ``upgrade'' of the classical average (\ref{eq:48TE}) to the time-normal average (\ref{eq:68UB}). 
\subsection{Summation of the linear Perel-Keldysh series and verification of conjectures of \protect\mbox{Sec.\ \ref{ch:CQ}}}%
\label{ch:KST}
\subsubsection{Preliminary remarks}%
\label{ch:KSTPr}
The quantum Wyld series we recovered on response transformation of the Perel-Keldysh series is structurally identical to the classical Wyld series analysed in appendix \ref{ch:Z}. The critical properties---retardation of \mbox{$\Pi _{\text{R}}$} and reality and symmetry of \mbox{$\Pi _{\mathrm{N}}$}---are shared by both series. Summation of both series reduced to solving Eq.\ (\protect\ref{eq:43SZ}) for the dressed retarded propagator \mbox{$\mathcal{D}_{\text{R}}$}. However, we still have two questions to answer. The first is the interpretation of \mbox{$\mathcal{D}_{\text{R}}$} as Kubo's linear response function (\ref{eq:64TX}). The second one is whether, with redefinitions (\ref{eq:72KR}), (\ref{eq:4SU}), the Wyld cumulant (\ref{eq:58TR}) coincides with the time-normal cumulant (\ref{eq:65TY}). 
\subsubsection{Radiation of the external source}%
\label{ch:RadE}
To answer the first question, we restore the external source \mbox{$J_{\mathrm{e}}$} and verify Eq.\ (\protect\ref{eq:66TZ}). The one-pole cumulants \mbox{$\protect \langle 
T_{\pm}\protect{\hat{\mathcal A}}
 \rangle $} (cf.\ appendix \ref{ch:RL}) reduce to two-pole ones, 
\begin{align} 
\begin{aligned}\raisebox{-0.6ex}{\includegraphics{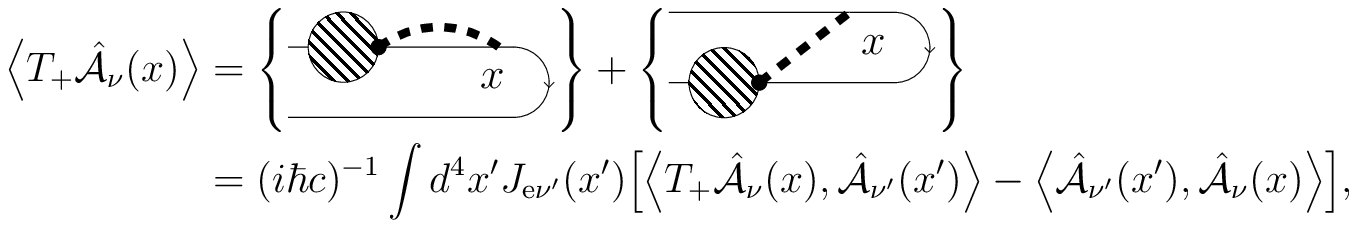}} 
\end{aligned} 
\label{eq:7MC} 
\end{align}%
and 
\begin{align} 
\begin{aligned}\raisebox{-0.6ex}{\includegraphics{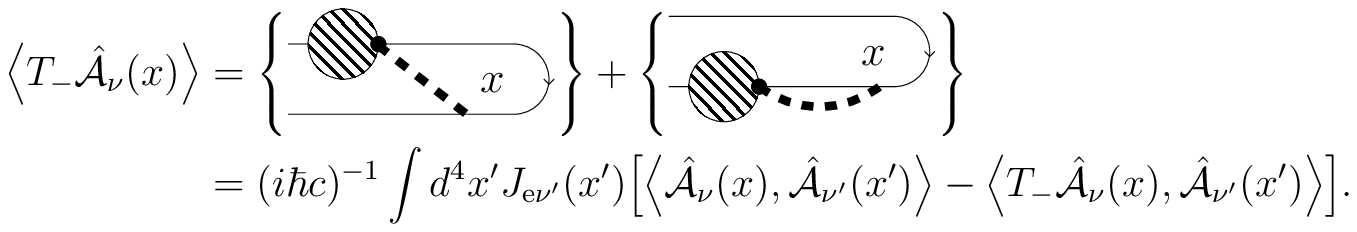}} 
\end{aligned} 
\label{eq:8MD} 
\end{align}%
Recalling Eqs.\ (\protect\ref{eq:15AK}), we see that Eqs.\ (\protect\ref{eq:7MC}) and Eq.\ (\protect\ref{eq:8MD}) coincide with Eq.\ (\protect\ref{eq:66TZ}). 

We have recovered Eq.\ (\protect\ref{eq:66TZ}) for the average potential directly from the Perel-Keldysh series. 
For the record, we also consider response transformation of the regular-source vertices (\ref{eq:63KF}). In diagrams, they occur ``fastened'' to ends of lines (\ref{eq:89WM}). After transformation (\ref{eq:93SJ}), retarded propagators may either start from or end on regular-source vertices. Similar to Eq.\ (\protect\ref{eq:94SK}), diagrams with retarded propagators ending on regular-source vertices pairwise cancel due to the property, 
\begin{align} 
\begin{aligned}\raisebox{-0.6ex}{\includegraphics{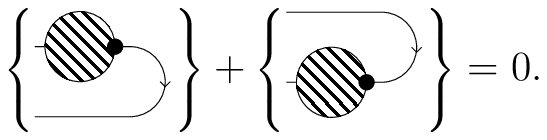}} 
\end{aligned} 
\label{eq:75KU} 
\end{align}%
As a side effect, this warrants cancellation of another class of vacuum bubbles: diagrams with regular-source vertices (\ref{eq:63KF}) at both ends of a chain (as expected of all bubbles in the closed-time-loop\ formalism). 
Regular-source vertices ``fastened'' to ``earlier'' ends of retarded propagators give rise to the {\em causal regular-sorce vertex\/}, 
\begin{align} 
\begin{aligned}\raisebox{-0.6ex}{\includegraphics{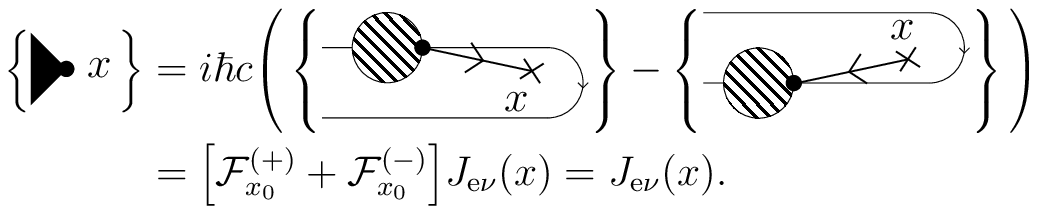}} 
\end{aligned} 
\label{eq:51TJ} 
\end{align}%
For polarised media, one should apply replacement (\ref{eq:96VH}). 

\subsubsection{Two-pole closed-time-loop\ cumulants}%
\label{ch:TPC}
To answer the second question, we derive explicit formulae for the dressed Keldysh cumulants (\ref{eq:94WS}), (\ref{eq:96WU}), and compare them to Eqs.\ (\protect\ref{eq:4LZ}), (\ref{eq:99BE}) with \mbox{$s=1$}. After the response transformation of propagators (appendix \ref{ch:RP}) and resumming the vertices (appendix \ref{ch:RV}), of all auxiliary lines (\ref{eq:66KK}) only those attached to external (free) ends of propagators survive, with their factors \mbox{$\pm i\hbar c$}. All other factors are absorbed by the vertices \mbox{$\Pi _{\text{R}}$} and \mbox{$\Pi _{\mathrm{N}}$}. Whence for the cumulants (\ref{eq:94WS}) and (\ref{eq:96WU}) we obtain, 
\begin{align} 
\begin{aligned}\raisebox{-0.6ex}{\includegraphics{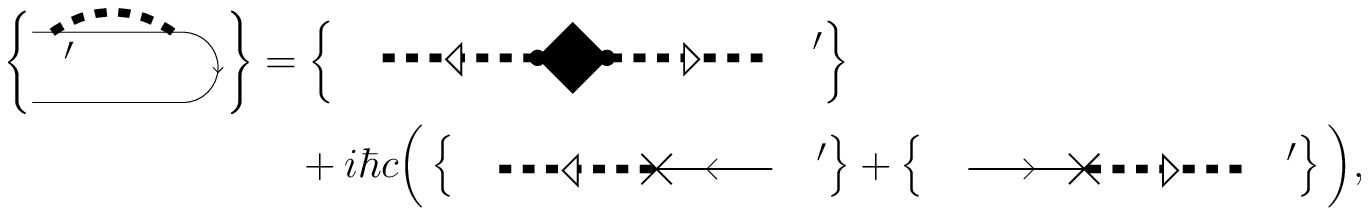}} 
\end{aligned} 
\label{eq:3LY} 
\end{align}%
and 
\begin{align} 
\begin{aligned}\raisebox{-0.6ex}{\includegraphics{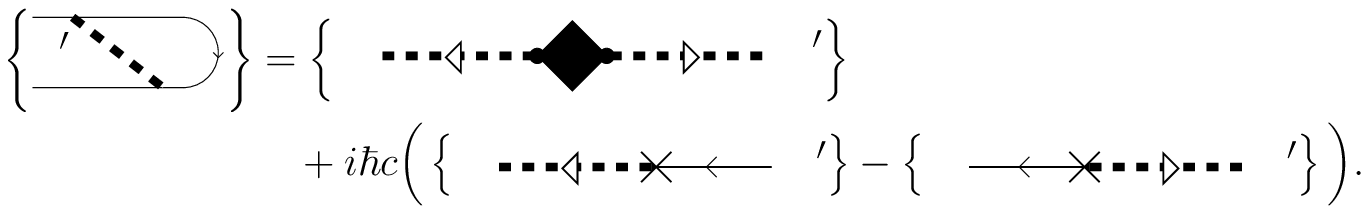}} 
\end{aligned} 
\label{eq:83UT} 
\end{align}%
In analytical terms, these relations amount to, 
\begin{align} 
\protect\big \langle 
T_+\protect{\hat{\mathcal A}}_{\nu }(x),\protect{\hat{\mathcal A}}_{\nu '}(x')
 \big \rangle 
 &
= \int d^4\bar x d^4\bar x'
\mathcal{D}_{\mathrm{R}\nu }^{\bar \nu }(x,\bar x) 
\mathcal{D}_{\mathrm{R}\nu '}^{\bar \nu '}(x',\bar x')
\Pi _{\mathrm{N}\bar \nu \bar\nu '}(\bar x,\bar x') . 
\nonumber\\ &\quad
+ i\hbar c\protect\big [ 
\mathcal{F}^{(-)}_{x_0'}\mathcal{D}_{\mathrm{R}\nu \nu '}(x,x') 
+ 
\mathcal{F}^{(-)}_{x_0}\mathcal{D}_{\mathrm{R}\nu' \nu }(x',x)
 \big ] , 
\label{eq:84UU} 
\\ 
\protect\big \langle 
\protect{\hat{\mathcal A}}_{\nu }(x),\protect{\hat{\mathcal A}}_{\nu '}(x')
 \big \rangle 
 &
= \int d^4\bar x d^4\bar x'
\mathcal{D}_{\mathrm{R}\nu }^{\bar \nu }(x,\bar x) 
\mathcal{D}_{\mathrm{R}\nu '}^{\bar \nu '}(x',\bar x')
\Pi _{\mathrm{N}\bar \nu \bar\nu '}(\bar x,\bar x') . 
\nonumber\\ &\quad
+ i\hbar c\protect\big [ 
\mathcal{F}^{(-)}_{x_0'}\mathcal{D}_{\mathrm{R}\nu \nu '}(x,x') 
- 
\mathcal{F}^{(+)}_{x_0}\mathcal{D}_{\mathrm{R}\nu' \nu }(x',x)
 \big ] . 
\label{eq:85UV} 
\end{align}%
Comparing them to Eqs.\ (\protect\ref{eq:4LZ}), (\ref{eq:99BE}) with \mbox{$s=1$} shows that the cumulant (\ref{eq:56TP}) in the quantum Wyld series indeed coincides with the time-normal cumulant (\ref{eq:65TY}). In turn, this proves Eq.\ (\protect\ref{eq:92VC}). This is the last piece in the ``jigsaw puzzle'' of verifying conjectures of \mbox{Sec.\ \ref{ch:CQ}}. 
\subsection{Summary of the formal argument}%
\label{ch:SumX}
Let us retrace the logic of the argument. In \mbox{Sec.\ \ref{ch:CQ}}, we conjectured that the classical stochastic theory of \mbox{Sec.\ \ref{ch:DQ}} may be upgraded to a quantum theory by expressing the susceptibilities (\mbox{$\Pi _{\text{R}}$} and \mbox{$\mathcal{D}_{\text{R}}$}) according to Kubo's linear response theory and replacing classical averages by time-normal averages while preserving all dynamical relations. We then showed that the classical theory amounts to a Wyld series (appendix \ref{ch:Z}), and that the Perel-Keldysh series in quantum theory (appendix \ref{ch:F}) may be transformed into a structurally identical Wyld series (appendices \ref{ch:R} and \ref{ch:RadE}). The open question was however if the dressed cumulants in the quantum Wyld series afford the expected interpretation. Equation (\protect\ref{eq:66TZ}) follows trivially (appendix \ref{ch:RadE}), with Kubo's formula for \mbox{$\mathcal{D}_{\text{R}}$} emerging ``on the run''. The hard part was to prove identity of the graphical cumulant (\ref{eq:56TP}) with the time-normal cumulant (\ref{eq:65TY}). This was achieved by deriving graphical Eqs.\ (\protect\ref{eq:3LY}), (\ref{eq:83UT}) and comparing them to the analytical Eqs.\ (\protect\ref{eq:4LZ}), (\ref{eq:99BE}) with \mbox{$s=1$}. 

\section{Linear susceptibility of the Dirac sea}\label{ch:O}%
\subsection{The Dirac field basics}%
\label{ch:DB}
In this appendix, we outline details of the calculation of the commutator (\ref{eq:30PQ}) and of the regularised microscopic susceptibility (\ref{eq:49JR}). For this calculation, we need explicit formulae for the frequency-positive and frequency-negative\ parts of the anticommutator of the Dirac fields, (with $\alpha ,\beta $ being the spinor indices) 
\begin{align} 
\begin{aligned} 
\protect\big [ 
\hat\psi_{\alpha }(x),\hat{\bar\psi}_{\beta } (x')
 \big ]_+ &= \protect\big \langle 0\big| 
\hat\psi_{\alpha }(x)\hat{\bar\psi}_{\beta } (x')
 \big |0\big\rangle 
+ \protect\big \langle 0\big| 
\hat{\bar\psi}_{\beta } (x')\hat\psi_{\alpha }(x)
 \big |0\big\rangle, \\ 
\protect\big \langle 0\big| 
\hat\psi_{\alpha }(x)\hat{\bar\psi}_{\beta } (x')
 \big |0\big\rangle &= i\hbar c\Delta_{\alpha \beta }^{(+)}(x-x'), 
\quad 
\protect\big \langle 0\big| 
\hat{\bar\psi}_{\beta } (x')\hat\psi_{\alpha }(x)
 \big |0\big\rangle = i\hbar c\Delta_{\alpha \beta }^{(-)}(x-x') 
, 
\\ 
i\hbar c\Delta_{\alpha \beta }^{(\pm)}(x-x') 
 &= \pm\int\frac{d^4k}{(2\pi )^3}\text{e}^{-ik(x-x')}\theta (\pm k_0) 
\delta(k^2-\mu_0^2)(\mu_0\delta_{\alpha \beta }+k_{\nu }
\gamma^{\nu }_{\alpha \beta }), 
\end{aligned} 
\label{eq:24PJ} 
\end{align}%
where $\mu_0$ is the scaled mass of the electron given by Eq.\ (\protect\ref{eq:81WC}). These formulae may be found in any texbook \cite{Schweber,Bogol,Itzykson}. We also use the relations, 
\begin{align} 
\begin{aligned} 
\text{Tr}\gamma^{\lambda }\gamma^{\nu } &= 4 g^{\lambda\nu }, \quad 
\text{Tr}\gamma^{\lambda }\gamma^{\rho }\gamma^{\nu } = 0, \\ 
\text{Tr}\gamma^{\lambda }\gamma^{\rho }\gamma^{\nu }\gamma^{\sigma } &= 4\big(
g^{\lambda\rho }g^{\nu \sigma } + 
g^{\lambda\sigma }g^{\nu\rho } - g^{\lambda\nu }g^{\rho\sigma }
 \big) , 
\end{aligned} 
\label{eq:26LT} 
\end{align}%
cf., e.g., Eq.\ (A-29) in the appendix in \mbox{Ref.\ \protect\cite{Itzykson}}. 
\subsection{The commutator of currents}%
\label{ch:OE}
By making use of Eqs.\ (\protect\ref{eq:24PJ}) and (\ref{eq:26LT}), for the commutator of currents we obtain, 
\begin{align} 
 &\protect\big \langle 0\big| 
\protect\big [ 
\hat J^{\mu }(x),\hat J^{\nu }(x')
 \big ] 
 \big |0\big\rangle 
= e^2c^2\int \frac{d^4k}{(2\pi )^4}\text{e}^{-ik(x-x')}C^{\mu \nu }(k),
\label{eq:32TZ} 
\end{align}%
where 
\begin{align} 
C^{\mu \nu }(k) 
 &= - 4\pi^2\varepsilon (k_0)\int \frac{d^4p}{(2\pi )^4}
U(p,k)V^{\mu \nu }(p,k) , 
\label{eq:3UE} 
\end{align}%
and 
\begin{align} 
U(p,k) &= \theta\Big(
\frac{k_0^2}{4}-p_0^2 
 \Big) 
\delta\bigg(
\Big(
\frac{k}{2}+p
 \Big)^2-\mu_0^2
 \bigg) 
\delta\bigg(
\Big(
\frac{k}{2}-p
 \Big)^2-\mu_{0}^2
 \bigg) 
, 
\label{eq:34UB} 
\\ 
V^{\mu \nu }(p,k) &= 
\big(
4 \mu_0^2 + k^2 - 4p^2
 \big) g^{\mu\nu } 
+ 8 p^\mu p^\nu - 2 k^{\mu} k^{\nu } 
. 
\label{eq:35UC} 
\end{align}%
Since \mbox{$
\delta (a) \delta (b) = \delta (a-b)\delta(
\frac{a+b}{2} 
 )
$},
the product of the delta-functions may also be written as, 
\begin{align} 
 &\delta\Big(
\Big(
\frac{k}{2}+p
 \Big)^2-\mu_0^2
 \Big) 
\delta\Big(
\Big(
\frac{k}{2}-p
 \Big)^2-\mu_{0}^2
 \Big) 
= 
\delta\Big(
p^2+\frac{k^2}{4}
-\mu_0^2
 \Big) 
\delta\big(
2pk \big) 
\label{eq:33UA} 
\end{align}%
One can therefore replace, 
\begin{align} 
p^2\to \mu_0^2-\frac{k^2}{4},\quad pk \to 0, 
\label{eq:39UH} 
\end{align}%
whenever these quantities appear. 

The integrand in (\ref{eq:3UE}) is subject to three observations: 
\\ $\bullet$\ {\em $U(p,k)$ is a relativistic scalar.\/}
Indeed, while not a scalar by itself, the theta-function behaves as such when multiplied by the delta-functions. It is nonzero if $k_0/2 +p_0$ and $k_0/2 -p_0$ are either both positive or both negative. The delta-functions assure that these quantities are time components of time-like vectors, hence their signs do not change under special Lorentz transformations. Invariance of the theta-function under space and time inversions is obvious. 
\\ $\bullet$\ {\em $U(p,k)$ is nonzero only if\/} 
\begin{align} 
k^2\geq 2\mu_0^2 . 
\label{eq:36UD} 
\end{align}%
To see this, note the inequalities, 
\begin{align} 
\begin{aligned} 
 &\frac{k_0^2}{4}\geq p_0^2 , &
 &
\frac{k_0^2}{4}+p_0^2\geq \mu_0^2 . 
\end{aligned} 
\label{eq:45MP} 
\end{align}%
The first one originates in the theta-function, the second one in the first delta-function on the RHS of (\ref{eq:33UA}). Summing these inequalities we find that \mbox{$
k_0^2\geq 2\mu_0^2
$}. This condition must survive Lorentz transformations. It is straightforward to bring the assumption that it holds in an arbitrary frame, while (\ref{eq:36UD}) does not, to contradiction. Eq.\ (\protect\ref{eq:36UD}) is weaker than condition (\ref{eq:42ML}) we recover in the end, but it suffices for the algebra. 
\\ $\bullet$\ {\em $V_{\mu \nu }(p,k)$ is 4-transverse when multiplied by $U(p,k)$ (as expected).\/} 
Indeed, 
\begin{align} 
k_{\mu }V^{\mu \nu }(p,k) &=\big(
4 \mu_0^2 - k^2 - 4p^2 
+ 8 pk \big) k^{\nu } \to 0, 
\label{eq:37UE} 
\end{align}%
cf. Eqs.\ (\protect\ref{eq:39UH}). 

Following these observations, we can replace, 
\begin{align} 
V^{\mu \nu }(p,k) 
 &\to \frac{k^2g^{\mu \nu }-k^{\mu }k^{\nu }}{3k^2}\, 
V_{\sigma }{}^{\sigma }(p,k) 
\to \frac{4}{3}\big(
k^2g^{\mu \nu }-k^{\mu }k^{\nu }
 \big)\Big(
1 + \frac{2\mu_0^2}{k^2}
 \Big) , 
\label{eq:38UF} 
\end{align}%
where the final result again employs (\ref{eq:39UH}). Thus, 
\begin{align} 
C^{\mu \nu }(k) 
 &= \varepsilon (k_0)\frac{k^{\mu }k^{\nu } - k^2g^{\mu \nu }}{3\pi^2}\,\Big(
1 + \frac{2\mu_0^2}{k^2}
 \Big)\int d^4p
U(p,k) . 
\label{eq:40UJ} 
\end{align}%
The integral here is a scalar. We calculate it in the coordinate frame where 
\begin{align} 
k = \{\sqrt{k^2},{\bf 0}\} . 
\label{eq:41UK} 
\end{align}%
In this frame, 
\begin{align} 
U(p,k) = \theta\Big(
\frac{k^2}{4}
 \Big) \delta\Big(
\frac{k^2}{4}-{\bf p}^2
-\mu_0^2
 \Big) 
\delta\big(
2p_0\sqrt{k^2} \big) . 
\label{eq:42UL} 
\end{align}%
Quantity (\ref{eq:42UL}) is nonzero only if, 
\begin{align} 
\begin{aligned} 
k^2 \geq 4\mu_0^2. 
\end{aligned} 
\label{eq:42ML} 
\end{align}%
This condition is stronger than (\ref{eq:36UD}). The theta-function is thus redundant and may be replaced by unity. With the theta-function gone, the integration is trivial. The result reads, 
\begin{align} 
\int d^4p U(p,k) = \frac{\pi }{2}\,\theta\big(
k^2 - 4\mu_0^2
 \big) \sqrt{1-\frac{4\mu_0^2}{k^2}}\, .
\label{eq:43UM} 
\end{align}%
Putting Eqs.\ (\protect\ref{eq:32TZ}), (\ref{eq:40UJ}) and (\ref{eq:43UM}) together we arrive at Eq.\ (\protect\ref{eq:30PQ}). 

\subsection{Regularization of the commutator}%
\label{ch:OR}
We write \mbox{$
K^{\mathrm{reg}}(x)
$} introduced by Eq.\ (\protect\ref{eq:32PS}) in K\"all\'en-Lehmann style as, 
\begin{align} 
K^{\mathrm{reg}}(x) = \frac{i}{2\pi }\int_{4\mu_0^2}^{\infty} d\mu^2 
K^{\mathrm{reg}}(\mu^2) D(x,\mu^2), 
\label{eq:35PV} 
\end{align}%
where \mbox{$
D(x,\mu^2)
$} is the Pauli-Jordan function of the Klein-Gordon field with 
mass $\mu $, (in units where \mbox{$
\hbar =c=1
$})
\begin{align} 
D(x,\mu^2) &= -i\int \frac{d^4k}{(2\pi )^3}\text{e}^{-ikx}
\varepsilon (k_0)
\delta(\mu^2-k^2) 
\nonumber\\ &
= -\frac{\varepsilon (t)\delta(x^2)}{2\pi } 
+ \frac{\mu \varepsilon (t)\theta(x^2)}{4\pi \sqrt{x^2}}J_1\big(
\mu \sqrt{x^2}
 \big) . 
\label{eq:36PW} 
\end{align}%
The sign of \mbox{$
D(x,\mu^2)
$} is adopted from Itzykson and Zuber \cite{Itzykson}, and the explicit expression is taken from Bogoliubov and Shirkov \cite{Bogol} (where it occurs with opposite sign). 
Recalling the series expansion of the Bessel function, 
\begin{align} 
J_1(y) = \frac{y}{2}\sum_{l=0}^{\infty}\frac{1}{l!(l+1)!}\Big(
-\frac{y^2}{4}
 \Big)^l , 
\label{eq:37PX} 
\end{align}%
we see that the condition, 
\begin{align} 
\int_{4\mu_0^2}^{\infty}d\mu^2 \mu^{2n}K^{\mathrm{reg}}(\mu^2) = 0, 
\quad n = 0,1,\cdots,M+1, 
\label{eq:38PY} 
\end{align}%
warrants that \mbox{$
K^{\mathrm{reg}}(x)
$} is \mbox{$M$} times continuously differentiable everywhere in space-time including the light cone. 

Analyses of conditions (\ref{eq:38PY}) are postponed till appendix \ref{ch:AR}. For the time being, it suffices to know that they amount to the system of linear equations, 
\begin{align} 
\mathcal{A}(2n): & &\sum_{l=0}^{N}(-1)^ld_l\mu_l^{2n} &=0,
 &n &=0,\cdots,M+2,
\label{eq:44QE} 
\\ 
\mathcal{B}(2n): & &\sum_{l=1}^{N}(-1)^ld_l\mu_l^{2n}\ln\frac{\mu_l^2}{\mu _0^2} &=0, &n &=2,\cdots,M+2,
\label{eq:43QD} 
\end{align}%
where the labels \mbox{$
\mathcal{A}
$} and \mbox{$
\mathcal{B}
$} are used to refer to specific equations, and that their solutions stay bounded in the limit, 
\begin{align} 
\begin{aligned} 
\mu _0\ll \mu _1\ll \cdots \ll \mu _N . 
\end{aligned} 
\label{eq:77EP} 
\end{align}%
For details see appendix \ref{ch:AR}. 
\subsection{Regularized response and separation of the divergent constant $ R_0$}%
\label{ch:OA}
One effect of regularization is that, after multiplying (\ref{eq:30PQ}) by the step-function, the latter may be commuted with the differential operator. Indeed, the correction term produced by this commutation is a linear combination of \mbox{$
\delta'(x_0-x_0')K^{\mathrm{reg}}(x-x')
$} and 
\mbox{$
\delta(x_0-x_0')[K^{\mathrm{reg}}(x-x')]'_{x_0}
$} \cite{endSchwT}. Both are zero because \mbox{$
K^{\mathrm{reg}}(x-x')
$} and 
\mbox{$
[K^{\mathrm{reg}}(x-x')]'_{x_0}
$} are, firstly, continuous everywhere, and, secondly, equal zero outside of the light cone and hence on it, {\em including the origin\/}. 
With this observation we can write the regularized susceptibility as, 
\begin{align} 
 &\mu_{\mathrm{vac}}\Pi^{\mathrm{reg}}_{\mathrm{R}\mu \nu }(x-x')
\nonumber\\ &\quad
= -\frac{i e^2 c\mu_{\mathrm{vac}}}{\hbar}\big(
g_{\mu \nu }\Box - \partial_{\mu }\partial_{\nu }
 \big) 
\theta(t-t')
\int \frac{d^4k}{(2\pi )^4}\text{e}^{-ik(x-x')}\varepsilon (k_0)
K^{\mathrm{reg}}\big(
k^2
 \big) 
\nonumber\\ &\quad
= -\big(
g_{\mu \nu }\Box - \partial_{\mu }\partial_{\nu }
 \big) 
\int \frac{d^4k}{(2\pi )^4}\text{e}^{-ik(x-x')}
R^{\mathrm{reg}}\big(
k
 \big) 
, 
\label{eq:47QJ} 
\end{align}%
where 
\begin{align} 
R^{\mathrm{reg}}(k) = 2\alpha \int_{4\mu_0^2}^{\infty}\frac{d\mu^2}{2\pi }\, 
\frac{K^{\mathrm{reg}}(\mu^2)}
{\mu^2-k^2-i 0^+ \operatorname{sign} k_0} , 
\label{eq:49QL} 
\end{align}%
and $\alpha $ is the fine structure constant given by Eq.\ (\protect\ref{eq:83WE}).

Without regularization, quantity (\ref{eq:49QL}) is logarithmically divergent. One subtraction suffices to make it convergent, 
\begin{align} 
R^{\mathrm{reg}}(k) = R_0 + R^{\mathrm{obs}}(k), 
\label{eq:52QP} 
\end{align}%
where \mbox{$
R^{\mathrm{obs}}(k)
$} is given by Eq.\ (\protect\ref{eq:53QQ}), and 
\begin{align} 
 R_0 = R^{\mathrm{reg}}(0) 
= 2\alpha \int_{4\mu_0^2}^{\infty} 
\frac{d\mu^2}{\mu^2}K^{\mathrm{reg}}(\mu^2) 
. 
\label{eq:55QS} 
\end{align}%
In order to calculate $ R_0$ we rewrite this as, 
\begin{align} 
 R_0 = \frac{\alpha }{3\pi }\lim_{M\to\infty}
\sum_{l=0}^N(-1)^ld_l\int_{1}^{M^2/\mu_l^2} 
\frac{dy}{y}F(y) 
, 
\label{eq:58QV} 
\end{align}%
and use the formula, 
\begin{align} 
 &\int_1^{M^2/\mu_l^2}\frac{dy}{y}F(y) 
 = \int_1^{\infty}\frac{dy}{y}\protect\big [ 
F(y) - 1
 \big ] 
+ 
\ln \frac{M^2}{\mu_0^2} - \ln \frac{\mu_l^2}{\mu_0^2}
+ o\Big(
\frac{\mu_l^2}{M^2}
 \Big) . 
\label{eq:57QU} 
\end{align}%
The first two terms here ``perish'' in summation in (\ref{eq:58QV}) due to condition (\ref{eq:44QE}) for \mbox{$
n=0
$}. The third term leads to Eq.\ (\protect\ref{eq:59QW}). 

\subsection{Explicit solution for regularization parameters}%
\label{ch:AR}
\subsubsection{Equations for regularization parameters}%
\label{ch:ARB}
To calculate the integrals (\ref{eq:38PY}), we use the series expansion, 
\begin{align} 
\begin{aligned} 
F(y) &= F_n(y) + O\Big(
\frac{1}{y^{n+1}} \Big), 
\quad y \gg 1 , \end{aligned} 
\label{eq:60DV} 
\end{align}%
where 
\begin{align} 
\begin{aligned} 
F_n(y) &= \sum_{m=0}^{n}\frac{c_m}{y^m}. 
\end{aligned} 
\label{eq:89FB} 
\end{align}%
For the record, 
\begin{align} 
\begin{aligned} 
c_0 &= 1, \quad c_1 = 0, \quad c_2 = - \frac{3}{8}, \qquad
\frac{c_{m+1}}{c_m} = \frac{m(2m-3)}{2(m^2-1)}, 
\quad m\geq 2. 
\end{aligned} 
\label{eq:62DX} 
\end{align}%
Of consequence is only cancellation of \mbox{$c_1$}, because it reduces the number of conditions to be satisfied. 

Consider now the integral, 
\begin{align} 
\begin{aligned} 
\int_1^Y dy\, y^n F(y) &= \int_1^Y dy\, y^n F_{n+1}(y) 
+ \int_1^Y dy\, y^n \protect\big [ 
F(y) - F_{n+1}(y)
 \big ] . 
\end{aligned} 
\label{eq:68ED} 
\end{align}%
The second integral on the rhs here converges as \mbox{$Y\to\infty$}, 
\begin{align} 
\begin{aligned} 
\int_1^{\infty} dy\, y^n \protect\big [ 
F(y) - F_{n+1}(y)
 \big ] \equiv a_n , 
\end{aligned} 
\label{eq:69EE} 
\end{align}%
while the first one is readily evaluated, 
\begin{align} 
\begin{aligned} 
 &\int_1^Y dy\, y^n F_{n+1}(y) 
= \sum_{m=0}^n \frac{c_m}{n-m+1}\big(
Y^{n-m+1} - 1
 \big) 
+ c_{n+1}\ln Y . 
\end{aligned} 
\label{eq:70EF} 
\end{align}%
Using Eq.\ (\protect\ref{eq:68ED})--(\ref{eq:70EF}), Eq.\ (\protect\ref{eq:38PY}) may be written as, 
\begin{align} 
\begin{aligned} 
 &\lim_{\mu \to\infty}\int_{0}^{4\mu ^2}dk^2 k^{2n}K^{\mathrm{reg}}(k^2) 
\\ &\quad
= 
2^{2(n+1)}\lim_{\mu \to\infty}\sum_{l=0}^N(-1)^ld_l 
\protect\bigg \{ 
\sum_{m=0}^n\frac{c_m}{n-m+1}\protect\Big [ 
\mu ^{2(n-m+1)}\mu _l^{2m}-\mu_l ^{2(n+1)}
 \Big ] 
\\ &\qquad
+ c_{n+1}\mu_l ^{2(n+1)}\ln\frac{\mu ^2}{\mu _l^2} + a_n\mu_l ^{2(n+1)} 
 \bigg \} = 0 . 
\end{aligned} 
\label{eq:71EH} 
\end{align}%
Thus the equations (\ref{eq:44QE}), (\ref{eq:43QD}) indeed ensure that \mbox{$
K^{\mathrm{reg}}(x)
$} given by Eq.\ (\protect\ref{eq:32PS}) is $M$ times continuously differentiable. 
Equation \mbox{$\mathcal{B}(2)$} is absent because of \mbox{$
c_1=0
$}.
The total number of conditions (\ref{eq:44QE}), (\ref{eq:43QD}) is \mbox{$
2M+4
$}, so that one needs at least as many regularization masses, (recall that \mbox{$
d_0=1
$}) 
\begin{align} 
\begin{aligned} 
N\geq N_{\mathrm{min}}=2M+4. 
\end{aligned} 
\label{eq:72EJ} 
\end{align}%
With \mbox{$
N> N_{\mathrm{min}}
$} one may impose additional conditions, such as Eq.\ (\protect\ref{eq:81QJ}). In the \mbox{$\mathcal{A}/\mathcal{B}$} nomenclature, this is equation \mbox{$\mathcal{B}(0)$}. The minimal number of regularization masses then increases to \mbox{$
2M+5
$}. 
\subsubsection{The linear system}%
\label{ch:AREL}
\begin{figure*}
\includegraphics[width=\textwidth]{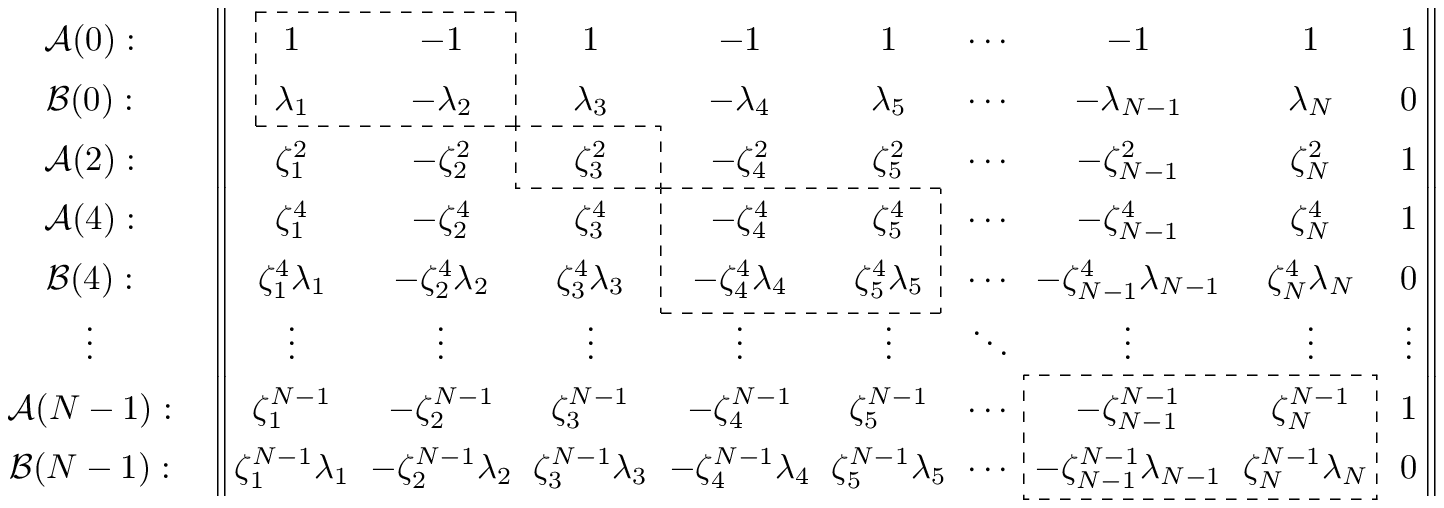}
\caption{The extended matrix of the linear system (\ref{eq:95FJ}); \mbox{$
\zeta _l = \mu _l/\mu _0
$}, \mbox{$
\lambda_l = \ln \zeta _l
$}. The column on the left lists equation labels. When calculating the system determinant in the leading order in condition (\ref{eq:77EP}), elements outside the dashed rectangles may be set to zero.}
\label{fig:matr}
\end{figure*}
With suitable rescaling the system of linear equations (\ref{eq:44QE}), (\ref{eq:43QD}) supplemented by \mbox{$\mathcal{B}(0)$} acquires the form, 
\begin{align} 
\begin{aligned} 
\sum_{l=1}^N\mathcal{M}_{ml}c_{l} = \rho _m, \quad m=1,\cdots,N, 
\end{aligned} 
\label{eq:95FJ} 
\end{align}%
where \mbox{$
\rho_m=0,1 
$}. The elements of the system matrix read, 
\begin{align} 
\begin{aligned} 
\mathcal{M}_{ml} = (-1)^{l+1}\Big(
\frac{\mu _l}{\mu _0}
 \Big)^{\nu_m} \Big(
\ln\frac{\mu _l}{\mu _0}
 \Big)^{1-\rho _m} . 
\end{aligned} 
\label{eq:90FC} 
\end{align}%
The integers \mbox{$
\nu_m
$} and \mbox{$
\rho _m
$} are specified by Eq.\ (\protect\ref{eq:91FD}) below. 

The extended matrix of system (\ref{eq:95FJ}) may be seen in Fig.\ \protect\ref{fig:matr}. The equation labels \mbox{$
\mathcal{A}(0)
$}, \mbox{$
\mathcal{B}(0)
$}, etc., will also be used to refer to rows of the system matrix; the terms $\mathcal{A}$-rows and $\mathcal{B}$-rows are self-explanatory. For the $\mathcal{A}$-rows and $\mathcal{B}$-rows, respectively,
\begin{align} 
\begin{aligned} 
\mathcal{A}(2n): \ \nu _m = 2n,\ \rho _m=1, \qquad 
\mathcal{B}(2n): \ \nu _m = 2n,\ \rho _m=0. 
\end{aligned} 
\label{eq:91FD} 
\end{align}%
For the order of equations as in Fig.\ \protect\ref{fig:matr}, explicit formulae for \mbox{$
\nu _m
$} and \mbox{$
\rho _m
$} as functions of the row number \mbox{$
m
$} may be easily worked out. We have no use for them, because all equations are both fully determined and conveniently referred to by their labels. Of importance is that the way we order the equations in Fig.\ \protect\ref{fig:matr} ensures that, 
\begin{align} 
\begin{aligned} 
\nu _m \geq \nu _{m'}, \quad m>m' . 
\end{aligned} 
\label{eq:92FE} 
\end{align}%
Equality occurs only for adjacent
\mbox{$
\mathcal{A}(2n)$}--\mbox{$\mathcal{B}(2n)
$} pairs ($\mathcal{A}\mathcal{B}$ pairs, for brevity). For them, 
\begin{align} 
\begin{aligned} 
\nu _m=\nu _{m+1} , \quad \rho _m = 1, \quad \rho _{m+1} = 0 , 
\end{aligned} 
\label{eq:3FR} 
\end{align}%
where \mbox{$m,m+1$} are the row numbers occupied by the pair. 

\subsubsection{The system determinant}%
\label{ch:ARED}
\begin{figure}
\begin{center}
\includegraphics{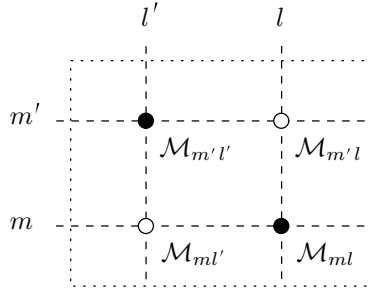}
\end{center}
\caption{Visualisation of Eq.\ (\protect\ref{eq:93FF}). Bold dark and light dots symbolize, respectively, the elements in the numerator and denominator of Eq.\ (\protect\ref{eq:93FF}). The product with the ``diagonal'' choice of the elements (dark dots) is large compared to that with the ``nondiagonal'' choice (light dots), except when the rows with numbers \mbox{$m,m'=m+1$} form an $\mathcal{A}\mathcal{B}$ pair.}
\label{fig:matr1} 
\end{figure}
Hereinafter we assume the limit (\ref{eq:77EP}). 
In the leading order in this condition the linear system in Fig.\ \protect\ref{fig:matr} is easily solved by Cramer's rule. Consider two products of matrix elements contributing to the system determinant, which differ only in the choice of a particular pair of elements: 
\mbox{$
\mathcal{M}_{ml},\mathcal{M}_{m'l'}
$} {\em versus\/} 
\mbox{$
\mathcal{M}_{ml'},\mathcal{M}_{m'l}
$}, where \mbox{$
m>m'
$} and \mbox{$
l>l'
$} (cf.\ Fig.\ \protect\ref{fig:matr1}). Their ratio equals, (ignoring signs)
\begin{align} 
\begin{aligned} 
\left|\frac{\mathcal{M}_{ml}\mathcal{M}_{m'l'}}{\mathcal{M}_{ml'}\mathcal{M}_{m'l}}\right| = \Big(
\frac{\mu _l}{\mu _{l'}}
 \Big)^{\nu_m-\nu _{m'}} \protect\Big [ 
\frac{\ln(\mu _l/\mu _0)}{\ln(\mu _{l'}/\mu _0)}
 \Big ]^{\rho _{m'}-\rho _m} . 
\end{aligned} 
\label{eq:93FF} 
\end{align}%
Under condition (\ref{eq:77EP}), 
\begin{align} 
\begin{aligned} 
\left|\frac{\mathcal{M}_{ml}\mathcal{M}_{m'l'}}{\mathcal{M}_{ml'}\mathcal{M}_{m'l}}\right| \gg 1, 
\quad \nu _m>\nu _{m'}. 
\end{aligned} 
\label{eq:97FL} 
\end{align}%
We assume that the logarithmic factor does not contribute to the scaling \cite{endScale}. E.g., for a geometric sequence of masses, (with \mbox{$Y$} being a large parameter)
\begin{align} 
\begin{aligned} 
\mu _l=\mu _0Y^l, \quad 
\frac{\ln(\mu _l/\mu _0)}{\ln(\mu _{l'}/\mu _0)} = \frac{l}{l'}. 
\end{aligned} 
\label{eq:96FK} 
\end{align}%
Thus the product with the nondiagonal choice of elements (shown by light dots in Fig.\ \protect\ref{fig:matr1}) is small compared to that with the diagonal choice (shown by dark dots). The exception are $\mathcal{A}\mathcal{B}$ pairs of rows, for which \mbox{$
\nu _m=\nu _{m'}
$}. Using this it is straightforward to show that the leading contribution to the system determinant comes from the products of elements encirled by the dashed rectangles in Fig.\ \protect\ref{fig:matr}. All other elements may be set to zero. The determinant of the remaining block-diagonal matrix is easily calculated, resulting in, 
\begin{align} 
\begin{aligned} 
\operatorname{det} \mathcal{M}= Q\protect\Big [ 
\ln\frac{\mu _{2}}{\mu _{1}} + o(1)
 \Big ] , 
\end{aligned} 
\label{eq:98FM} 
\end{align}%
where 
\begin{align} 
\begin{aligned} 
Q = (-1)^{\frac{N-1}{2}}
\frac{\mu _{3}^2}{\mu _0^2}
\prod_{l=2}^{\frac{N-1}{2}} 
\frac{\mu _{2l}^{2l}\mu _{2l+1}^{2l}}
{\mu _0^{4l}}
\ln\frac{\mu _{2l+1}}{\mu _{2l}} . 
\end{aligned} 
\label{eq:99FN} 
\end{align}%
(Recall that \mbox{$
N=2M+5
$} is odd.)
\subsubsection{Expansion in cofactors}%
\label{ch:AREC}
According to Cramer's rule, we have to calculate determinants emerging if replacing the $l$th column of the system matrix by constant terms. Such determinants are sums of cofactors \cite{endCofa} of the $l$th column, where the row index $m$ is limited to $\mathcal{A}$-rows. All matrices resulting from crossing out a column and a row from the system matrix remain subject to condition (\ref{eq:97FL}). Their determinants may be calculated in the leading order in condition (\ref{eq:77EP}) the same way as the system determinant has been. 

The problem simplifies drastically if we are only interested in $d_l$'s which do not vanish in the limit (\ref{eq:77EP}). Crossing out rows and columns reduces the maximal available power of regularization masses, so that the corresponding cofactors are small compared to the system determinant. The exception are cofactors of the first two rows and first two columns of the system matrix. Since the second row of the system matrix is a $\mathcal{B}$-row, only the cofactors of the first row and first two columns remain. We denote them \mbox{$
C^{(11)}
$} and \mbox{$
C^{(12)}
$}. They contribute, respectively, to $d_1$ and $d_2$. All other coefficients vanish in the limit (\ref{eq:77EP}). By the same means as Eq.\ (\protect\ref{eq:98FM}) was obtained we find, 
\begin{align} 
\begin{aligned} 
C^{(11)} = Q\protect\Big [ 
\ln\frac{\mu _{2}}{\mu _{0}} + o(1)
 \Big ] , \quad
C^{(12)} = Q\protect\Big [ 
\ln\frac{\mu _{1}}{\mu _{0}} + o(1) 
 \Big ] . 
\end{aligned} 
\label{eq:1FP} 
\end{align}%
This way, 
\begin{align} 
\begin{aligned} 
d_0 = 1, \quad 
d_1 = 1+d_2, \quad
d_2 = \frac{\ln(\mu _1/\mu _0)}{\ln(\mu _2/\mu _1)} , 
\qquad 
d_l \ll 1, \quad l=3, \cdots, N. 
\end{aligned} 
\label{eq:56XW} 
\end{align}%
Note that boundedness of \mbox{$d_2$} in the limit (\ref{eq:77EP}) is not automatic. For instance, let 
\begin{align} 
\begin{aligned} 
\frac{\mu _1}{\mu _0} = \text{e}^{X},\quad 
\frac{\mu _2}{\mu _1} = X,\quad 
d_2 = \frac{X}{\ln X} . 
\end{aligned} 
\label{eq:63YD} 
\end{align}%
If \mbox{$X\to\infty$}, condition (\ref{eq:77EP}) is satisfied while \mbox{$d_1,d_2\to\infty$}. Boundedness of \mbox{$d_l$'s} is thus an additional condition to be imposed on the regularisation masses. 

\begin{thebibliography}{10}
\bibitem{endF}
With \mbox{$ D_{\text{R}}$} defined by Eq.\ (\protect\ref{eq:59PK}), \mbox{$ j(t) $} is in fact minus the
 driving force. The sign in (\ref{eq:59PK}) is chosen to suit relativistic
 QED, cf.\ appendix \ref{ch:FA}.
\bibitem{endElim}
It goes without saying that no formal transformation can fully eliminate the
 difference between quantum and classical mechanics. There exist both quantum
 systems impossible in classical mechanics (as witnessed by experimentally
 verified violation of Bell's inequalities), and classical systems impossible
 in quantum mechanics (a coherent noiseless quantum amplifier \cite{Caves}, to
 think of one example).
\bibitem{endAvC}
Nonzero average current in the medium signifies either spontaneous symmetry
 breaking or presence of a strong driving field (while \mbox{$J_{\mathrm{e}}$}
 is a weak probe field).
\bibitem{endSobs}
Strictly speaking, Eq.\ (\protect\ref{eq:53QQ}) only follows on removal of regularization. Under
 regularization, the quantity occuring in the integrand should be not
 $F=6\pi K$ but $6\pi K^{\mathrm{reg}}$, cf.\ Eqs.\ (\protect\ref{eq:33PT}) and (\ref{eq:32PS}).
\bibitem{endAlpha}
Not to be confused with the complex amplitude \mbox{$\alpha $} appearing in
 \mbox{Sec.\ \ref{ch:A}}.
\bibitem{endScale}
This is in fact an additional requirement, cf.\ Eq.\ (\protect\ref{eq:63YD}).
\end{thebibliography}

\end{document}